\documentclass[trackchanges]{aastex701}

\usepackage{multirow}
\usepackage{natbib}
\usepackage{amsmath}

\begin{document}

   \title{QUIJOTE scientific results - XX. Commissioning and First Results from the Thirty and Forty Gigahertz Instrument (TFGI)}

    \shorttitle{QUIJOTE-TFGI commissioning \& first results}

    \shortauthors{Fernández-Torreiro, M. et al.}

\newcommand{\LPSC}{\affiliation{Laboratoire de Physique Subatomique et de Cosmologie, 
Universit\'{e} Grenoble Alpes, CNRS/IN2P3, 53 Avenue des Martyrs, Grenoble, France }}
\newcommand{\IACULL}{\affiliation{Instituto de Astrof\'{\i}sica de Canarias, E-38200 La Laguna, Tenerife, Spain}\affiliation{Departamento de Astrof\'{\i}sica, Universidad de La Laguna, E-38206 La Laguna, Tenerife, Spain}}
\newcommand{\UDCCITIC}{\affiliation{Universidade da Coruña, CITIC, Campus de Elviña, E-15071 A Coruña, Spain}}
\newcommand{\KavliTokyo}{\affiliation{Kavli Institute for the Physics and Mathematics of the Universe (Kavli IPMU, WPI), UTIAS, The University of Tokyo, Kashiwa, Chiba 277-8583, Japan}}
\newcommand{\IFCA}{\affiliation{Instituto de Fisica de Cantabria (IFCA, CSIC-UC), Avenida los Castros SN, 39005, Santander, Spain}}
\newcommand{\Imperial}{\affiliation{Imperial College London, Blackett Lab, Prince Consort Road, London SW7 2AZ, UK}}
\newcommand{\CSIC}{\affiliation{Consejo Superior de Investigaciones Científicas, E-28006 Madrid, Spain}}

\newcommand{\maybe}[1]{\textbf{\textcolor{blue}{[#1 (?)]}}}
\newcommand{\TODO}[1]{\textbf{\textcolor{red}{TODO : #1}}}
\newcommand{\NOTE}[1]{\textbf{\textcolor{orange}{NOTE : #1}}}
\newcommand{\changes}[1]{#1} % undefined changes
\newcommand{\HEALPix}{{\tt HEALPix}}

% constants
\newcommand{\kb}{k_{\useplaintextmode{B}}}
\newcommand{\h}{\useplaintextmode{h}}
\newcommand{\lightspeed}{\useplaintextmode{c}}
\newcommand{\e}[1][]{\useplaintextmode{e}^{#1}}

\newcommand{\FWHM}{\useplaintextmode{FWHM}}
\newcommand{\Nhits}{N_{\useplaintextmode{hits}}}

%this alows to use multiples of units
\newcommand{\useplaintextmode}[1]{\ifmmode {\rm #1}\else#1\fi}
\newcommand{\unitmultiples}[1]{%
	\expandafter\newcommand\csname mu#1\endcsname{\ifmmode\,\useplaintextmode{\mu#1}\else\,\useplaintextmode{$\mu$#1}\fi }
	\expandafter\newcommand\csname m#1\endcsname{\,\useplaintextmode{m#1}}
	\expandafter\newcommand\csname c#1\endcsname{\,\useplaintextmode{c#1}}
  	\expandafter\newcommand\csname k#1\endcsname{\,\useplaintextmode{k#1}}
	\expandafter\newcommand\csname M#1\endcsname{\,\useplaintextmode{M#1}}
	\expandafter\newcommand\csname G#1\endcsname{\,\useplaintextmode{G#1}}
    \expandafter\newcommand\csname n#1\endcsname{\,\useplaintextmode{n#1}}
    \expandafter\newcommand\csname T#1\endcsname{\,\useplaintextmode{T#1}}
}

% Parsec
%\newcommand{\pc}[1][]{\ifmmode{\rm #1pc}\else{#1pc}\fi}
\newcommand{\pc}{\useplaintextmode{\,pc}}
\unitmultiples{pc}

%Meter
\newcommand{\m}{\useplaintextmode{\,m}}
\unitmultiples{m}

\newcommand{\s}{\useplaintextmode{\,s}}
\unitmultiples{s}

\newcommand{\yr}{\useplaintextmode{\,year}}
\unitmultiples{yr}

\newcommand{\kms}{\useplaintextmode{\,km\,s$^{-1}$}} % kilometres per second

\newcommand{\Hz}{\useplaintextmode{\,Hz}}
\unitmultiples{Hz}

\newcommand{\K}{\useplaintextmode{\,K}}
\unitmultiples{K}

\newcommand{\V}{\useplaintextmode{\,V}}
\unitmultiples{V}
% stereoradian
\newcommand{\sr}{\useplaintextmode{\,sr}}
% pc / cm^6
\newcommand{\pccmEM}{\useplaintextmode{\,pc\,cm^{-6}}}
% degree (with letters)
% \newcommand{\deg}{\useplaintextmode{\,deg}}

% jansky
\newcommand{\Jy}{\,\useplaintextmode{Jy}}
\unitmultiples{Jy}

% dB
\newcommand{\dB}{\,\useplaintextmode{dB}}
\unitmultiples{dB}

% electronvolts
\newcommand{\eV}{\,\useplaintextmode{eV}}
\unitmultiples{eV}

\author[0000-0002-6805-9100]{M.~Fern\'{a}ndez-Torreiro}
\altaffiliation{Corresponding author: mateo@fernandeztorreiro.com}
\LPSC{} \IACULL{} \UDCCITIC{}
\email{mateo@fernandeztorreiro.com}

\author[]{J.~A.~Rubi\~{n}o-Mart\'{\i}n}
\IACULL{}
\email{jalberto@iac.es}

\author[]{R.~T.~G\'{e}nova-Santos}
\IACULL{}
\email{rgs@iac.es}

\author[]{G.~Pascual-Cisneros}
\KavliTokyo{} \IFCA{}
\email{guillermo.cisneros@ipmu.jp}

\author[]{A.~Fasano}
\IACULL{}
\email{alessandro.fasano@iac.es}

\author[]{F.~J.~Casas}
\IFCA{}
\email{casas@ifca.unican.es}

\author[]{R.~J.~Hoyland}
\IACULL{}
\email{roger.hoyland@iac.es}

\author[]{M.~W.~Peel}
\Imperial{}
\email{m.peel@imperial.ac.uk}

\author[]{C.~H.~L\'{o}pez-Caraballo}
\IACULL{}
\email{lopezcaraballoch@gmail.com}

\author[]{U.~Bose}
\IACULL{}
\email{utsavbose96@gmail.com}

\author[]{R.~Rebolo}
\IACULL{} \CSIC{}
\email{rrl@iac.es}

\author[]{K.~Aryan}
\IACULL{}
\email{kumar.aryan@iac.es}

\author[]{R.~B.~Barreiro}
\IFCA{}
\email{barreiro@ifca.unican.es}

\author[]{R.~Cepeda-Arroita}
\IACULL{}{}
\email{roke.cepeda@iac.es}

\author[]{D.~Herranz}
\IFCA{}{}
\email{herranz@ifca.unican.es}

\author[]{E.~Martínez-González}
\IFCA{}{}
\email{martinez@ifca.unican.es}

\author[]{F.~Poidevin}
\IACULL{}{}
\email{frederick.poidevin@iac.es}

\author[]{R.~Puddu}
\IACULL{}{}
\email{roberto.puddu@iac.es}

\author[]{P.~Vielva}
\IFCA{}{}
\email{vielva@ifca.unican.es}

%% Use the \collaboration command to identify collaborations. This command
%% takes an optional argument that is either a number or the word "all"
%% which tells the compiler how many of the authors above the command to
%% show. For example "\collaboration[all]{(DELVE Collaboration)}" wil include
%% all the authors above this command.
%%
%% Mark off the abstract in the ``abstract'' environment. 
\begin{abstract}
We present the commissioning and first results of the Thirty and Forty Gigahertz Instrument (TFGI), which observes the sky at 31 and 41\,GHz with angular resolutions of $21^\prime$ and $18^\prime$ from the second QUIJOTE telescope at the Teide Observatory. 
Its primary goal is to conduct a deep cosmological survey in selected regions of the Northern sky with high-sensitivity polarization measurements.
The commissioning phase covered from November 2021 to October 2022, during which the instrument operated with a configuration of seven receivers, four at 31\,GHz and three at 41\,GHz. Over this period, approximately 1200\,h of data were acquired. 
Of these, 380\,h were dedicated to calibration sources, used to characterize the instrumental properties of TFGI, including the pointing model, beam response, gain stability, polarimetric performance, and instantaneous sensitivity.
We provide a detailed characterization of these properties and describe how they are being improved for future observing runs.
We use 230\,h of observations from
bright Galactic regions (Cygnus, W43, W44, and W47) to further validate the instrument performance. As an illustrative example, we present the intensity and polarization spectral energy distributions of W44, finding good agreement with existing measurements. 
From the noise map of these observations, we measure a polarization sensitivity of $\sim {8.3}\,\mu$K\,deg$^{-1}$ after an effective observing depth of $0.57\,\mathrm{h}\,\mathrm{deg}^{-2}$. This performance, achieved considering only two detectors at 31\,GHz, is already comparable to that achieved by WMAP (with almost three times the integration time per unit area, $1.61\,\mathrm{h}\,\mathrm{deg}^{-2}$). Extrapolating these results to the full TFGI array, with up to 29 detectors, we show that the instrument is expected to reach the target sensitivity of $\sim 1\,\mu$K\,deg$^{-1}$ at both 31 and 41\,GHz over three cosmological fields covering a total area of $3600\,\mathrm{deg}^2$ after an effective integration time of \changes{5.7} years.

\end{abstract}

%% Keywords should appear after the \end{abstract} command. 
%% PASP uses Unified Astronomy Thesaurus (UAT) concepts:
%% https://astrothesaurus.org
%% You will be asked to selected these concepts during the submission process
%% but this old "keyword" functionality is maintained in case authors want
%% to include these concepts in their preprints.
%%
%% You can use the \uat command to link your UAT concepts back its source.

%% From the front matter, we move on to the body of the paper.
%% Sections are demarcated by \section and \subsection, respectively.
%% Observe the use of the LaTeX \label
%% command after the \subsection to give a symbolic KEY to the
%% subsection for cross-referencing in a \ref command.
%% You can use LaTeX's \ref and \label commands to keep track of
%% cross-references to sections, equations, tables, and figures.
%% That way, if you change the order of any elements, LaTeX will
%% automatically renumber them.

\section{Introduction}
    Observations of the cosmic microwave background (CMB) have granted some of the
    most important cosmological insights of the last decades. From the first detection
    of its anisotropies \citep{smoot1992} and blackbody shape \citep{mather1994} by the 
    COBE satellite or its acoustic peaks \citep[e.g.,][]{deBernardis2000}, its current
    description has been a combined effort of space (WMAP, \citealt{bennett2013} and 
    \textit{Planck}, \citealt{planckPR3}), balloon-borne (e.g. BOOMERANG, \citealt{boomerang},
    or SPIDER, \citealt{spider}) and ground-based (e.g. ACT, \citealt{ACT}, BICEP/Keck,
    \citealt{bicep} or CLASS \citealt{CLASS}) instruments. These discoveries have supported 
    significant scientific findings, summarized on the consolidation of the 
    $\Lambda$CDM model \citep{planck2018VI}.
    
    In addition, the Simons Observatory \citep{simons} is already observing and the
    LiteBIRD \citep{LiteBIRD_PV} satellite is expected to be deployed in the 2030s.
    % several other experiments are planned to be built or deployed in the near
    % future, some of the most important being the LiteBIRD 
    % satellite \citep{litebird} together with Simons Observatory .
    These aim to continue using the analysis of the CMB to fulfill new and exciting 
    scientific prospects, the most promising one being the detection of imprints from
    primordial gravitational waves (GW) in the large-scale domain of the polarized CMB power 
    spectrum \citep{Bmodes}. The figure of merit for this search is the ratio between
    the amplitudes of the tensor and scalar modes from the polarized CMB related to GW during inflation \citep{Bmodes_separation}\footnote{Tensor modes
    can also arise because of lensing of the scalar ones by massive objects along the path 
    of the CMB photons. Separating these modes from the primordial ones, or delensing, is 
    also one of the main concerns from CMB experiments.}. 
    The tightest current upper limit on the tensor-to-scalar ratio is $r < 0.028$, obtained from a combination of CMB data and gravitational-wave interferometry experiments \citep{tensorratio_PV}. Using CMB data alone, the tightest constraint is $r < 0.032$ at 95 per cent confidence level \citep{tristram2022}.

    The main advantage from ground-based experiments over those from space primarily lies with their lower cost, but also on their higher
    flexibility, modularity and scalability. On the other hand, some of their drawbacks
    are additional noise sources such as the atmosphere and ground pick-up. The former 
    increases not only with the depth of atmospheric layers, but also with its turbulence, so high
    altitude observatories with stable atmospheric conditions are preferred
    \citep[e.g.,][]{ngEHT_pwv}. This explains
    the prominence of the Atacama desert, in Chile, and the South Pole as the main
    current ground observatories for CMB observations. However, both of them are located
    in the Southern Hemisphere, which leaves a gap for an instrument in the
    Northern Hemisphere.

    One of the few CMB observatories in the Northern Hemisphere is the Teide Observatory,
    which has hosted experiments since the 1980s. As an example, data from the Tenerife
    instrument were already used to measure the CMB anisotropies confirming COBE results
    \citep{tenerife}. After several other instruments (e.g. COSMOSOMAS, \citealt{cosmosomas},
    and VSA, \citealt{vsa}), the QUIJOTE telescopes were installed at the beginning of
    the 2010s \citep{Rubino10}. Their first instrument, the Multi-Frequency Instrument
    \citep[MFI,][]{mfiwidesurvey} observed the sky between 10 and 20\GHz{} with the main
    goal of improving our understanding of the low frequency foregrounds of the CMB. In
    particular, the accurate characterization of the polarized synchrotron signal 
    \citep{MFIcompsep_pol} carried special interest, as it is a key requirement to
    accurately subtract such a component from a possible B-mode signal.
    
    The MFI was optimized as a true polarimeter from its first design stages in order
    to achieve such a goal: this means that each  detector (horn) was able to produce a pair of polarimetric measurements (normally referred to as $Q$ and $U$), by changing the orientation of the polar modulator (equivalent to a half-wave plate).
%    and then computing their differential measurements. 
This minimizes the impact of
    bandpass mismatch \citep{bandpass_mismatch} and consequently possible leakage from intensity
    to polarization. As the priority from the \textit{Planck} satellite were the intensity 
    measurements, $Q$ and $U$ were computed as differential 
    measurements between detectors. The data would then be post-processed in order to correct 
    for the mismatch. This presented an opportunity for an instrument, observing at 
    similar frequencies as WMAP or \textit{Planck}, completely focused on polarization
    measurements from the Northern Hemisphere. That is the ambition of the Thirty-and-Forty 
    Gigahertz Instrument, TFGI, first introduced in \cite{TGI_SPIE2014} and 
    \cite{TFGIreceivers_artal}. The complete configuration of the TFGI will consist of 29
    detectors observing either at 31 or 41\GHz{}, with 10\GHz{} bandwidth. In this configuration, the TFGI is
    designed to achieve an instantaneous sensitivity of 60$\muK{}\s^{1/2}$. The goal for the 
    instrument was defined as being capable of putting an
    upper limit on the tensor-to-scalar ratio of  $r\leq0.05$. To do so, observations
    with $1\muK\deg^{-1}$ of three different cosmological fields (with $1200\deg^2$ in size
    each) are required.
    
    In this paper we describe the results from the commissioning of the TFGI, which was
    installed in the second QUIJOTE telescope from November 2021 to October 2022. In 
    Sect.~\ref{sec:description}
    we present concise descriptions of the instrument, its acquisition system, the data
    reduction, and the external tools used in the following sections. In 
    Sect.~\ref{sec:commissioning} we elaborate on the commissioning tests that were run to characterize
    the properties of the instrument, such as its beam, gain, polarization
    properties and instantaneous sensitivities. In Sect.~\ref{section:galactic_regions} we
    show additional tests, now run on several bright Galactic regions, in order to cross-check
    the result from the previous analyses (which used data from calibrators and skydips). In
    Sect.~\ref{sec:sensitivities} we present the sensitivities achieved in the noise maps
    obtained from those same regions. We use these to forecast the performance
    that would be achieved by the
    TFGI after its nominal observing run, showing that it would be sufficient to fulfill its objectives.
    Finally, in Sect.~\ref{sec:conclusions} we summarize and present the conclusions from
    the work.

%--------------------------------------------------------------------

\section{Description of the instrument and methodology}
\label{sec:description}

The QUIJOTE CMB experiment \citep{Rubino10} is a collaboration between the Instituto de Astrofísica de Canarias (IAC), the Instituto de Física de Cantabria (IFCA), the Universities of Manchester and Cambridge in the UK, the Departamento de Ingeniería de Comunicaciones (DICOM) at the Universidad de Cantabria, and the IDOM company. QUIJOTE is located at the Teide Observatory (latitude $28\degr18\arcmin04\arcsec$ N, longitude $16\degr30\arcmin38\arcsec$ W, altitude 2400\,m a.s.l). The site offers excellent conditions for radio and millimetre observations, with a median precipitable water vapour (PWV) of 3.5\,mm \citep{pwv_OT}, and has been extensively validated for CMB observations over the past 40 years.

The QUIJOTE experiment \citep{RubinoSPIE12} consists of two identical telescopes with an offset crossed-Dragone design \citep{dragone}, with projected apertures of 2.25 and 1.89\,m for the primary and secondary mirrors, respectively, providing optimal cross-polarization properties. The main difference between the two telescopes is that the mirror surface of the second unit was prepared for observations at higher frequencies (up to 200\,GHz, \citealt{status2016SPIE}). The first QUIJOTE telescope is primarily dedicated to the study of low-frequency \citep[10--20\,GHz;][]{mfiwidesurvey} foregrounds of the CMB (mainly synchrotron, free-free and anomalous microwave emission), while the second telescope is designed for cosmological surveys at higher frequencies. 

The TFGI combines, within a single cryostat, the Thirty-Gigahertz Instrument \citep[TGI;][]{TGI_SPIE2014} and the Forty-Gigahertz Instrument \citep[FGI;][]{TFGIreceivers_artal}. 
The TGI was fully assembled with 29 receivers in 2016, achieving a technical first light on 12 May 2016 \citep{status2016SPIE, Rubino17}. However, observations were soon interrupted due to a cryostat leak. The FGI detectors became available in 2018, and a joint integration of the TGI and FGI receivers into the same cryostat was carried out. Technical observations with a configuration of 15 TGI and 14 FGI detectors were performed in 2019, but operations were suspended during the COVID-19 pandemic. After resolving the cryostat leak problem and carrying out several improvements in the cryostat connectors, a preliminary version of the TFGI with 7 detectors was installed in 2021. This is the reference configuration analyzed in this paper. Currently, a new configuration with 19 detectors (10 from the TGI and 9 from the FGI) is ready to begin observations in 2026.

\subsection{TFGI commissioning phase with 7 receivers}
\label{sec:properties_commissioning}
The observations used in this work were carried out during the commissioning phase of the TFGI, spanning the period from November 2021 to October 2022
    \footnote{Data from June to October 2022 showed a significant degradation
    of the noise properties, so only the data up to May 2022 (inclusive)
    are discussed in this paper.}.
    Only 7 detectors were installed in the focal plane of the instrument during 
    these months, out of a total of 29 available slots.

Table~\ref{table:pixels} 
lists the identification numbers and basic characteristics of the seven pixels.
Pixel 19, observing at 31\,GHz, exhibited multiple issues with one of its low-noise amplifiers (LNA), which prevented any scientific use of its data. 
In this work, we focus on TGI pixels 23 and 26, together with FGI pixel 63. TGI pixel 5 and FGI pixels 41 and 42 exhibited highly variable polarization angles and lower-than-expected polarization efficiencies. These issues are discussed in detail in Sect.~\ref{sec:pol}. 
Although the intensity responses from all pixels were adequate, we choose to present a comprehensive and complete analysis of only those pixels with consistently good overall performance both in intensity and polarization (i.e. pixels 23, 26, and 63), while explaining the origin of the issues affecting the remaining pixels.
    
No calibration diode was installed in the telescope during this commissioning phase. 
As a result, a relative gain calibration could not be performed, which is different from MFI \citep{mfiwidesurvey}. 
Therefore, in this paper we rely on a simplified gain model in which average gains are derived for each month using on-sky calibrators, primarily Tau A (see Sect.~\ref{section:gain_calibration}). This simplified model does not account for day/night variations in the gain, which typically arise from temperature changes in the back-end modules (BEM). Consequently, the overall calibration error for this commissioning campaign is estimated to be around 10 per cent, as detailed in Sect.~\ref{section:gain_calibration}.

    \begin{table}
        \centering
        \caption{Summary of the TFGI pixels installed on the telescope during
        the commissioning phase described in this paper. 
        }
        \begin{tabular}{cccccccc}
\hline \hline
%   FP  & Pixel \#}  & \multirow{2
%   }{*}{FEM} & BEM} & DAS
%   \#} & TGI} & FGI} \\
%   position & & & & & &\\ \hline
  FP & Pixel  & FEM & BEM & DAS  & TGI & FGI     & Used    \\ \hline
   1 & 23 & 11/12 &  8 & 24 & \checkmark &            & \checkmark \\
   2 & 19 &   7/8 &  7 & 23 & \checkmark &            &            \\
   3 & 26 &   5/6 &  4 & 21 & \checkmark &            & \checkmark \\
   4 & 41 & 13/14 &  4 &  4 &            & \checkmark &            \\
   5 & 63 & 15/16 &  5 &  5 &            & \checkmark & \checkmark \\
   6 & 42 & 17/18 &  6 &  6 &            & \checkmark &            \\
   7 &  5 &  9/10 & 11 & 25 & \checkmark &            &            \\ \hline
\end{tabular}
        \tablecomments{Focal plane position
        \#1 corresponds to the central pixel, followed by the first ring of detectors, with the position number increasing clockwise. 
        The Pixel column refers to the physical pixel numbering, while the DAS value corresponds to the numbering assigned in the digital acquisition system.
        FEM and BEM columns indicate the front-end module bias card and the back-end module to which each pixel is connected, respectively. The following two columns specify whether the pixels observe at 31 or 41\,GHz. 
        Finally, the last column identifies the pixels used in this work.}
        \label{table:pixels}
    \end{table}

    \subsection{TFGI data acquisition system (DAS)}
    \label{sect:acquis}
    The TFGI observing system uses series of phase switches to modify the
    polarization states of the channels of the instrument \citep{TGI_SPIE2014, TFGIreceivers_artal}. Similar methods have been already used by previous
    experiments \citep{quiet}, and will be used by future projects at the Teide Observatory \citep[STRIP, ][]{stripjcap}. The difference with respect to a classic half-wave 
    plate 
    is that the phase switches allow
    for much faster changes in the polarization states. These are also normally 
    referred to as phase states and are equivalent to $I\pm Q$ and $I\pm U$,
    so their differences directly trace the polarization parameters\footnote{In the
    reference frame of the instrument. An additional rotation is required to transform
    the values to sky coordinates.}. 
    
    While half-wave plates typically rotate every few seconds \citep[e.g.,][]{SO_HWP}, the TFGI (electronic) phase switches can change polarization states at a
    frequency up to 160\kHz{}. The faster sampling improves the subtraction of $1/f$
    noise when building polarization maps.

    Although the data are sampled at 160\kHz{}, in the TFGI scientific operations described here we switch over the 16 possible phase states at 4\kHz{}, resulting in a complete cycle of all phase states every 4\ms{}, equivalent to a sampling rate from the scientific data of 
    250\Hz{} \citep[see][]{tfgifasano}. 
    Four values are simultaneously saved after that complete cycle, which account for the
    four effective different phase states, namely 0\degr, 90\degr, 180\degr and 270\degr.
    The dependence of the polarization response
    with respect to the phase state changes between the channels of the pixels from 
    the instrument. However, this change between pixels can be corrected \textit{a 
    posteriori} together with the polarization angle. Because of that, and for
    the sake of simplicity, we assume the same equation for the outputs of all
    channels. The output of channel $i$ is described by:
    \begin{equation}
        V_i = \frac{1}{2}\left[I\frac{g_1^2+g_2^2}{2} + g_1g_2(Q\cos{\delta}+U\sin{\delta})\right],
    \label{eq:det_response}
    \end{equation}
    where $g_1$ and $g_2$ are the gains from the LNAs;
    in the following we will assume $g_1=g_2$. $\delta$ is defined as
    $\delta = PS + 2\psi = PS + 2(q - \gamma_{\rm ref})$, where $q$ stands for the
    parallatic angle and $\gamma_{\rm ref}$ is the reference polarization angle to be calibrated for
    each channel. $PS$ takes the values available from the phase states, as
    discussed in the following.
    
    We must highlight an important difference between TGI and FGI acquisition 
    systems, which is the order in which the phase states are recorded into the
    time-ordered data (TOD). The phase switches change the angle of the 
    output in steps of $90\degr$: %, and each of them correspond to a phase state: 
    $0\degr$, $90\degr$, $180\degr$ and $270\degr$. The FGI phase states are 
    stored in this increasing order, but TGI phase states are stored as $0\degr$, 
    $180\degr$, $90\degr$ and $270\degr$.
\changes{In both cases, the total intensity ($I$) can be obtained by combining the four phase-state measurements, thereby cancelling the polarization terms on the right-hand side of Eq.~\ref{eq:det_response}:}
    \begin{equation}
        I \equiv \frac{1}{2}\sum_{PS}V_i . 
    \end{equation}
\changes{In contrast, the Stokes parameters $Q$ and $U$ in the local reference frame of the orthomode transducers \citep[OMT, ][]{TFGIreceivers_artal} can be obtained from the differences between phase-state measurements with phase angles separated by $180^\circ$. These differential measurements also suppress the correlated ($1/f$) noise component owing to the pseudo-correlation architecture of the TFGI. Since the differences are formed within each detector, bandpass leakage is intrinsically mitigated, unlike in experiments where polarization is recovered by differencing signals from independent detectors with different bandpasses, as was the case for WMAP or \textit{Planck}. For TGI, we define these differences, $\Delta_{\rm even}$ and $\Delta_{\rm odd}$, as:}
    % We will directly use the difference
    % between phase states with opposite angles, as polarization measurements
    % directly depend on them:
    % \begin{equation}
    %     \Delta_{\rm even}=(PS_1-PS_3) \quad, \quad
    %     \Delta_{\rm odd}=(PS_2-PS_4);
    % \end{equation}
    \begin{equation}
        \Delta_{\rm even}=V_i(PS_1)-V_i(PS_3) \quad, \quad
        \Delta_{\rm odd}=V_i(PS_2)-V_I(PS_4);
    \end{equation}
    where $(PS_1, PS_2, PS_3, PS_4)\equiv(0\degr, 90\degr, 180\degr, 270\degr)$.
    The calculation of $Q$ and $U$ from $\Delta_{\rm even}$ and $\Delta_{\rm odd}$
    is detailed in the next Sect.~\ref{sec:mapmaking}.
    The intensity, on the other hand, is computed as half the sum of the four
    phase states, so their storage order is irrelevant. Further details are
    available in \cite{tfgifasano}.

\subsection{Data flagging and map-making}
\label{sec:mapmaking}

The TFGI data processing pipeline inherits many of the characteristics of the MFI pipeline \citep{mfiwidesurvey, mfipipeline}. As a first step, all time-ordered data (TOD) are binned into 40\,ms samples. This binning allows a variance to be assigned to each sample, which is then used to define a weight (given by the inverse of the variance). A basic data flagging procedure, similar to that used for the MFI, is then applied by default to all observations. This includes flags based on voltage ranges, housekeeping parameters, and the emission from the Sun and the Moon, using a $10^{\circ}$ exclusion radius. After this, the TOD are evaluated to discard data with noise properties below an acceptable threshold.
    
After applying the basic flagging, 
a constant median value per scan is subtracted from the data in order to remove
the average system temperature contribution. This procedure also mitigates large-scale noise structures arising from atmospheric emission. We then apply two additional flagging steps.
The first one worked at the subscan level. QUIJOTE performed raster scans which consist of continuous back-and-forth drifts in azimuth, named subscans, at different elevation steps.
Our first flagging procedure took the RMS values computed along each subscan and compares them to a reference value. The subscans with RMS estimates higher than that reference value were removed. The second method took the final, individual maps for the different fields and placed a ``control'' aperture where no strong emission was present. We computed the RMS at the map level within that aperture, and built the distribution of values considering all the available observations for each field. We discarded those maps with RMS values below 0.5 and above 1.7 times the median RMS from such distribution.
Once we had run both flagging procedures, we calibrated the data (at the TOD level) using the values obtained in Sect.~\ref{section:gain_calibration}. 

Map-making for the TFGI data is applied in two different ways, depending on whether intensity ($I$) or polarization ($Q$, $U$) maps are being produced. For the polarization maps, a simple “naive” map-making approach is sufficient because of the small \changes{residual} contribution of $1/f$ noise in the polarization timelines (see Sect.~\ref{sec:sensitivites}); baselines are removed from the data using a median filter and the resulting timelines are projected onto the sky to produce $Q$ and $U$ maps according to the following equations: 
    $$
        a = \sum_i\left(\frac{\cos(\psi_i)^2}{\sigma_{{\rm even}, i}^2} + \frac{\sin(\psi_i)^2}{\sigma_{{\rm odd}, i}^2}\right),
        \quad
        b = \sum_i\left(\frac{\sin(\psi_i)^2}{\sigma_{{\rm even}, i}^2} + \frac{\cos(\psi_i)^2}{\sigma_{{\rm odd}, i}^2}\right),
        \quad
        c = \sum_i\cos(\psi_i)\sin(\psi_i)\left(\frac{1}{\sigma_{{\rm even}, i}} - \frac{1}{\sigma_{{\rm odd}, i}}\right);
    $$
    $$
        f = \sum_i\left(\frac{\cos(\psi_i)\Delta_{{\rm even},i}}{\sigma_{{\rm even}, i}^2} - \frac{\sin(\psi_i)\Delta_{{\rm  odd},i}}{\sigma_{{\rm odd}, i}^2}\right),
        \quad 
        g = \sum_i\left(\frac{\sin(\psi_i)\Delta_{{\rm even},i}}{\sigma_{{\rm even}, i}^2} + \frac{\cos(\psi_i)\Delta_{{\rm  odd},i}}{\sigma_{{\rm odd}, i}^2}\right);
    $$
    \begin{equation}
        \left(\begin{matrix}
            Q \\
            U
        \end{matrix}\right) = \frac{1}{ab-c^2}\left(\begin{matrix}
            b & -c \\
            -c & a
        \end{matrix}\right)\left(\begin{matrix}
            f \\
            g
        \end{matrix}\right),
    \end{equation}
    where $\sigma_{{\rm even},i}$ and $\sigma_{{\rm odd},i}$ are the weights in the binned TOD
    for $\Delta_{{\rm even},i}$ and $\Delta_{{\rm odd},i}$. In the case that the weights
    for both coincide, we recover the simplest case scenario:
    \begin{equation}
            Q = \sum_i \left(\frac{\cos{(\psi_i)\Delta_{{\rm even}, i}} 
                - \sin{(\psi_i)\Delta_{{\rm odd}, i}}}{\sigma_i^2}\right),
            \quad
            U = \sum_i \left(\frac{\sin{(\psi_i)\Delta_{{\rm even}, i}} 
                + \cos{(\psi_i)\Delta_{{\rm odd}, i}}}{\sigma_i^2}\right),
    \end{equation}
    where the rotation defined by the difference $\psi=q_i - \gamma_{\rm ref}$
    transforms our data from the instrument to the sky reference frame. These equations are a generalization of the MFI equations \citep{tesisalba,destriper} for the case of two polarization timelines. 
    
However, the intensity maps require the use of a destriper map-making algorithm, due to the stronger correlated $1/f$ noise in the intensity timelines and the potential artifacts in baseline determination caused by the high brightness of the sky emission. For this purpose, we employed an adapted version of the PICASSO destriper code \citep{destriper}, previously used for QUIJOTE-MFI, which takes the calibrated TODs as direct inputs. 
In the TFGI implementation, the sum of the four phase states (divided by two to recover the total intensity) is treated as a single timeline by the PICASSO code. 
\changes{Using numerical simulations, we verified that the configuration parameters adopted for MFI, in particular a baseline length of 2.5\,s, also provides satisfactory performance for TFGI data. A more detailed discussion of the destriper configuration and its implementation for TFGI polarization map-making is deferred to future work (Bose et al., in prep.).}

\changes{As demonstrated for the QUIJOTE MFI instrument, the PICASSO destriping code preserves the sky signal over the relevant angular scales ($20 < \ell < 200$), with a signal error below 0.001\,\%. Since the same implementation, adapted for TFGI, is used in this work, no additional transfer-function correction is required beyond that already validated for the MFI map-making pipeline.}
\changes{Consequently, the TFGI instrument transfer function is expected to be accurately recovered on angular scales of $20^\prime$ and larger, which are those of primary interest for the present analysis. An example illustrating the recovery of diffuse emission with this configuration is presented in Sect.~\ref{sec:recovery_diffuse_emission}.}

%\changes{We leave for future work the optimization of the map-making parameters used in the destriper algorithm (Bose et al., in prep.), based on numerical simulations. Although, preliminary work has shown that the same parameters adopted for the MFI data reduction grant satisfactory performance. \cite{destriper} showed that the map-making recovers the scales between $20<l<200$ with a signal error lower than 0.001\%. When comparing our input (raw) and output maps from the map-making code, there are not apparent changes, aside for the stripes the algorithm is tailored to reduce. Such result probes that the transfer function from the instrument works fine at 20$^\prime$ and larger angular scales, which are the ones we are more interested in. These parameters include a baseline length of 2.5\s{}, together with 40\ms{} binning. An example of the recovery of diffuse emission with this configuration is shown in Section~\ref{sec:recovery_diffuse_emission}.}

In Table~\ref{tab:obs_time}, we show the objects and fields observed during the commissioning 
period with their total observing time. We also quote the total number of scans and the percentage 
of flagged data for the objects described in this paper. We only considered the months for
which we computed calibration estimates (Sect.~\ref{section:gain_calibration}): November 2021 
to February 2022, and then April and May 2022. We have obtained more than 1200\,hours of 
astrophysical data during this part of our commissioning period, with an observing
efficiency of 35\,\%. We stress that this number refers to a preliminary observation 
period, and will increase once the survey observations begin. 

    \begin{table}[]
        \centering
        \caption{Objects and fields observed by the TFGI during its commissioning period.}
        \begin{splittabular}{ccccccccBccccccccc}
    \hline \hline
    Object & 3C273 & 3C274 & 3C279 & TAU A & CAS A & JUPITER & LOCALMAP & Object & M42 & MOON & SKYDIPS & PERSEUS & VENUS & W44 & W63 & Total \\ \hline
    $t_{\rm obs}$ (hours) & 33.7 & 133.9 & 65.3 & 179.1 & 58.9 & 92.4 & 2.5 & $t_{\rm obs}$ (hours) & 54.1 & 6.2 & 67.2 & 304.3 & 48.0 & 201.2 & 27.8 & 1207.4 \\
    Data flagged (\%) & ---& ---& --- & 26.9 & --- & --- & --- & Data flagged (\%) & --- & --- & --- & --- & --- & 22.2 & 13.4 & --- \\ 
    \hline
\end{splittabular}
        \tablecomments{Commissioning data acquired between November 2021 and May 2022.
        We only quote the percentage of 
        flagged data for those sources whose data is described in this paper (Tau A, W44
        and W63). We leave the detailed analysis of the rest of the regions for future work.}
        \label{tab:obs_time}
    \end{table}

    \subsection{SED MCMC fitting}
    \label{section:SED_AP}
        After describing the tests performed during the commissioning phase to verify the correct functioning of the instrument in Sect.~\ref{sec:commissioning}, we present in Sect.~\ref{section:galactic_regions} the \changes{analysis of the Galactic region W44, and in Appendix~\ref{appendix:all_SED_regions} similar ones for W43 and W47}. These analyses also serve as cross-checks that further validate the instrument performance. 
        In particular, we derive spectral energy distributions (SEDs) by combining TFGI data with more than 30 ancillary survey maps spanning a wide range of frequencies, both at $30\arcmin$ and $1\degr$ scales. The external data used in these
        analyses are described in Appendix~\ref{appendix:SED_data}.
        
        Those photometry estimates for the SEDs in Sect.~\ref{section:galactic_regions}
        were computed in a similar way to previous papers, using the aperture photometry (AP) method \citep[e.g.,][]{planckAMEsources, MP_Planck, Perseus, Taurus, AMEwidesurvey, snrwidesurvey, M31quijotemfi}. 
        This method consists of integrating the signal within a circular aperture around the
        studied object, and then subtracting the signal from the surrounding region to
        estimate the impact of the background. We used a circular aperture with radius 
        $r_1=60^\prime$ and a ring with radii $r_2,r_3=80^\prime,100^\prime$, 
        respectively. We chose these values to keep consistency with the analyses from \cite{W44} and \cite{perseus_w43_rhooph}.
        
        In the case of the intensity SEDs, the set of flux density values were fitted to the same
        parametric model described in Section 3 of \cite{ameplanewidesurvey}, which consisted
        of synchrotron (only for W44), free--free, anomalous microwave emission, thermal
        dust and CMB anisotropies. In the case of polarized intensity ${(P=\sqrt{Q^2+U^2})}$,
        % \footnote{The only object with a polarized 
        % counterpart is W44, for which we fitted its polarized intensity, $P=\sqrt{Q^2+U^2}$.},
        the data were fitted to the combination of synchrotron and thermal dust, as polarized 
        free-free, AME and CMB components are expected to be negligible. For synchrotron we 
        used the same model as in the intensity case, while for thermal dust we replaced the 
        modified blackbody model by another toy power law. This was done because we do not have 
        polarized data above the \textit{Planck} 353\GHz{} band, and thus we were not able to
        properly reconstruct the peak of the thermal dust emission, i.e. its emissivity and/or 
        temperature.
        
%--------------------------------------------------------------------

\section{Commissioning tests}
    \label{sec:commissioning}
    In this Section we review the tests run on the TFGI data to ensure that it achieved a stable
    observing configuration. We also describe in detail the properties of the acquired data in
    order to perform astrophysical analyses with it. A summary of the results from these
    tests is shown in Table~\ref{tab:TFGI_summary}.

    \begin{table*}[]
        \centering
        \caption{Summary of the basic TFGI performance parameters derived from the  
        commissioning tests.}
        \begin{tabular}{lccc}
\hline
                         Parameter  & Pixel 23 & Pixel 26 & Pixel 63 \\ \hline
    Nominal central frequency (GHz) &       31 &       31 &       41 \\
                    Bandwidth (GHz) &       10 &       10 &       12 \\
        %   Pointing offset (arcmin) &     0.39 &     0.36 &     0.79 \\
           % Pointing offset (arcmin) &     0.44 &     0.75 &     0.22 \\
                %  Beam FWHM (arcmin) &     20.4 &     20.3 &     16.4 \\
                 % Beam FWHM (arcmin) &     21.6 &     21.7 &     18.0 \\ % mambrinobefore2022
                 Beam FWHM (arcmin) &     22.0 &     21.9 &     18.2 \\
%              Main beam solid angle, 
%   $\Omega_{\rm mb}$ ($10^{-5}\sr$) &     4.00 &     3.95 &     2.58 \\
             Main beam solid angle, 
   $\Omega_{\rm mb}$ ($10^{-5}\sr$) &     4.51 &     4.54 &     3.11 \\
            %   Beam ellipticity, $e$ &    0.100 &    0.043 &    0.011 \\
              % Beam ellipticity, $e$ &    0.079 &    0.063 &    0.107 \\
              Beam ellipticity, $e$ &    $<0.091$ &    $<0.066$ &    $<0.165$ \\
  %     Antenna sensitivity, $\Gamma$ 
  % (${\rm \mu K}_{\rm CMB}\Jy^{-1}$) &      ? &      ? \\
     White-noise level in timelines 
         ($\muK_{\rm CMB}\s^{1/2}$) &      314 &      365 &     376 \\
        %  Knee frequency $f_{\rm K}$ 
        %       in intensity (mHz)    &          &          &         \\
        % $1/f$ slope in polarization &          &          &         \\
        %   Intensity-to-Polarisation 
        %               leakage (\%) &     <0.30 &     <0.42 &    <0.63 \\ \hline
          Intensity-to-Polarisation 
                       leakage, $\alpha$ (\%) &     $<0.19$ &     $<0.19$ &    $<0.36$ \\ \hline
                    %   leakage (\%) &  $0.36\pm0.30$ & $0.39\pm0.29$ & $0.54\pm0.50$ \\ \hline
\label{table:tfgi_parameters}
\end{tabular}
        \label{tab:TFGI_summary}
    \end{table*}
    
    \subsection{Pointing model}
         The telescope requires a model that transforms the local coordinates from
         its azimuth and elevation encoders to the correct on-sky position. In order to do so, we observed
         several objects with well-known positions and accounted
         for the differences between those expected positions and the ones actually
         recorded. 
         
         We used the 7-parameter pointing model described in \cite{wallace2008}, 
         in the same way that it was done for the first QUIJOTE telescope for the MFI 
         \citep{mfiwidesurvey} and KISS \citep{kiss} instruments. We fitted a Gaussian to the observation of the source in the TOD.

         We used a Markov Chain Monte Carlo (MCMC) maximum likelihood estimator (MLE) with
         {\tt emcee} \citep{emcee} to search iteratively for the best-fit parameters of 
         the pointing model that minimize the difference between the simulated
         and the real data. The FWHM values were fixed to those expected from the 
         telescope properties: 17 and 21\,arcmin.

         We began fitting the model to observations from the Moon. With that first version
         of the model, we further refined it by simultaneously fitting the data from
         Tau A, Cas A, Venus and Jupiter. We also fitted their amplitudes, in contrast
         with the Moon analysis where we simply normalized the real TOD. The final values 
         are shown in Appendix~\ref{appendix:pointing_model_parameters}.

\subsection{Beam characterization}
\label{sec:beam}

A detailed comparison between the TFGI beam profiles and the predicted numerical models is deferred to future work. In this commissioning phase, we focus on confirming two basic main-beam parameters: widths and ellipticities. For this, 
we fitted different versions of Tau A maps to a 2D elliptical Gaussian, using a MCMC MLE code.

        We used co-added maps from different data combinations, ranging from individual Tau A observations to monthly maps, and finally to the full dataset from November 2021 to May 2022, amounting to 131\,hours %145\,hours
        of observations (Table~\ref{tab:obs_time}).
        In all cases, we subtracted in quadrature the intrinsic size of the observed
        source ($\sim6\arcmin$) to the fitted FWHM value. 

\begin{figure}
            \centering
            \includegraphics[width=0.65\linewidth, clip, trim=1.7cm 0 0 0]{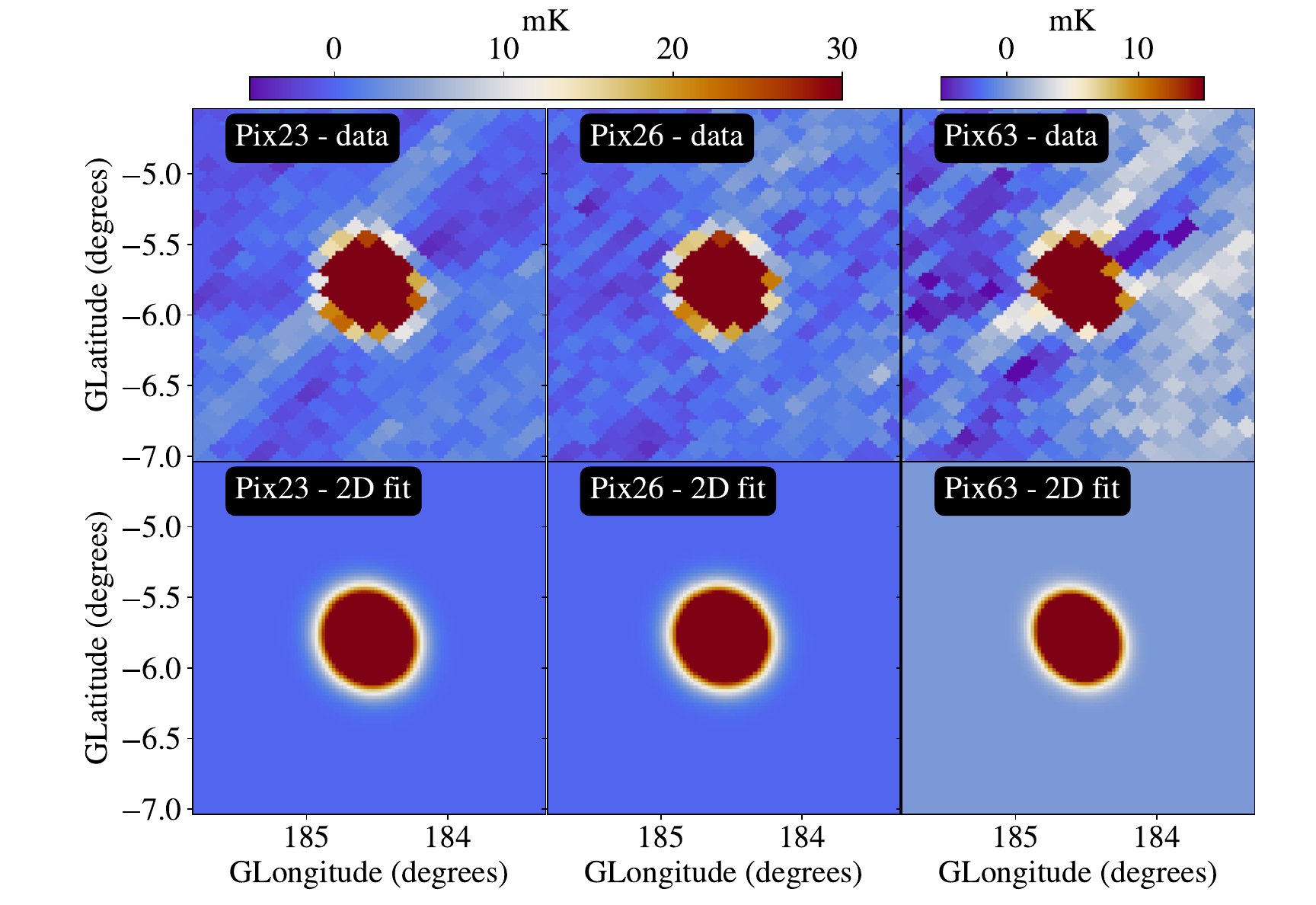}
            \caption{Left column: Maps of Tau~A obtained by combining all observations from November 
            2021 to May 2022. Right column: Best-fitting elliptical beam model. From top to bottom, pixels 23, 26 and 63, computed as the
            average of their four channels. The beam is slightly narrower for FGI pixel 63 because of its higher frequency, 
            as expected. }
            \label{fig:beams}
        \end{figure}

Monthly maps of the Tau~A observations yield characteristic ellipticities in the range 0.01--0.1. These strong variations are most likely due to residual striping in the maps caused by $1/f$ noise from some observations. As an example of this, Fig.~\ref{fig:beams} shows the co-added maps of all Tau~A observations, where low-level $1/f$ stripes along the scanning direction are visible for pixels 23 and 63. The corresponding fitted ellipticities from these maps are 0.091, 0.066, and 0.165 for pixels 23, 26, and 63, respectively. However, for pixel 63 we obtain significantly lower ellipticity values for some months; for example, we find 0.011 when using only the November data maps. The values reported in Table~\ref{tab:TFGI_summary} are therefore quoted as upper limits derived from the combined maps in Fig.~\ref{fig:beams}, and will be refined with additional observations in future campaigns. 
        
Numerical simulations using GRASP\footnote{The GRASP software was developed by TICRA (Copenhagen, DK) for analysing general reflector antennas (\url{http://www.ticra.com}).} indicate that the expected main-beam ellipticities for the TFGI in the second QUIJOTE telescope are of the order of $e \approx 0.01$ for the central pixel, and $e \la 0.02$ for the other six pixels analysed here. Indeed, both QUIJOTE telescopes have identical optical configurations, and the measured beam ellipticites for the MFI instrument range between 0.01 and 0.04 \citep{mfiwidesurvey}. On top of that, Tau A shows an intrinsic elongation that could be responsible for part of the measured ellipticity of the beam. Its $7^\prime\times5^\prime$ size implies intrinsic ellipticities $e_0$ of $0.023$ and $0.034$ for the TGI and FGI cases, respectively.

Finally, we also investigate the potential impact of temporal variations of the pointing model parameters on the measured beam properties, as we detect small systematic offsets between the reconstructed Tau~A location in jack-knife maps (Sect.~\ref{section:gain_calibration}). 
        However, the limited number of Tau A observations  during the final months of the commissioning period prevents a robust assessment. We nevertheless estimated the impact of such pointing offsets on our photometric measurements. 
        Even an offset of $3^\prime$, which is about twice the largest offset measured, induces only a $\sim$2\,\% error, well below the uncertainty associated with our gain model (10\,\%, see Sect.~\ref{section:gain_calibration}). Thus, we can safely use the current pointing model solution for the following analyses.

    \subsection{Gain calibration}
    \label{section:gain_calibration}
        As mentioned in Sect.~\ref{sec:properties_commissioning}, a 
        thermally stabilized calibration diode was used during the QUIJOTE-MFI observations in order to trace the relative gain variations with respect to the mean gain in a given observing period
        \citep{mfiwidesurvey}. We did not have such a diode during the TFGI
        commissioning phase described in this work, so these variations
        were not traced as regularly. Instead, we used monthly coadded maps from Tau A (each accounting between 20--45\,hours of observations, exact values
        in Table~\ref{tab:obs_time_monthly_TauA}) to calibrate
        our response. Thus, here we assume that the gain was stable during these monthly periods. A more precise gain model will be elaborated in future observations once the calibration diode is in place.

        \begin{table}[]
            \centering
            \caption{Observation times used to produce the monthly maps of Tau A, which are then used to 
            estimate the average gain. November and December 2021 show a similar percentage of flagged data,
            while this percentage increases during January and February 2022. April and May 2022 again show good values.}
            \begin{tabular}{ccc}
    \hline \hline
    Month  & $t_{\rm obs}$ (hours) & Data flagged (\%) \\ \hline
    November 2021 & 19.4 & 24.5 \\
    December 2021 & 26.4  & 17.1 \\
    January 2022 & 29.5 & 57.6 \\
    February 2022 & 36.5 & 41.2 \\
    April 2022 & 25.2 & 11.3 \\
    May 2022 & 43.9 & 10.2 \\ \hline
    Total & 179.1 & 26.9 \\
    \hline
\end{tabular}

            \label{tab:obs_time_monthly_TauA}
        \end{table}
        
        We worked with the maps at the 
        original resolution, and applied beam fitting (BF) photometry on them.
        In order to do so, an ellipse in Galactic coordinates with the shape fixed to that obtained
        in Sect.~\ref{sec:beam} was fitted to the Tau A monthly maps,
        fitting its amplitude as the only free parameter.
        
        We calibrated this result with the model from \cite{mfiwidesurvey}:
        \begin{equation}
            S_\nu = 358.3\left(\frac{\nu}{22.8\GHz{}}\right)^{-0.297}\Jy{},
        \end{equation}
        for $t_0=2016.3$. We took into account the secular decrease
        factor obtained by \cite{weiland2011}, which is $-0.218\,\%\,$year$^{-1}$,
        to transfer the model to the TFGI epoch ($t=2022.3$).

        \begin{figure}
            \centering
            \includegraphics[width=0.65\linewidth, clip, trim=1.7cm 0 0 0]{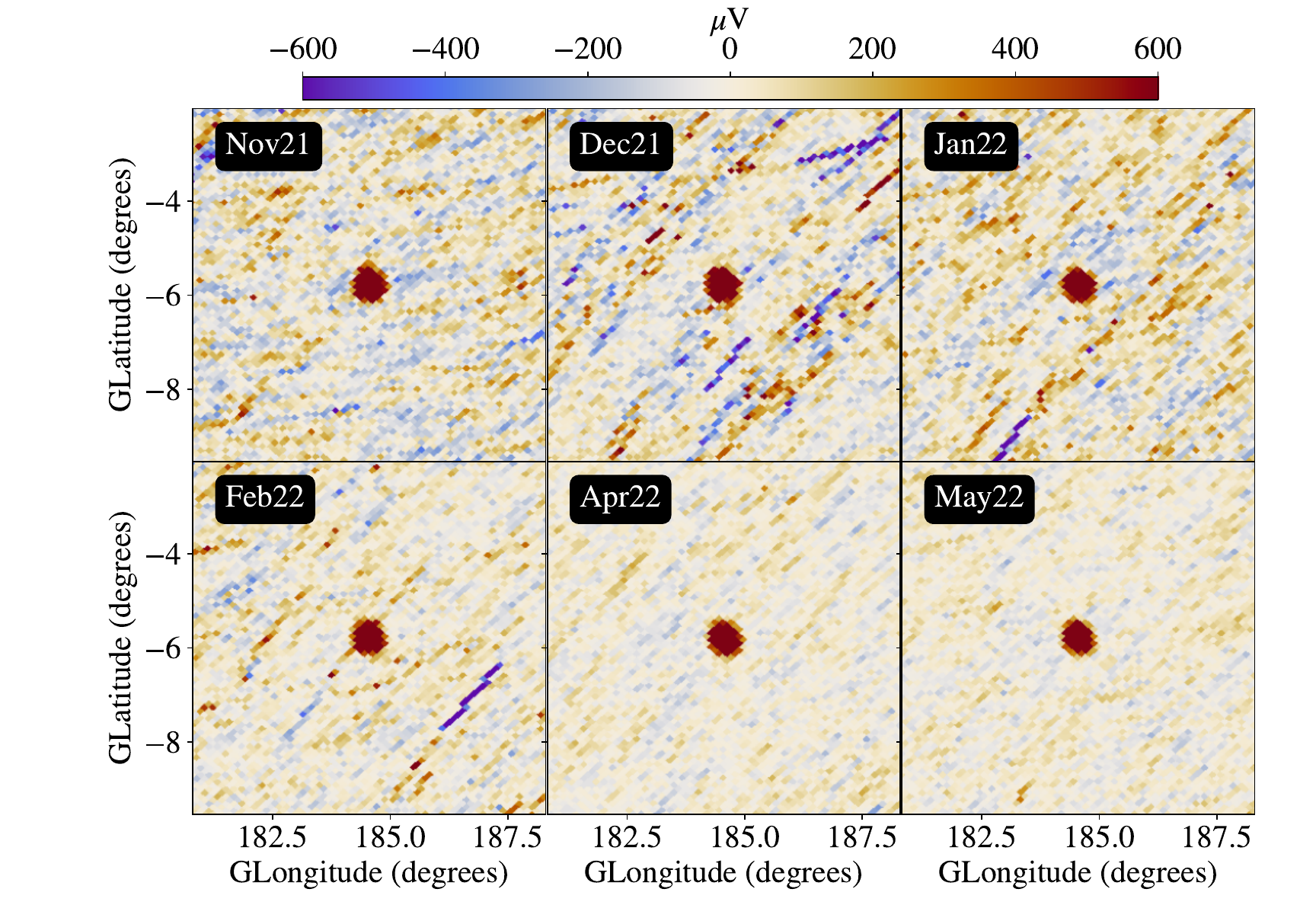}
            \caption{Monthly Tau A maps from November 2021 to May 2022 used to study the temporal gain variations of the instrument. The maps correspond to channel 1 of pixel 23 at 31\,GHz. They are shown in voltage units ($\mu$V), as no absolute calibration has been applied here. }
            \label{fig:gain_calibration_maps}
        \end{figure}
        
\begin{figure}
            \centering
            \includegraphics[clip, trim=0 0 0 4.35cm, width=0.49\linewidth]{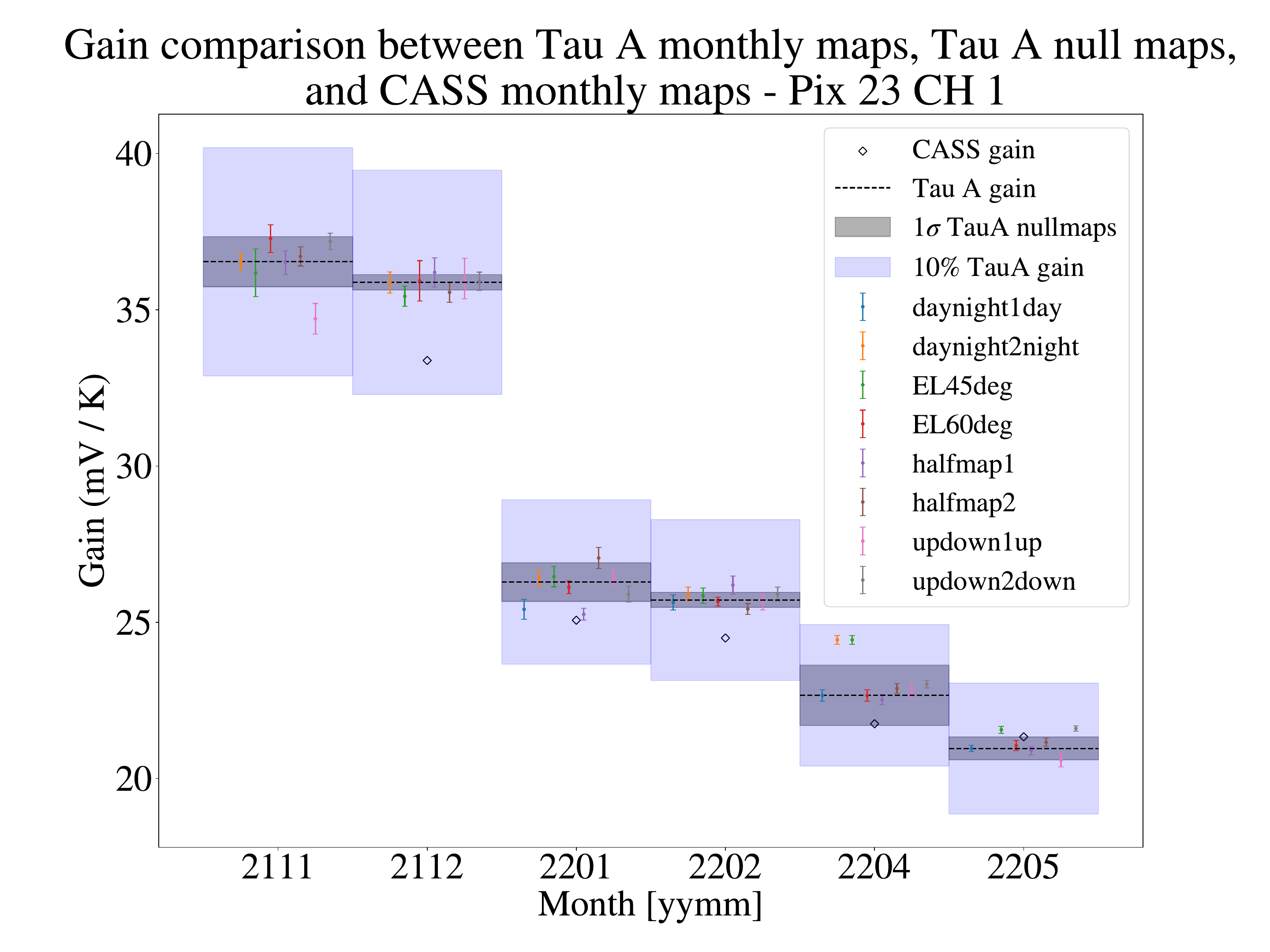}
            \includegraphics[clip, trim=0 0 0 4.35cm, width=0.49\linewidth]{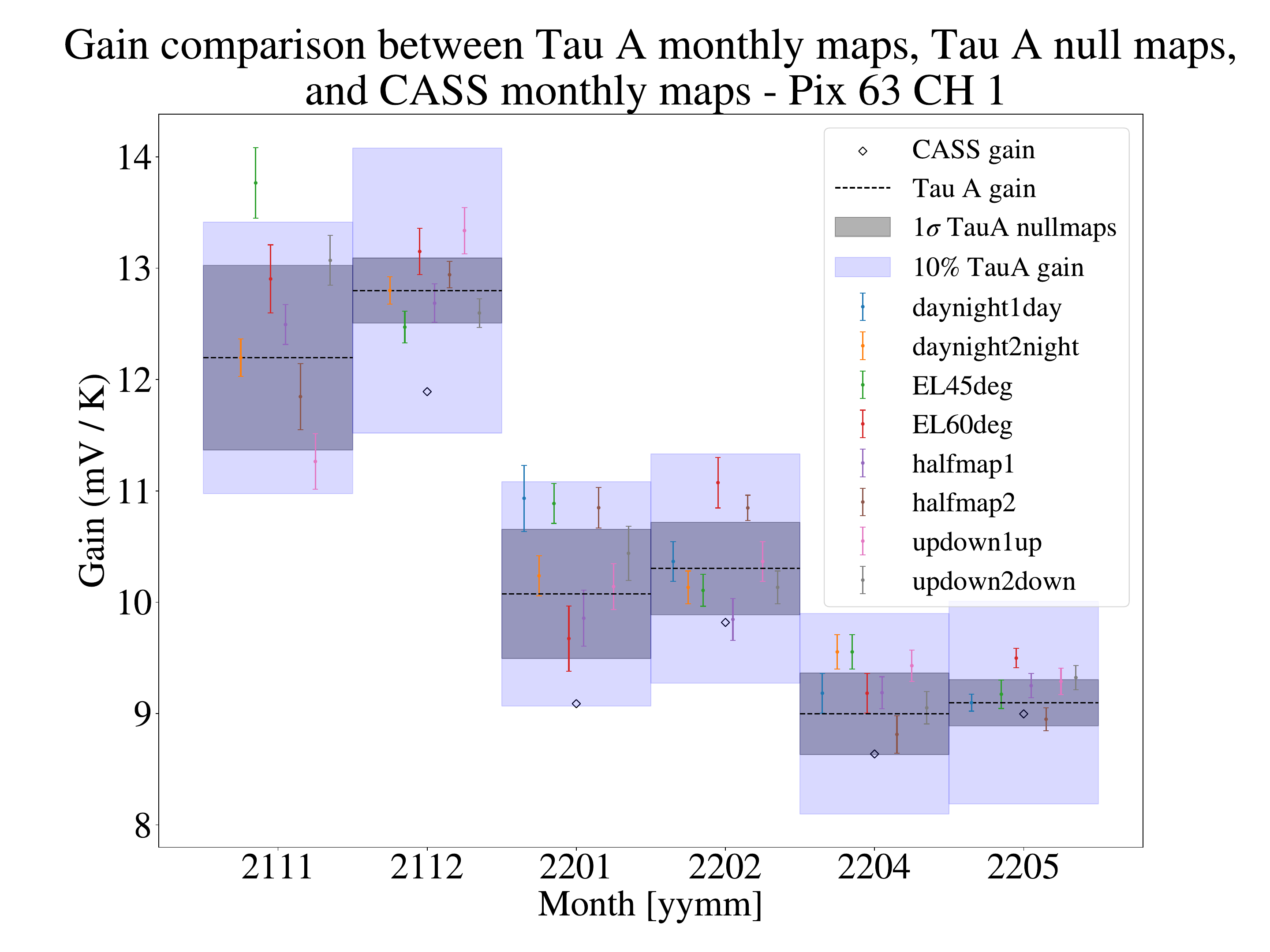}
            \caption{Dependence of the gain with time for the first channel of TGI pixel 23 (left panel)
            and FGI pixel 63 (right panel). 
            We show the gain value computed from the Tau A map taking into 
            account all monthly measurements (after flagging) as dashed horizontal
            bars. The colored points show the values computed from data split maps for Tau A (see text), while the filled
            black regions show the standard deviation from those points. Finally, the
            gain factors computed from CasA data are also shown, and the filled blue regions
            show the level of the assumed calibration uncertainty, 10\,\%. See 
            Sect.~\ref{section:gain_calibration} for a detailed discussion on
            this value.}
            % We see two different behaviors, before and after the beginning of 2022.}
            \label{fig:gain_calibration_drift}
        \end{figure}

        We show the Tau A monthly maps for the first channel of pixel 23 in 
        Fig.~\ref{fig:gain_calibration_maps}, and in 
        Fig.~\ref{fig:gain_calibration_drift} we show the gain estimates computed from 
        those maps. In general, we see a trend of decreasing gain with time.
        
        In Appendix~\ref{appendix:all_gain_calibration} we show the monthly gain
        calibration values for all the pixels. 
        To estimate the uncertainty on these calibration factors, we constructed a set of jack-knife map pairs by reorganising the Tau~A database into two independent subsets, and deriving the calibration factor for each subset separately.
        In particular, the data were split in four different ways: according to the time of observation (daytime, UTC 07:00–19:00, or night-time), the source elevation at the time of observation (typically 45\degr\ or
        60\degr\ in our observing schedule), the direction of source motion (rising or setting), and, finally, by alternating observations in time. We computed the gain for each split map following the same procedure used for the full monthly map
        (we show them in Fig.~\ref{fig:gain_calibration_drift} as horizontal bars). 
        We then computed the standard deviation among the eight split maps for each month as an estimate of the uncertainty. We find median and mean uncertainties across all months and channels
        of 2.4 and 3.2\,\%, respectively. We detected a larger value for FGI pixel 63 (median 3.9\,\%)
        when compared to TGI pixels 23 and 26 (2.1 and 2.0\,\%, respectively). 
        
        Furthermore, we also computed the gain calibration factor from monthly Cas~A maps; however, there were not enough Cas A observations as to generate all the jack-knife map pairs discussed above. We compared the calibration factors derived from Tau~A and Cas~A\footnote{For Cas A we used the 
        power-law model from \cite{weiland2011}, assuming a secular decrease of 
        $-0.54\,\%\yr{}^{-1}$.} 
        in order to 
        account for uncertainties in the source models, as well as variations arising from the different observing conditions (due to the different RA of the two sources, since Tau A and Cas A are observed at different times on a given day). We found a systematic shift of the gain towards lower values when
        using the Tau A data, with mean and median values $-5.6$ and $-5.2\,\%$. This offset
        changed significantly between pixels, with mean values $-3.8$, $-7.2$ and $-5.8\,\%$
        for pixels 23, 26 and 5, respectively. No clear trend with time was observed.
        We combined both sources of uncertainty (statistical and systematic) into a probably overly%one
        conservative 10\,\% error bar. This is the current best estimate of the calibration
        uncertainty, and the nominal value we used when building the SEDs in following 
        Sect.~\ref{section:galactic_regions}. 
        We expect to significantly improve this value and to achieve performance comparable to MFI \changes{\cite[5\% absolute calibration uncertainty, ][]{mfiwidesurvey}} once the calibration diode is installed on the second QUIJOTE telescope and relative gain measurements are routinely performed.

        Regarding the gains of those pixels not used in this work
        (TGI pixel 5 and FGI pixels 41 and 42), 
        they presented comparable performance (high gains) to the ones described here in depth.
        However, we identified issues in their polarization behavior. These are discussed 
        in the following Sect.~\ref{sec:pol}.
        
    \subsection{On-sky polarization properties}
    \label{sec:pol}

        In this Section we present the analyses performed on the astrophysical data 
        from the commissioning period to characterize the TFGI polarization performance.
        A complementary analysis focused on the TFGI performance in the laboratory
        before its installation on the telescope has been presented in \cite{tfgifasano}.
        
\subsubsection{Polarization fraction}
\label{section:cal_eff}

We mainly rely on the analysis of Tau A data to verify that the polarization response of the instrument was stable. In order to do so, we first produced Stokes $Q$ and $U$ maps as described in 
Sect.~\ref{sec:mapmaking}. We then estimated the Tau A flux in both with BF, as we did during the gain calibration  (Sect.~\ref{section:gain_calibration}).
            The $Q$ and $U$ estimates are combined in quadrature and
            divided by the intensity counterpart to obtain the polarized intensity
            fraction from Tau A, $P/I=\sqrt{Q^2+U^2}/I$. This test is independent
            from both the gain calibration, as the dependence from $(Q,U)$ and $I$
            cancels out; and the polarization angle, as $P$ depends on the module
            of $(Q,U)$ alone. However, $P$ can be biased towards positive values %when we are
            in the low signal-to-noise regime, so it must be treated with care \citep{Vaillancourt_poldebiased, jarm_poldebiased}.
    
\begin{figure}
                \centering
                \includegraphics[width=0.65\linewidth, clip, trim=1.7cm 0 0 0]{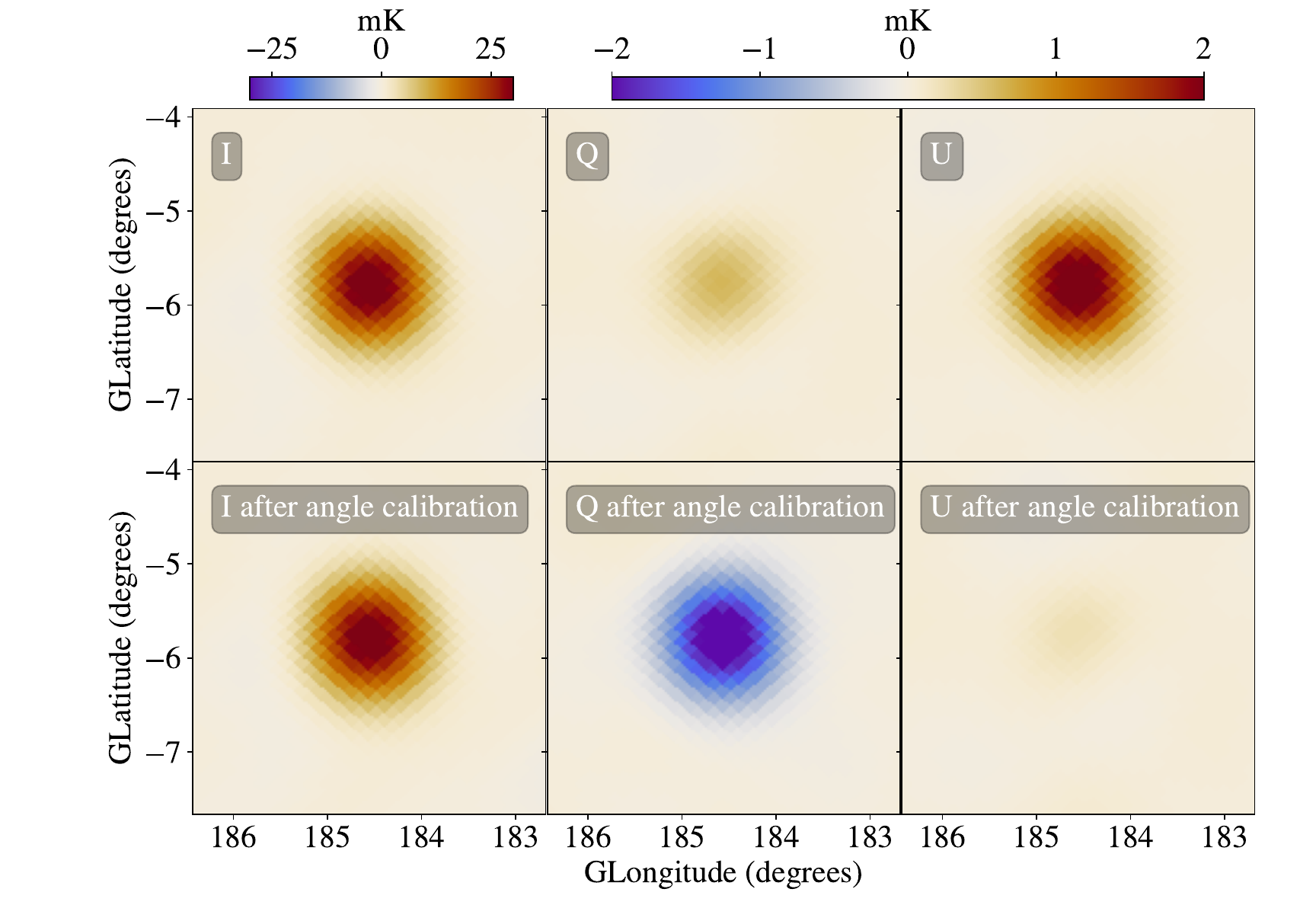}
                \caption{
                Example of Tau~A Stokes $I$, $Q$, and $U$ maps (top, middle, and bottom rows) before (left) and after (right) applying the $\phi_c$ rotation as 
                described in Eq.~\ref{eq:required_polarangle}
                to recover the expected polarization angle of Tau~A, $\gamma_{\rm Tau\, A}$. After applying the rotation, most of the polarized signal is contained in $Q$ with negative sign, as expected from ancillary data. The data correspond to the first channel of pixel 23.
                }
                \label{fig:crab_polangle}
            \end{figure}

            The polarization fraction from Tau A is approximately constant between microwave and
            millimeter frequencies \citep[see e.g. Fig.~39 from][]{mfiwidesurvey},
            and we use the average value between 20 and 350\,GHz from \cite{ritacco_crab}, $p=6.95\pm0.03\,\%$. We
            only found consistent polarization fractions to this for TGI pixels 23 and 26,
            plus FGI pixel 63.  As previously mentioned, for TGI pixel 5 
            and FGI pixels 41 and 42 we detected lower-than-expected polarization 
            fractions: we explain this issue in further depth in the following 
            section. We show an example of $IQU$ maps for Tau A in
            Fig.~\ref{fig:crab_polangle}. We also show the obtained
            polarization fraction values in Table~\ref{tab:crab_polfrac}.
            These values are used to correct the TFGI data for the polarization
            efficiency.

            \begin{table}[]
                \centering
                \caption{Measured polarization fractions for Tau A (in \%), and derived polarization efficiencies
                (computed as the mean of the conversion factors to recover the expected Tau~A polarization fraction  for the four channels) for the three pixels showing acceptable 
                values.}
                \begin{tabular}{cccccc}
                    \hline
                    \multirow{2}{*}{Pixel} & \multicolumn{4}{c}{Pol. fraction (\%)} & Pol.\\
                     & CH1 & CH2 & CH3 & CH4 & efficiency (\%)\\
                    \hline
                    23 & 6.76 & 6.68 & 6.74 & 6.69 & 96\\
                    26 & 6.15 & 5.77 & 5.88 & 6.03 & 85\\ \hline
                    \multirow{2}{*}{63} & 3.22 & 2.27 & 2.28 & 2.9 & 38 \\ 
                     & 4.60 & 5.80 & 5.20 & 4.71 & 73\\ \hline
                \end{tabular}
                \tablecomments{The two sets of values shown for pixel 63 correspond to the data 
                taken before and after we changed the BEM it was connected to in April 2022.
                There is a significant improvement in the latter period, further supporting that the BEMs are responsible for the low polarization fractions achieved in the other pixels. Laboratory
                measurements showed that the efficiencies for pixels 23, 26 and 63 
                were at $90$--$95$\,\%, $85$--$90$\,\% and $70$--$75$\,\% levels,
                respectively. We see that the three are roughly consistent, except
                for the pre-April data of pixel 63. The polarized data was rescaled 
                on a monthly basis, similar to the gain, to minimize the impact from that epoch.}             
                \label{tab:crab_polfrac}
            \end{table}
            
        \subsubsection{Calibration of the polarization angle}
        \label{sec:cal_pol_ang}
        The reference angle $\phi_c$ for each TFGI horn is calibrated using Tau~A observations,
        following the methodology described in Section 5.3 
        from \cite{tfgifasano}. Once the Tau~A maps are obtained, we use BF photometry to extract the Stokes Q and U parameters from those maps, and we use them to compute
            \begin{equation}
                \phi_c = \gamma_{\rm Tau A} - \gamma = \gamma_0 + {\rm RM}\lambda^2 - \frac{1}{2}\arctan\left(-\frac{U}{Q}\right),
            \label{eq:required_polarangle}
            \end{equation}
            where $\gamma_0=-88.31\pm0.25\,\deg$ and ${\rm RM} = -1406\pm12\,\deg\,{
            \rm m}^{-2}$ from \cite{mfiwidesurvey}. We estimate the $\phi_c$ uncertainty
            by taking $10^4$ random samples from $U$ and $Q$ distributions and recomputing
            Eq.~\ref{eq:required_polarangle}. We define the $Q,U$ distributions as
            Gaussians with average and $\sigma$ values being the results and uncertainties
            obtained with BF photometry.
            
            We reviewed the monthly maps from Tau A in order
            to detect possible changes of the calibration angle, $\phi_c$, with
            time. We found that only three pixels (TGI pixels 23, 26 and FGI pixel 63)
            showed stable $\phi_c$ values with time. We calibrated
            $\phi_c$ on a monthly basis, in a similar manner as the gain. 
            In this way we prevented biased estimations of the polarization angle
            in the data with respect to the calibrator\footnote{Data from Tau A maps and
            those from scientific observations mostly come from different epochs 
            (Nov21-Dec21 vs. Apr22-May22, respectively), so the calibration angle
            can change between the two, even if they were in principle consistent.}:
            we show the computed values for pixel 23 in Table~\ref{tab:phic_values_monthly_pix23}, and the rest in Appendix~\ref{appendix:all_gain_calibration}. \changes{The uncertainties in the determination of such angles lie in the 4--6\,deg range, similar to current instruments. Reducing these uncertainties well below the $1\,$deg threshold is one of the main challenges in current CMB instrumentation \citep[e.g.,][]{COSMOCAL}. For instance, future experiments targeting a sensitivity of $\delta r\sim10^{-3}$ will require a polarization-angle calibration accuracy better than $0.2$\,deg \citep{2020A&A...634A.100A}.}
            The other three pixels (TGI pixel 25 and
            FGI pixels 41 and 42) were highly variable, which translated into a
            lower-than-expected polarization response\footnote{This cannot be 
            attributed to a low polarization efficiency, as both FGI
            pixels 41 and 42 showed polarization efficiencies larger than 90\,\% in 
            laboratory measurements.}. 
            
            This issue was probably due to BEM performance:
            the BEMs are located outside the cryostat, so their working conditions on the
            telescope were not as stable as those in the laboratory.
            The difference between the gains from the two branches from the BEM seemed to
            vary with time, which implied changes of the polar angle. These variations
            translated as varying $QU$ values, which when integrated along several 
            observations partly cancelled out and returned a decreased response. 
            This hypothesis is further supported by the fact that the polarization efficiency from pixel 63 
            improved dramatically (from 38 to 73\,\%) at the middle of the commissioning period, after we changed the BEM channel it was connected to.

            This issue will be mitigated once the calibration diode is installed. 
            In that case, the sampling of the calibration signal 
            will be faster than the drifts in the BEM gain inbalances, so they will be
            properly traced. \changes{For TGI pixel 23, \citet{tfgifasano} already achieved an uncertainty below $1^\circ$ in the polarization-angle calibration from laboratory measurements with the diode. This is also comparable with the overall calibration uncertainty achieved by the previous QUIJOTE MFI instrument \citep[$0.6^\circ$ at 11\,GHz, ][]{mfiwidesurvey}. Taken together, these results indicate that the installation of the calibration diode is expected to significantly improve the polarimetric performance of the instrument, not only for the pixels analysed in this paper but also for the remaining detectors. We expect the polarization-angle calibration uncertainty to be below $1^\circ$.}
            
            \begin{table}[]
                \centering
                \caption{Values for the calibration angle, $\phi_c$, required to
                translate the coordinate system from raw measurements to that of the
                sky. This example corresponds to pixel 23 (TGI).}
                \begin{tabular}{ccccc}
\hline \hline  Month  &  CH1   &  CH2   &   CH3   &   CH4   \\
\hline  
Nov21   & $-$47.6 & $-$12.6 & $-$141.5 & $-$106.1 \\
Dec21   & $-$45.7 & $-$11.1 & $-$138.4 & $-$100.0 \\
Jan22   & $-$46.4 & $-$8.6  & $-$137.7 & $-$101.5 \\
Feb22   & $-$50.5 & $-$10.4 & $-$139.2 & $-$105.1 \\
Apr22   & $-$51.5 & $-$12.1 & $-$142.0 & $-$105.4 \\
May22   & $-$52.0 & $-$13.9 & $-$139.2 & $-$105.3 \\
\hline
\end{tabular}
                \tablecomments{Units are in degrees, and uncertainties lie in the 4--6$^\circ$
                range for each value. }
                \label{tab:phic_values_monthly_pix23}
            \end{table}

\subsection{Commissioning observations of the Cygnus region}
We carried out dedicated observations of the Cygnus region to test the instrument performance in recovering diffuse emission. The Cygnus region lies on the Galactic plane at $l \approx 80\degr$, and extends over 175\,deg$^2$. %It includes W63, a supernova remnant located at $(l,b) = (82.15^\circ, 5.31^\circ)$. 
Our TFGI observations were conducted in May 2022, comprising a total of 24.1 hours in 26 scans (see Table~\ref{tab:obs_time}). From these data, we generated a set of maps, reaching a sensitivity for the combined TGI pixels \footnote{The analysis of these maps is presented in Sect.~\ref{sec:sensitivities}.} of 257$\muK{}\deg{}^{-1}$ in intensity, and 25.3 and 25.1$\muK{}\deg{}^{-1}$ in Stokes $Q$ and $U$, respectively, at 31\GHz{}. The sensitivities worsen to 385, 35.8 and 35.1 $\muK{}\deg{}^{-1}$ for I, Q and U at 41\GHz{}, respectively. All observations were performed with the same telescope scanning strategy: azimuth drifts at constant elevation steps, and all at the same elevation in consecutive days. The lack of observations with different scanning directions or parallatic angles limit the impact of the destriper algorithm; therefore, the intensity maps are still significantly affected by $1/f$ noise stripes.

\subsubsection{Intensity-to-polarization leakage}
\label{sec:IPleak}
The integrated intensity SED of the Cygnus region is strongly dominated by free–free emission, which is essentially unpolarized. \changes{We assume that any hint of polarized emission in these maps cannot arise from genuine sky polarization, but instead reflects leakage of the total intensity signal into the polarization measurements. Such instrumental contamination is commonly referred to as intensity-to-polarization leakage, and these maps therefore provide a suitable dataset for assessing its level.} 

\begin{figure}
            \centering
            \includegraphics[width=1\linewidth]{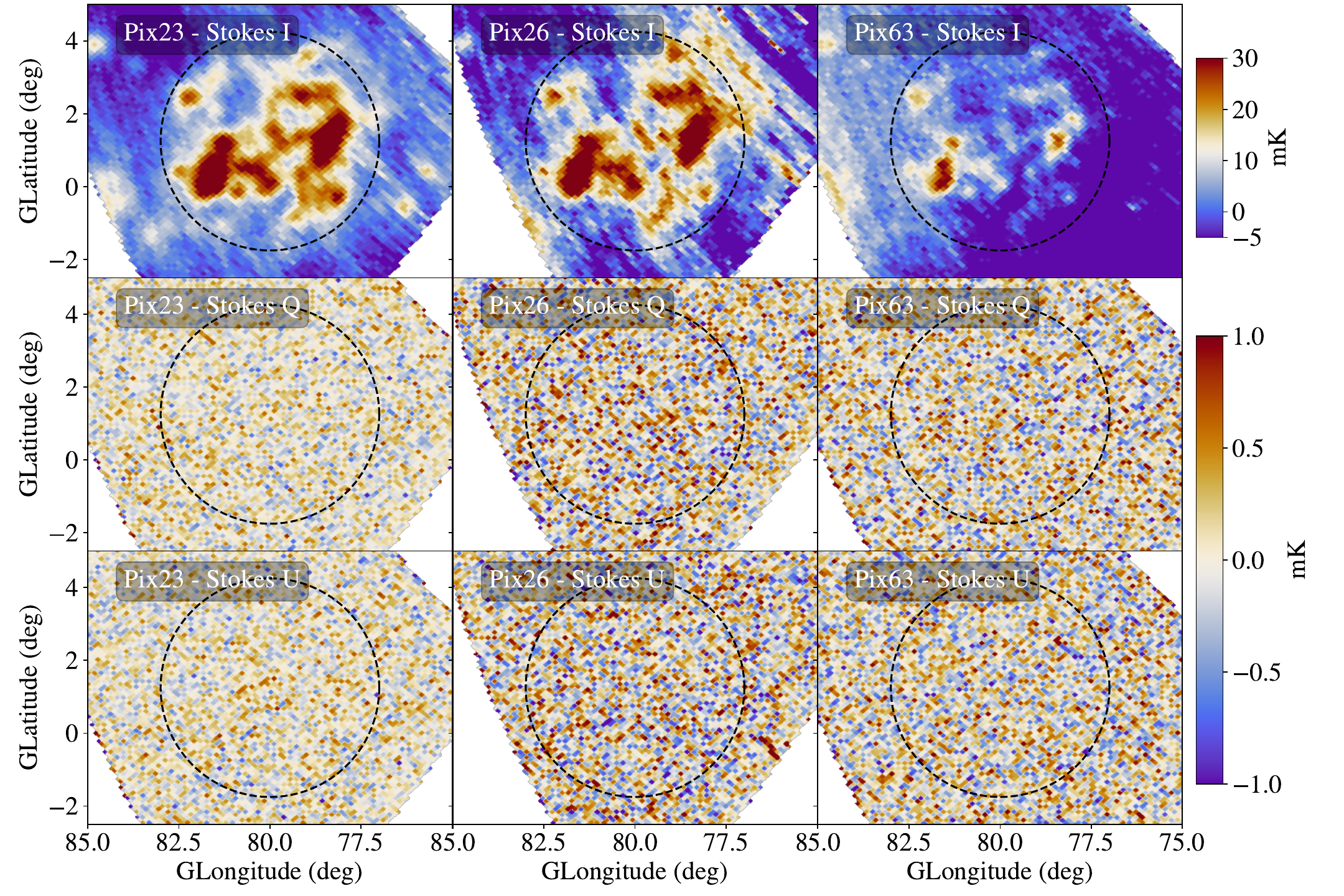}
            \caption{Stokes $I$, $Q$, and $U$ maps (top, middle, and bottom rows respectively) for the Cygnus region,
            as observed with TGI pixels 23 (left), 26 (center) and FGI pixel 63 (right column). We 
            clearly see the lower intensity at 41\GHz{} (pixel 63), because
            of the decreasing spectral index in temperature units from the dominant
            free-free emission in Cygnus ($\beta \approx -2.1$). \changes{The Cygnus region is known to hosts very little-to-none polarized emission, so the absent signal in $Q$ and $U$ visually indicate the proper functioning from the instrument. We show as } dashed line circles the apertures used in Sect.~\ref{sec:IPleak} to put an upper limit on the intensity-to-polarization leakage.}
            \label{fig:W63_maps}
        \end{figure}

We followed a similar approach to \cite{mfiwidesurvey}. We defined 
        a region with a radius equal to $3\degr$ and centered at
        $(l,b)=(80\degr,0\degr)$ and found the correlation coefficient $\alpha$
        that minimized $\alpha_{Q,U}I-(Q,U)$. Here, $IQU$ were computed as the total 
        signal from the maps within that region. In Fig.~\ref{fig:W63_maps} we show
        the Cygnus $IQU$ maps used in this analysis, while the $\alpha$
        results are quoted in Table~\ref{tab:TFGI_summary}, computed as
        $\alpha=\sqrt{\alpha_Q^2+\alpha_U^2}$. The estimates are consistent with
        zero for all cases; we provide upper limits to the leakage of less than $1\,\%$ for all the three pixels.

\subsubsection{Recovery of diffuse emission}
\label{sec:recovery_diffuse_emission}
\begin{figure}
            \centering
            \includegraphics[width=1\textwidth]{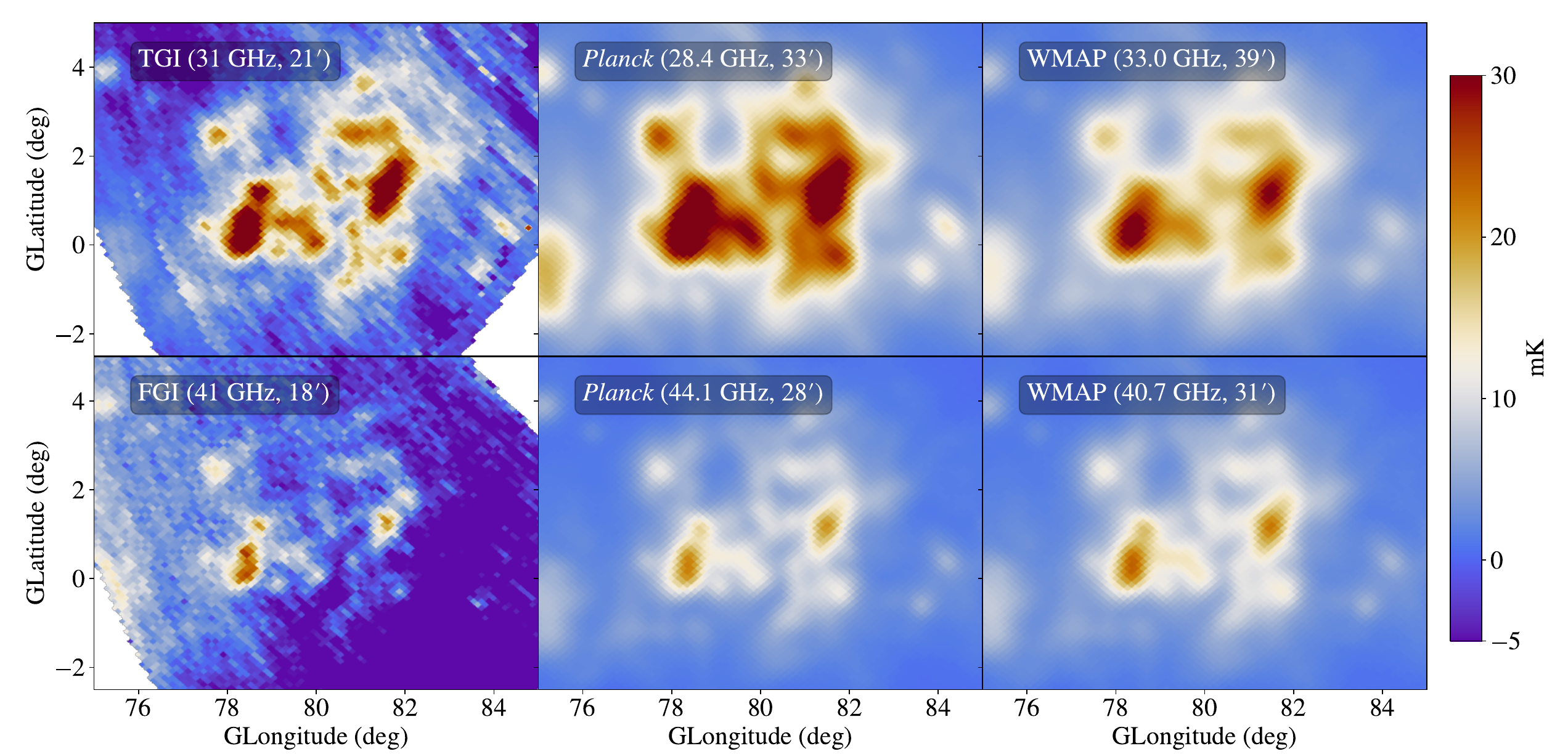}
            \caption{Intensity maps of the Cygnus region with QUIJOTE-TFGI, WMAP and \textit{Planck}.
            Top (from left to right): TGI combined map from two detectors (31\GHz{}), \textit{Planck} (28.4\GHz{}),
            WMAP (33\GHz{}). Bottom (from left to right): FGI (41\GHz{}), \textit{Planck}
            (44.1\GHz{}), WMAP (40.7\GHz{}). We computed the TGI map as the straight mean between
            pixel 23 and pixel 26 maps. The angular resolution of each map is indicated in the panels, illustrating the ability of TFGI to recover and resolve smaller-scale features.}
            \label{fig:w63_comparison}
        \end{figure}

We qualitatively assess the ability of TFGI to recover large-scale Galactic emission by comparing these Cygnus region intensity maps with those from WMAP and \textit{Planck} in Fig.~\ref{fig:w63_comparison}. Although the TFGI maps are noisier, the same structures are clearly detected in all cases. \changes{This example demonstrates the good performance of the map-making procedure adopted in this work, indicating that no significant transfer-function systematics are present over the angular scales of interest.}
%\changes{This proves that our selection of parameters\footnote{\changes{Mostly the length of the median filter and the size of the window function used to bin the data. Further optimization of such parameters will be discussed ina a future work (Bose et al., in prep.)}} during the map-making procedure has not impacted the transfer function in a negative way, and all physical scales are recovered without a loss of information.}

It is evident that the improved angular resolution of TFGI ($21^\prime$ at 31\,GHz vs. $33^\prime$ for \textit{Planck} at 28.4\,GHz and $39^\prime$ for \textit{WMAP} at 33\,GHz; and $18^\prime$ at 41\,GHz vs. $28^\prime$ for \textit{Planck} at 44.1\,GHz and $31^\prime$ for \textit{WMAP} at 40.7\,GHz) allows us to resolve some emission features that were previously blended within the beam.

The impact of the noise stripes in these TFGI intensity maps will be mitigated once complementary observations at different telescope elevations and scanning directions are included. We note that the Cygnus dataset presented here comprises only $\sim 20$ hours of observations, all recorded with similar telescope drift motions (i.e. at the same parallactic angles). This limitation reduced the ability of the destriper algorithm to converge to an optimal baseline solution. As demonstrated in Sect.~\ref{section:galactic_regions}, the reconstructed baselines improve significantly with additional observations obtained under different raster configurations and parallactic angles. 
        
    \subsection{Instantaneous sensitivities from skydip observations}
    \label{sec:sensitivites}
    In this section we evaluate the instantaneous sensitivity of the 
        instrument from the power spectra obtained from skydip observations.
        Skydips were taken by moving the telescope continuously in elevation ($\sim1^\circ\,$s$^{-1}$)
        within the full available range (30--90\degr) at a fixed azimuth position.
        In this case we only used the data from the end of one of the observations, when the telescope was pointing to a constant elevation of $30^\circ$.

        We took a total of 191
        skydips during the commissioning phase, accounting for 67.2\,hours of 
        observation. The data were weighted by $\Nhits$
        in order to get \muK{}\s$^{1/2}$ units. Once this was done, we computed 
        the power spectra both for intensity and polarization timelines. These power 
        spectra were then fitted to the combination of white and $1/f$ noise 
        components:
        \begin{equation}
            \sigma' = \sigma\left[1+\left(\frac{f}{f_k}\right)^\alpha\right],
        \end{equation}
where $\sigma$ is the white noise baseline and $f_k$ the knee frequency from $1/f$ noise. The correlated ($1/f$) noise component is expected to be negligible in polarization timelines due to the TFGI instrumental architecture and its differential nature on short timescales, which effectively suppresses correlated signals arising from the electronics or the atmosphere.

\begin{figure}
            \centering
            \includegraphics[width=1\linewidth]{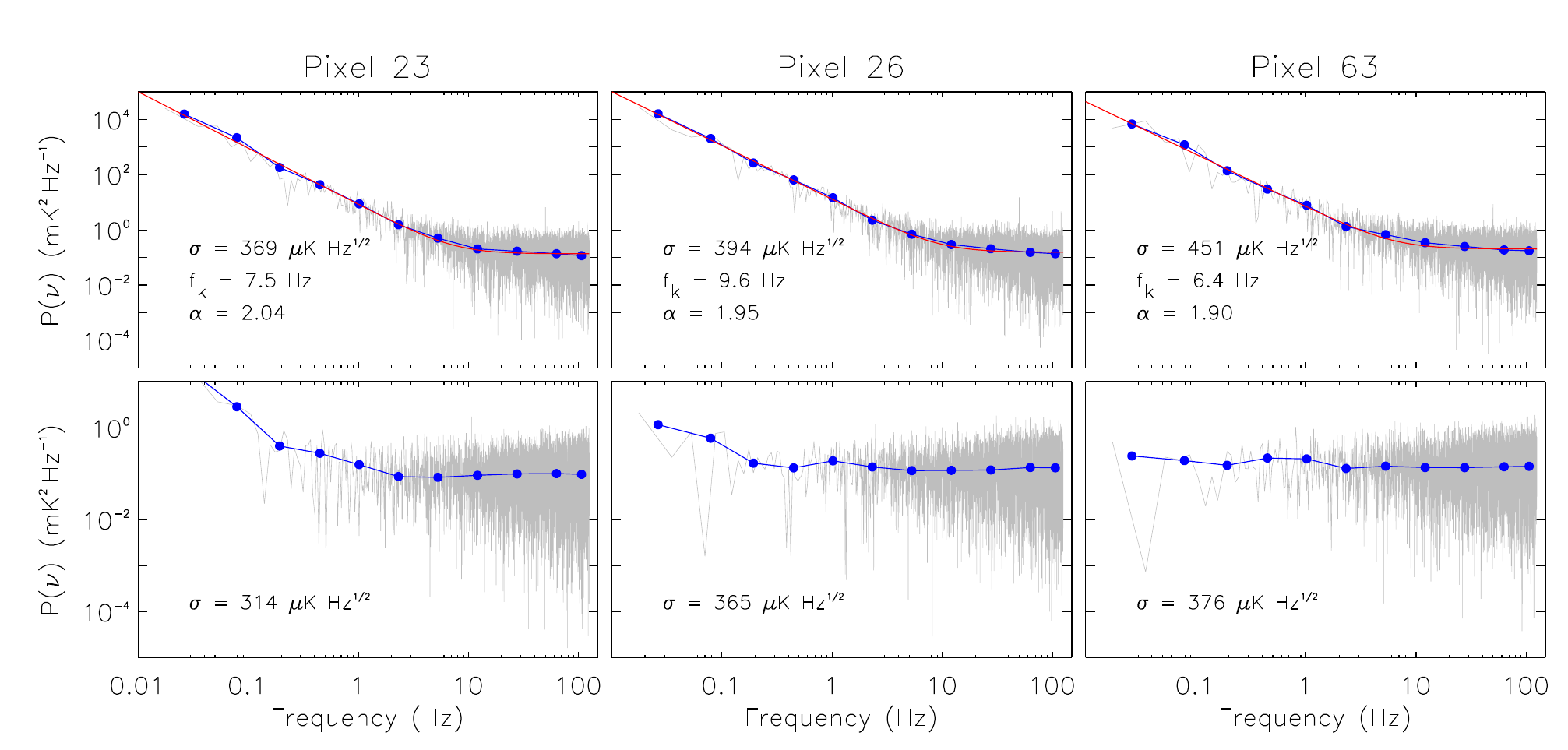}
            \caption{Power spectra from one skydip observation for the first
            channel of pixels 23, 26 and 63 (from left to right), both in
            intensity (top) and polarization (bottom). \changes{The measured data (computed as the Fourier transform of the TOD considered) is depicted in grey, while we show in blue the same data after being binned logarithmically. Red lines stand for the best-fit solution considering both a $1/f$ (exponential) and white noise (constant) contributions. Such combination was not used in the polarization case as the number of data points where $1/f$ is significant is too low to fit the component with enough significance.} We clearly see how
            the slope in the data is mostly (for $>1$\,Hz) absent in polarization, meaning
            no $1/f$ component is present in polarization data, as expected.
            White noise levels show consistent results with those expected 
            (see text).}
            \label{fig:power_spectra_skydips}
        \end{figure}

In Fig.~\ref{fig:power_spectra_skydips} we show the power spectra from pixels 23, 26 and 63 for one skydip. The white noise estimates can be compared to the expected sensitivity from the radiometer equation and the TFGI design temperature:
\begin{equation}
    \sigma(T)\sqrt{\tau} = \frac{T_{\rm sys}}{\sqrt{\Delta\nu}} = 350\muK\s^{1/2}, 
\end{equation}
where $T_{\rm sys}=35\K{}$ and $\Delta\nu=10\GHz{}$. We therefore 
find consistent results with the values computed from polarization data, with noise levels in the range 314--376\muK{}\s$^{1/2}$. In addition, the values of the knee frequency, $f_k$, in polarization are $\leq1\Hz{}$, demonstrating the effective cancellation of correlated noise.

%--------------------------------------------------------------------

\section{Observations of the Galactic plane ($25\degr < l < 50\degr$)}
\label{section:galactic_regions}

During the TFGI commissioning phase we also observed the Galactic plane region 
    between $l=25\degr$ and $l=50\degr$, which contains several bright features.
    We usually label this region after one of these features, W44. The field was 
    observed for more than 200\,hours, while our final maps account for 157.2\,hours
    after flagging (Table~\ref{tab:obs_time}).
    These regions have been studied already by various QUIJOTE works, such as \cite{W44}
    or \cite{perseus_w43_rhooph}. We focus this analysis in W43, W44 and W47, 
    whose Galactic coordinates and basic properties are summarized in 
    Table~\ref{table:w43_w44_w47}. While all of them host several HII regions that are
    unresolved at degree scales, W44 also contains a supernova remnant (SNR).
    This is the only object considered here where polarized emission is expected.
    
\begin{figure}[h!]
        \centering
        \includegraphics[width=0.45\linewidth, clip, trim=0 2.5cm 0 0]{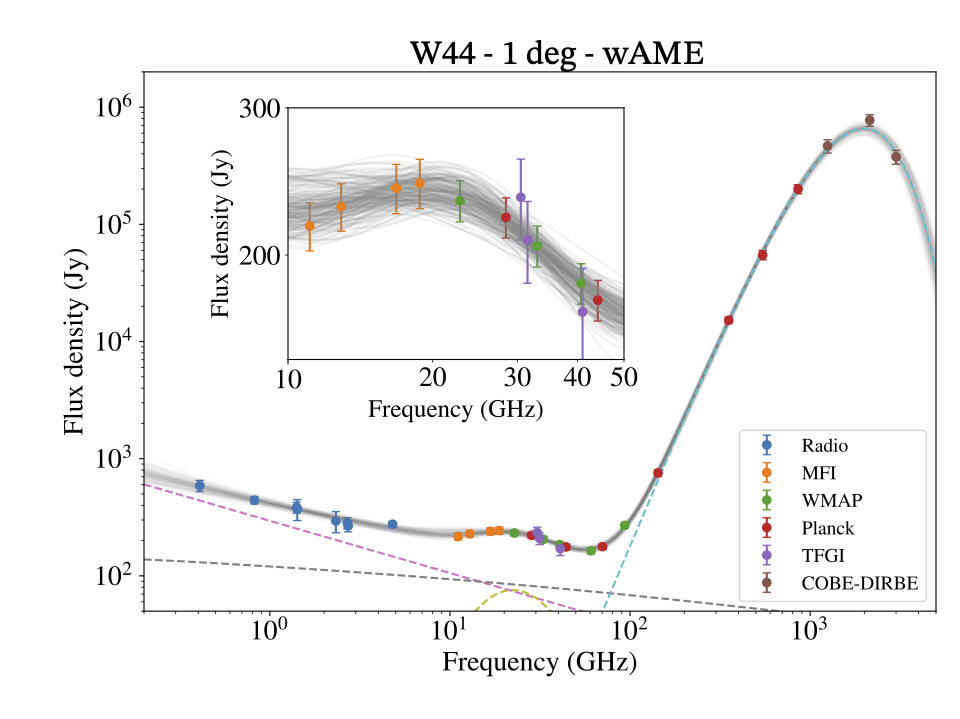}
        \includegraphics[width=0.45\linewidth, clip, trim=0 1.8cm 0 0]{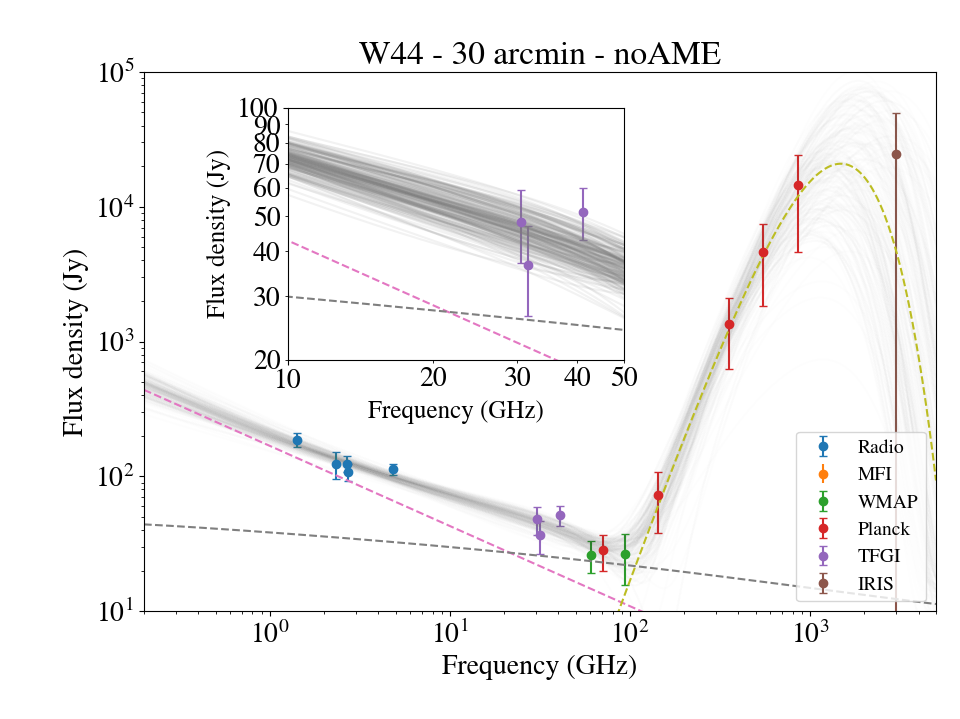}
        \includegraphics[width=0.45\linewidth, clip, trim=0 0 0 0.7cm]{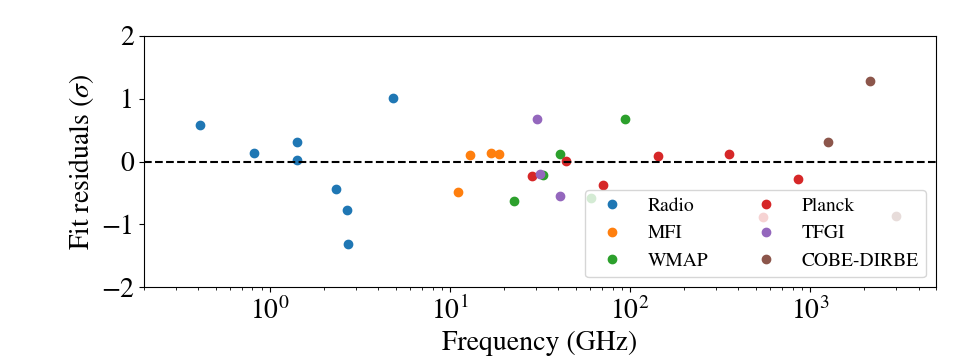}
        \includegraphics[width=0.45\linewidth, clip, trim=0 0 0 0.7cm]{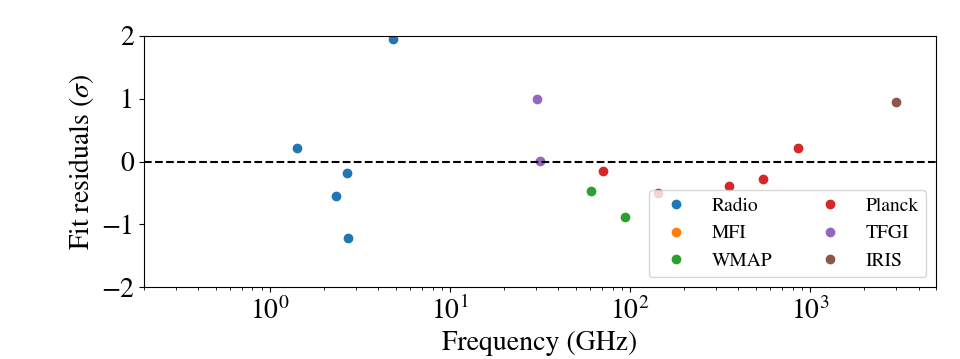}
\caption{(Left) Top: SED of W44 in intensity derived from maps at 1 degree resolution. The photometry is computed using an aperture of radius $r = 1^\circ$. The TFGI measurements are consistent with those from other experiments. The model includes synchrotron, free-free, AME, thermal dust and CMB components. All photometric data points have been colour-corrected, except those from TFGI.
Bottom: Residuals between the colour-corrected data and the best-fitting model (using the chain with maximum-likelihood value). The largest discrepancies are found at higher frequencies, where thermal dust emission dominates. 
These deviations likely reflect limitations of the adopted model: multiple free–free and dust-emitting regions with distinct spectral properties overlap within the aperture, and the model, based on a single modified blackbody component, does not fully capture this complexity.
(Right)  Same as left panels, but using maps at 30$^\prime$ resolution, enabled by the higher angular resolution of TFGI. The comparison between the two SEDs (left and right) indicates that both AME and free-free contributions become less significant when the aperture is reduced. This can be explained by the SNR being better resolved with respect to its surrounding region; as a result, the synchrotron emission from the SNR becomes more dominant, while the contribution from the   surrounding emission (AME and free-free) decreases due to reduced contamination from the larger aperture.}
    \label{fig:w44_SED_text}
    \end{figure}

\begin{figure}[h!]
        \centering
        \includegraphics[width=1\linewidth]{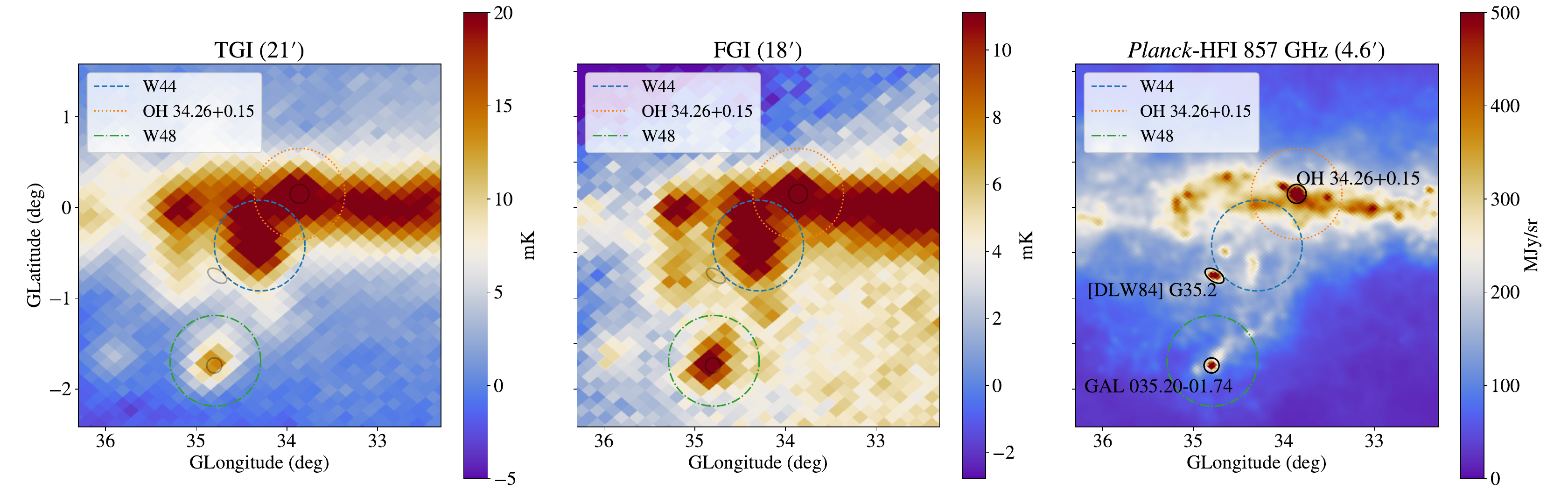}
        \caption{
        Details of the intensity maps centred on W44
        from the two TGI pixels 
        (left), FGI (middle), and \textit{Planck}-HFI 857\GHz{} (right), all shown at their native angular resolutions.
        Circular apertures with $r=30^\prime$ are overlaid to illustrate the integration regions used for the photometry of W44.
        Nearby sources not analyzed in this work, but clearly visible in the TFGI maps (W48 and OH~34.26+0.15), are also indicated, as part of their flux is included in the $r=1^\circ$ aperture around W44.
        While W48 and OH~34.26+0.15 exhibit clear counterparts in the far-infrared, as seen in the 857\,GHz map, this is less evident for W44, suggesting differences in the associated dust emission. Finally, 
        the colour scale of the FGI map has been rescaled from that of TGI assuming a spectral index $\beta = -2.1$. Regions appearing brighter in the FGI map (mainly W48, but also the surrounding Galactic plane) are therefore likely to host strong AME emission.}
        \label{fig:w44_different_resolutions}
    \end{figure}

    We performed
    AP as described in Sect.~\ref{section:SED_AP}, with the maps smoothed to $30^\prime$
    and $1\degr$ (the maps at their original resolution will be presented in 
    Sect.~\ref{sec:sensitivities}). The latter are for consistency checks with
    previous analyses, while the higher resolution maps are used to test whether foregrounds behave
    differently with angular scale. In other words, the improved resolution from
    the TFGI with respect to WMAP and \textit{Planck} allows us to 
    resolve smaller regions that may have different emission mechanisms. We show the comparisons in 
    Fig.~\ref{fig:w44_SED_text}, highlighting the
    apparent absence of AME on W44 when using a smaller aperture. Similar comparisons for W43 and W47 are present in Appendix~\ref{appendix:all_SED_regions}. It is clear 
    when comparing TFGI and far-infrared maps, such as \textit{Planck}-HFI 857\GHz{} in
    Fig.~\ref{fig:w44_different_resolutions}, that several different regions are embedded in each of
    our apertures. Finer resolutions in mm-instruments allow for
    better distinction between regions. Such a result aligns with previous works claiming that AME is not present in SNR in general
    \citep{snrwidesurvey}.
    
    However, the lack of data between radio and TFGI frequencies with angular resolutions below $1\degr$ prevents
    a clear fit of the model at such frequencies. This mainly impacts the reconstruction
    of low frequency foregrounds (synchrotron, free-free and AME), so we leave a more
    precise discussion for future work. In $30^\prime$ analyses we use those surveys from
    Table~\ref{tab:ancillary_maps} (in Appendix~\ref{appendix:SED_data}) with FWHM lower than $30^\prime$, while for
    $1\degr$ we use all of them. In this latter case, we use COBE-DIRBE data instead 
    of IRIS. IRIS is calibrated with COBE-DIRBE, and this calibration actually dominates 
    IRIS uncertainty at large ($\geq1.25\degr$) scales \citep{iris}. Considering that
    both would then grant similar results on $1\degr$ analysis, we kept COBE-DIRBE for
    consistency with previous works.
        
\begin{figure}[h!]
        \centering
        \includegraphics[width=1\textwidth, clip, trim=0 8.15cm 0 0]{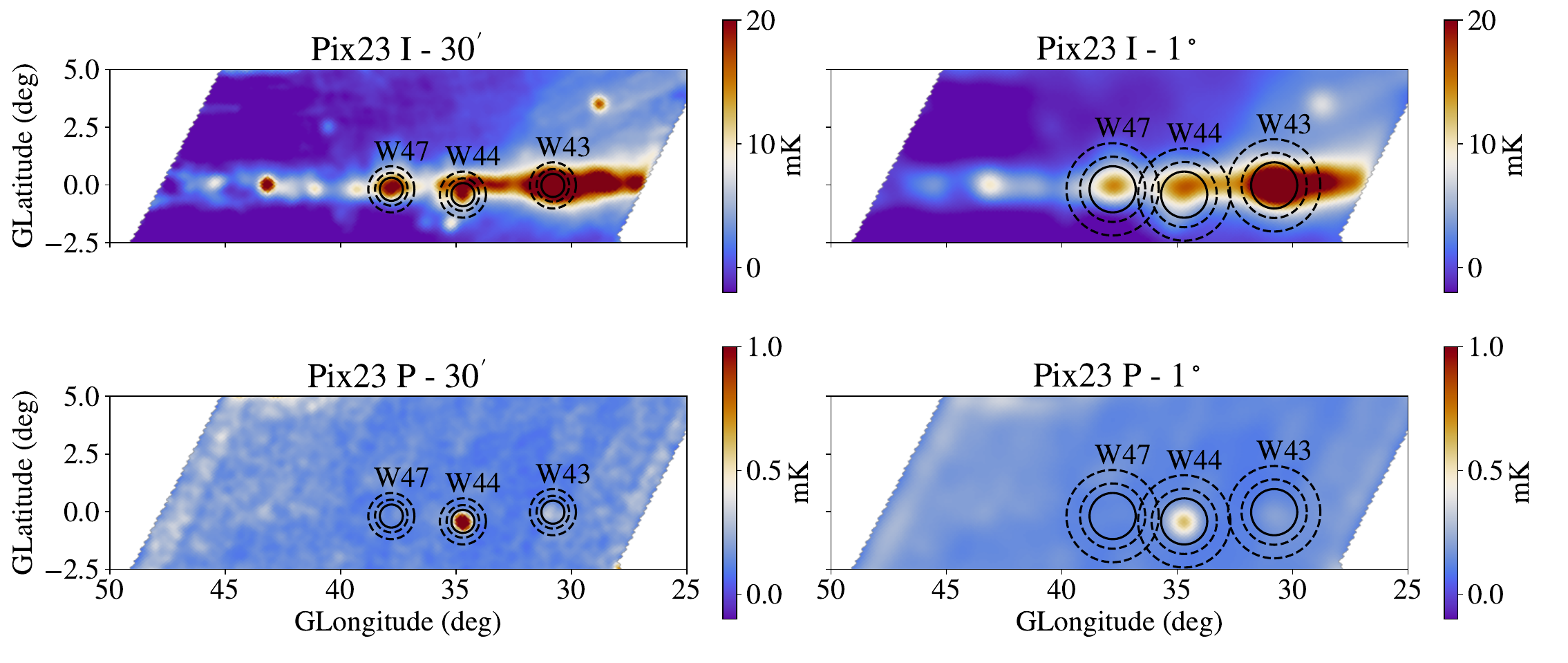}
        \caption{Intensity ($I$, top) and polarized intensity ($P$, bottom) maps
        at 31\GHz{} (from pixel 23 of the TGI) for the W44 Galactic plane region, both at $30^\prime$ (left) and $1\degr$ (right) scales, with the apertures used for photometry overlaid.}
        \label{fig:galactic_plane_I_and_P}
    \end{figure}

\begin{figure}[h!]
        \centering
        \includegraphics[width=1\linewidth, clip, trim=0 0 0 1.5cm]{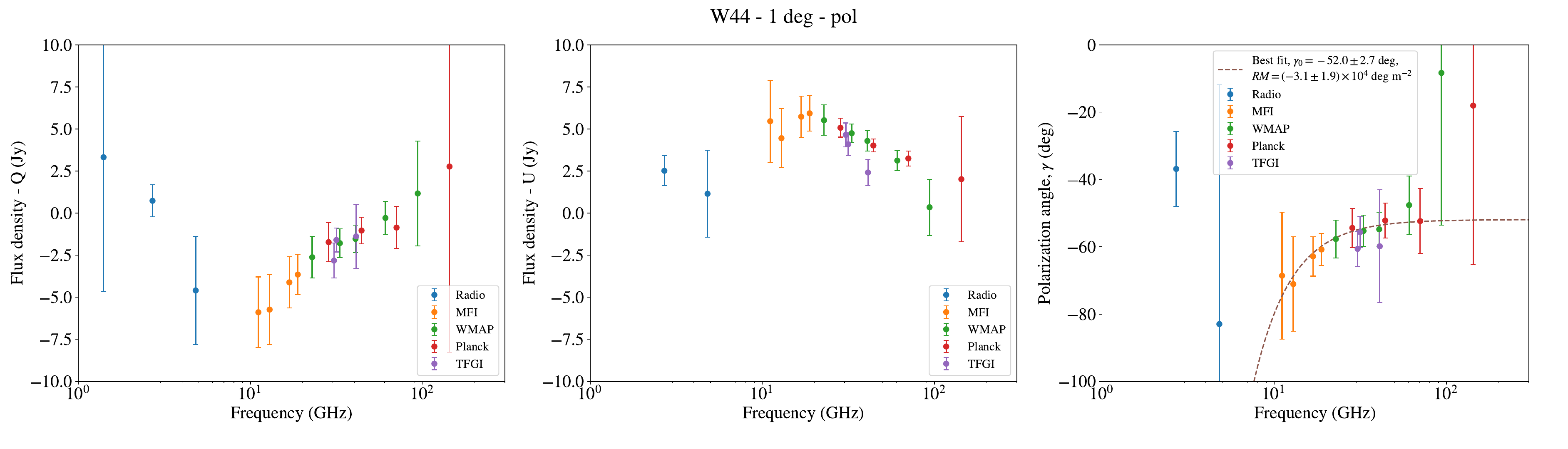}
        \caption{SED for W44 in polarization, extracted from the 1 degree maps. From left to right: Stokes $Q$, Stokes $U$ and the polarization angle. 
        % The TFGI measurements of $Q$ and $U$ are consistent with previous experiments. 
        \changes{The polarized intensity, $P$, is shown in Appendix~\ref{appendix:all_SED_regions}. 
        % However, the $U$ estimate from FGI pixel 63 shows a larger discrepancy.
        The TGI (pixels 23 \& 26) measurements are consistent with previous experiments, similar to the $Q$ measurement from FGI pixel 63. $U$ from the latter shows a $<2\sigma$ difference with respect to previous measurements. 
        In the right panel, we plot the best fit (following the model in Eq.~\ref{eq:RM_model}) obtained without the TFGI data, showing consistency with the latter. The larger errobars for FGI pixel 63 estimates are mostly due to the lower polarization efficiency.}}
        \label{fig:w44_Q_U}
    \end{figure}

    We show the intensity and polarized intensity maps from pixel 23 for the Galactic 
    plane region between $l=25\degr$ and $l=50\degr$ in 
    Fig.~\ref{fig:galactic_plane_I_and_P}, at $30^\prime$ and $1\degr$ resolution.
    We show the \changes{SEDs for Stokes I, Q and U} at $1\degr$ resolution for W44
    as an example in Figures~\ref{fig:w44_SED_text} and \ref{fig:w44_Q_U}. \changes{In the latter, we also combine $Q$ and $U$ to compute the polarization angle, $\gamma$. We fit this last parameter to the following model, taken from \cite{W44}:}
    \begin{equation}
        \gamma = \gamma_0 + {\rm RM}\left(\frac{c}{\nu}\right)^2
    \label{eq:RM_model}
    \end{equation}
    
    \changes{without taking into account any data from the TFGI. We also discarded the points with $\nu\leq10\GHz{}$ from the fit, as they are heavily affected by Faraday rotation. We see that the data from TGI pixels 23 and 26 are always consistent with the model (for $\gamma$) and the data (for $I$, $Q$, and $U$)
    from previous experiments.} On the other hand, the polarization data from pixel 63 
    shows a slightly lower-than-expected estimate, even after we corrected 
    the data with the
    polarization efficiencies from Sect.~\ref{section:cal_eff}. This suggests that these
    observations are more affected by the variations in the polarization angle than
    the ones for Tau A. As previously mentioned, we show the remaining SEDs at both
    resolutions in Appendix~\ref{appendix:all_SED_regions}, and in every case the 
    photometry remains consistent with expectations. This confirms that our gain 
    calibration is working properly. In Appendix~\ref{appendix:all_SED_regions} we also provide the best-fit 
    parameters for the intensity SEDs, following the description in 
    \cite{ameplanewidesurvey}.

    \begin{table}[]
        \centering
        \caption{Basic properties from W43, W44 and W47.}
        \begin{tabular}{cccc}
   \hline \hline
   Source & W43 & W44 & W47 \\ \hline
   Galactic Longitude ($\deg$) & 30.76 & 34.56 & 37.79 \\
   Galactic Latitude ($\deg$) & $-$0.02 & $-$0.50 & $-$0.19 \\
   Type & HII & SNR+HII & HII \\ \hline
\end{tabular}
        \label{table:w43_w44_w47}
    \end{table}
    
    Finally, we study the Stokes parameters $Q$ and $U$ of W44 separately, as an additional
    test of the calibration of the polarization angle. In Fig.~\ref{fig:w44_Q_U} we 
    show how the results remain consistent for TGI pixels, while being low for FGI,
    as seen in the polarized intensity SED. However, even though the differences of
    the pixels are small, their direction contribute together to a small ($\leq1\sigma$) 
    drift in the polarization angle with respect to the value expected from previous 
    data. This most probably comes from the uncertainty on the definition of the 
    polarization angle, stressing the need for the calibration diode. It is necessary
    not only to prevent a lower-than-expected polarization efficiency in those pixels 
    with rapidly varying angles, but also to achieve an optimal definition of the 
    calibration angle in the rest.
    
%--------------------------------------------------------------------

\section{Polarization sensitivities on Galactic patches}
    \label{sec:sensitivities}

    We evenly split the scans used to build the previous Galactic maps into two 
    half-mission difference maps (HMDM, $h_i\,|\, i\in(1,2)$), and compute their 
    \changes{weighted} difference, which acts as a noise map:
    % \begin{equation}
    %     n = \frac{h_1 - h_2}{2},
    % \label{eq:noise_map}
    \begin{equation}
        {n = \frac{h_1 - h_2}{\sqrt{(w_1+w_2)\left(\frac{1}{w_1} + \frac{1}{w_2}\right)}},}
    \label{eq:noise_map}
    \end{equation}
    where the \changes{factor in the denominator ensures that the resulting HMDM has the same noise level as the weighted sum of the two half maps \citep[see e.g.][]{2014A&A...571A...2P}.}
    \changes{We went over the individual scans in a chronological order, and assigned them alternatively to either $h_1$ and $h_2$:}
    \begin{equation}
        h_1 \propto \sum_{i=0, 2,4,...}^{N_{\rm scans}-1} m_i\quad;\quad h_1 \propto \sum_{i=1, 3,5,...}^{N_{\rm scans}} m_i
    \end{equation}
    \changes{In this way, both $h_1$ and $h_2$ contained data acquired under similar observing conditions. These conditions (e.g. atmospheric temperature or precipitable water vapour, PWV) are normally associated with additional correlated ($1/f$) atmospheric emission in intensity measurements on top of the instrumental contribution, which explains the residual striping still present in the intensity maps. We note that this effect is much less important in the polarization $Q$ and $U$ maps, as the correlated ($1/f$) noise is largely suppressed by the differencing of the phase states (see Section~\ref{sec:sensitivites}).}
    We built the HMDM for the three pixels
    studied throughout the paper, plus the combination of the two TGI pixels (23 and 26),
    which are used as the final 31\GHz{} map. 
    \changes{The two TGI pixels were combined with equal weights, with the aim of obtaining not the optimal noise level, but rather a characteristic sensitivity representative of the TGI instrument}.

\begin{figure}
        \centering
        \includegraphics[width=1\linewidth, clip, trim=0 2.5cm 0 2.5cm]{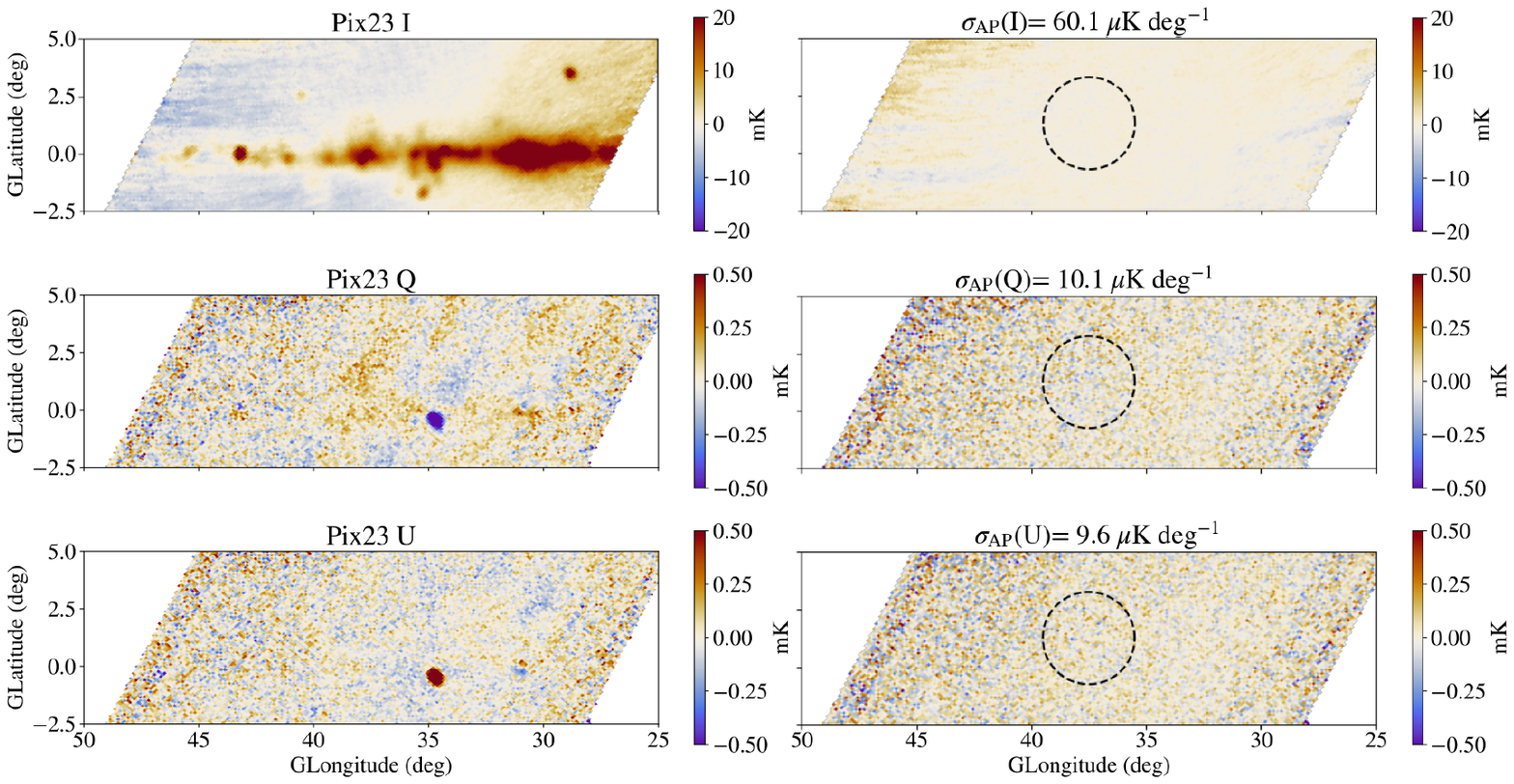}
        \caption{Left: Stokes $I$, $Q$, and $U$ maps (from top to bottom) of the region $l\in(25\degr, 50\degr)$ obtained with TGI pixel 23. 
        Small residuals associated with the strong unpolarized emission from W43 are visible in the $Q$ and $U$ maps, displaying the characteristic cloverleaf pattern produced by differences between the two co-polar beams, similar to that observed for Tau~A in the MFI $U$ maps (see
        Fig.~37 from \citealt{mfiwidesurvey}). 
        Right: HMDM for the same $IQU$ maps. We show with dashed lines the region used to compute the sensitivities in Section~\ref{sec:sensitivities}.}
        \label{fig:sens_w44_pix23}
    \end{figure}

\begin{figure}
        \centering
        \includegraphics[width=1\linewidth, clip, trim=0 2.5cm 0 2.5cm]{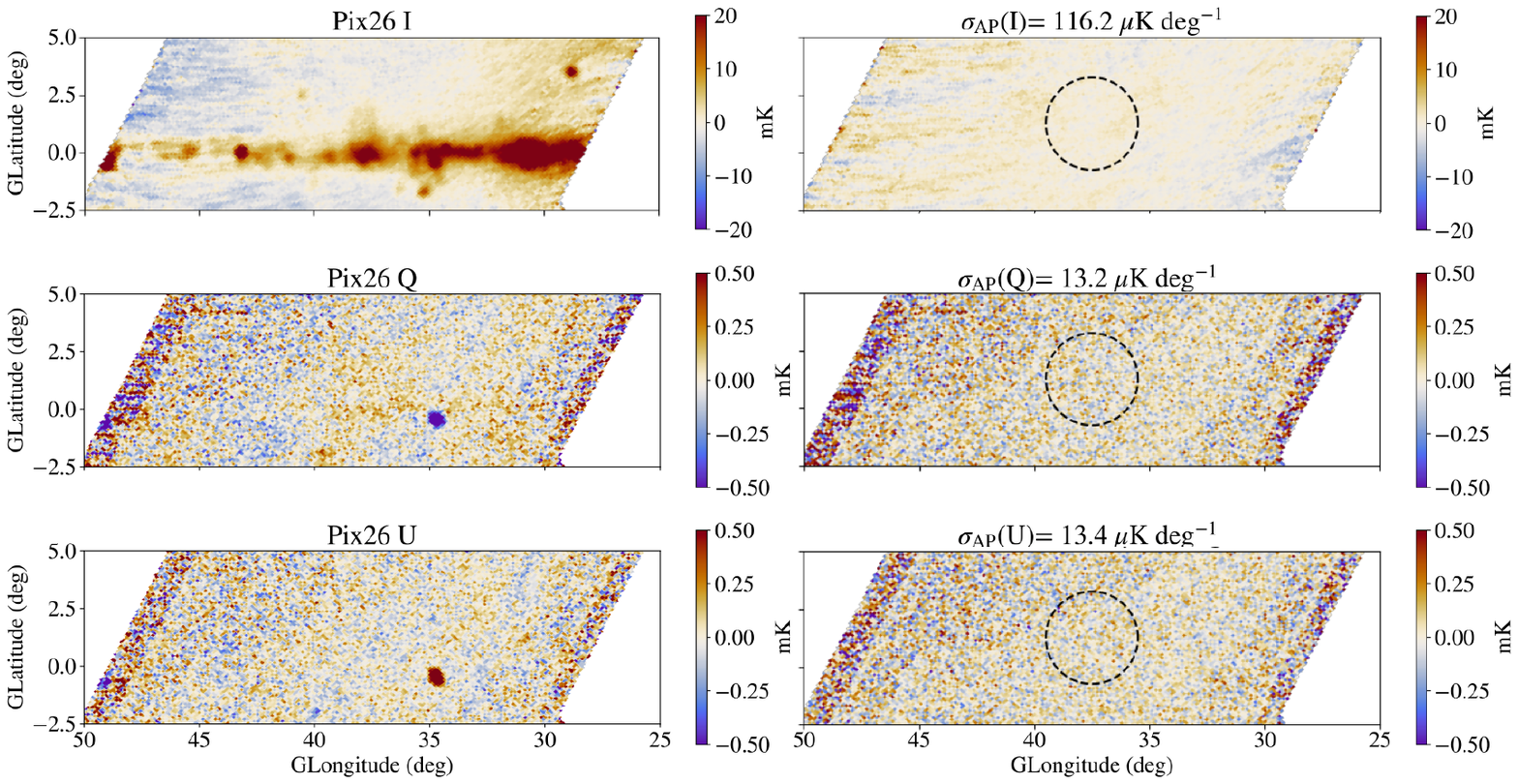}
        \caption{Same as Fig.~\ref{fig:sens_w44_pix23}, but for TGI pixel 26. W51, located in the easternmost part of the map, is only visible in this pixel owing to projection effects.}
        \label{fig:sens_w44_pix26}
    \end{figure}

\begin{figure}
        \centering
        \includegraphics[width=1\linewidth, clip, trim=0 2.5cm 0 2.5cm]{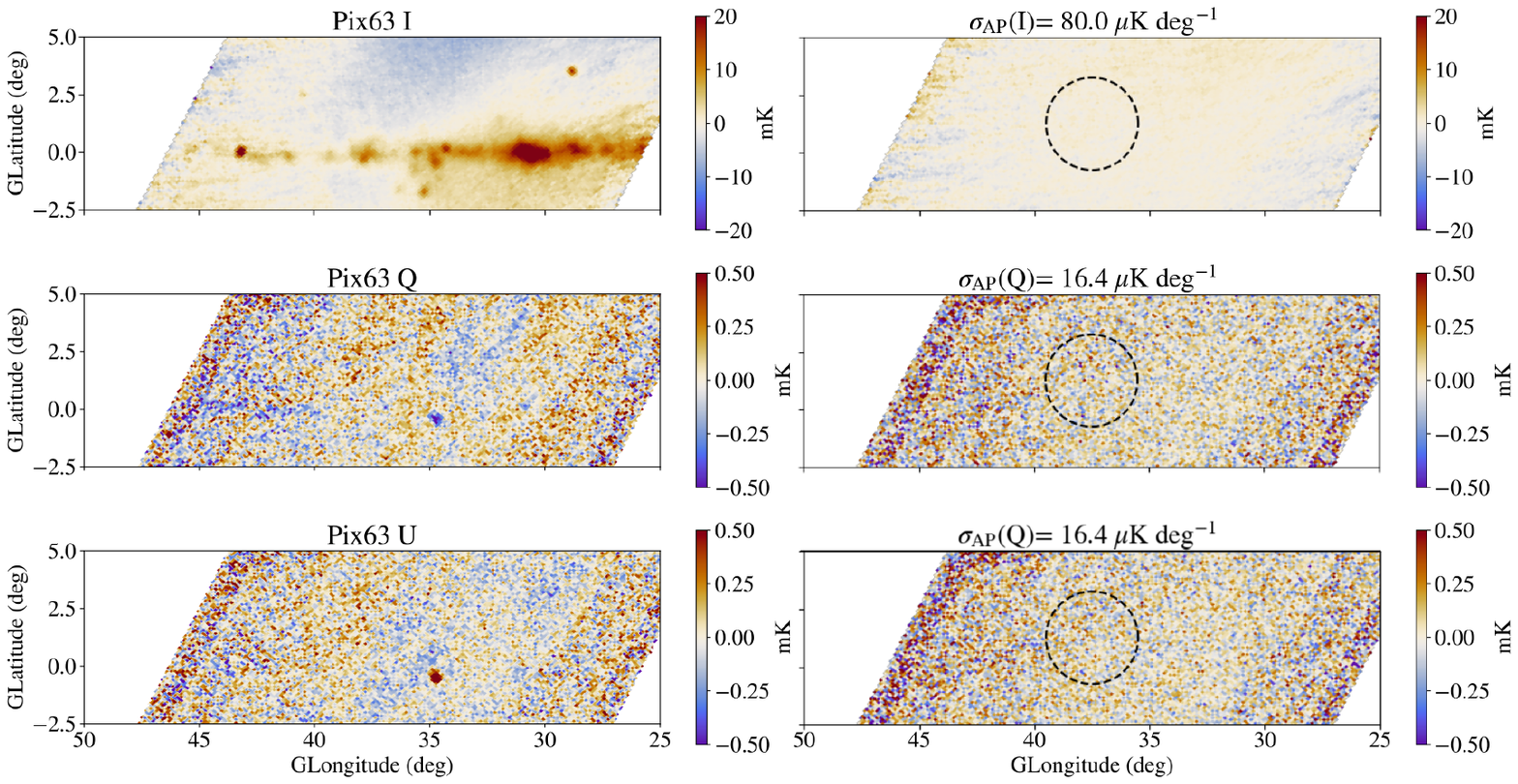}
        \caption{Same as Fig.~\ref{fig:sens_w44_pix23}, but for FGI pixel 63.}
        \label{fig:sens_w44_pix63}
    \end{figure}

\begin{figure}
        \centering
        \includegraphics[width=0.04625\linewidth, clip, trim=0 3.095cm 27.8cm 2.5cm]{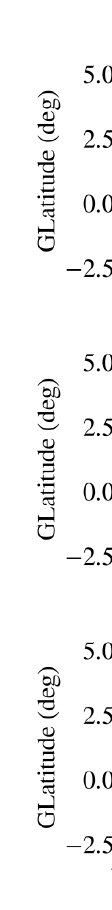}
        \includegraphics[width=0.28\linewidth, clip, trim=15.5cm 3.095cm 2.75cm 2.5cm]{figures/pol_sensitivities/W44_cumul_map_and_halfmaps_orig_das24+21_WEI_mK_RMS_limits_JARM_mambrino_9july2026_orig_reso_ADEPTEDFORMAT.pdf}
        \includegraphics[width=0.28\linewidth, clip, trim=25.4cm 0 4.6cm 0]{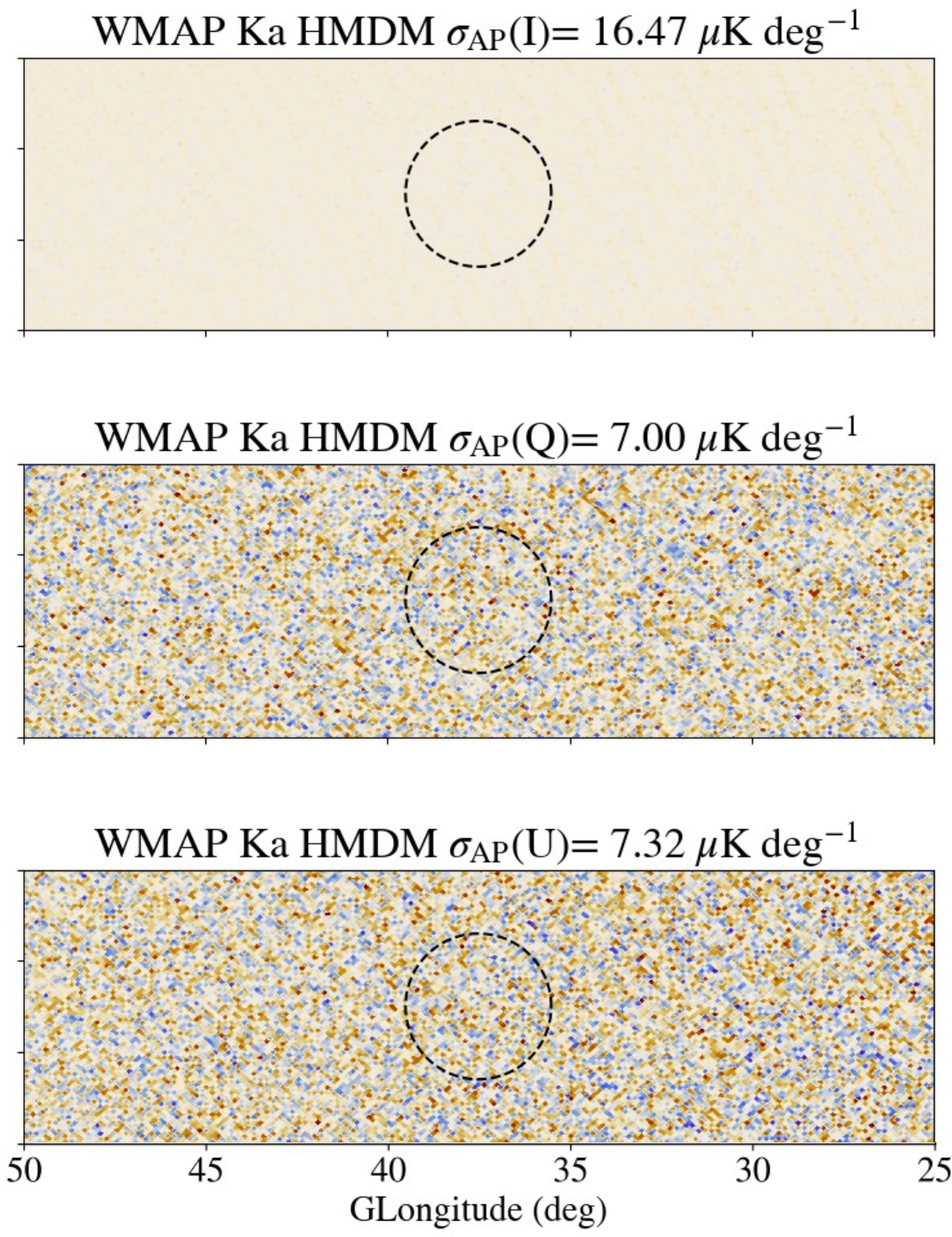}
        \includegraphics[width=0.28\linewidth, clip, trim=25.4cm 0 4.6cm 0]{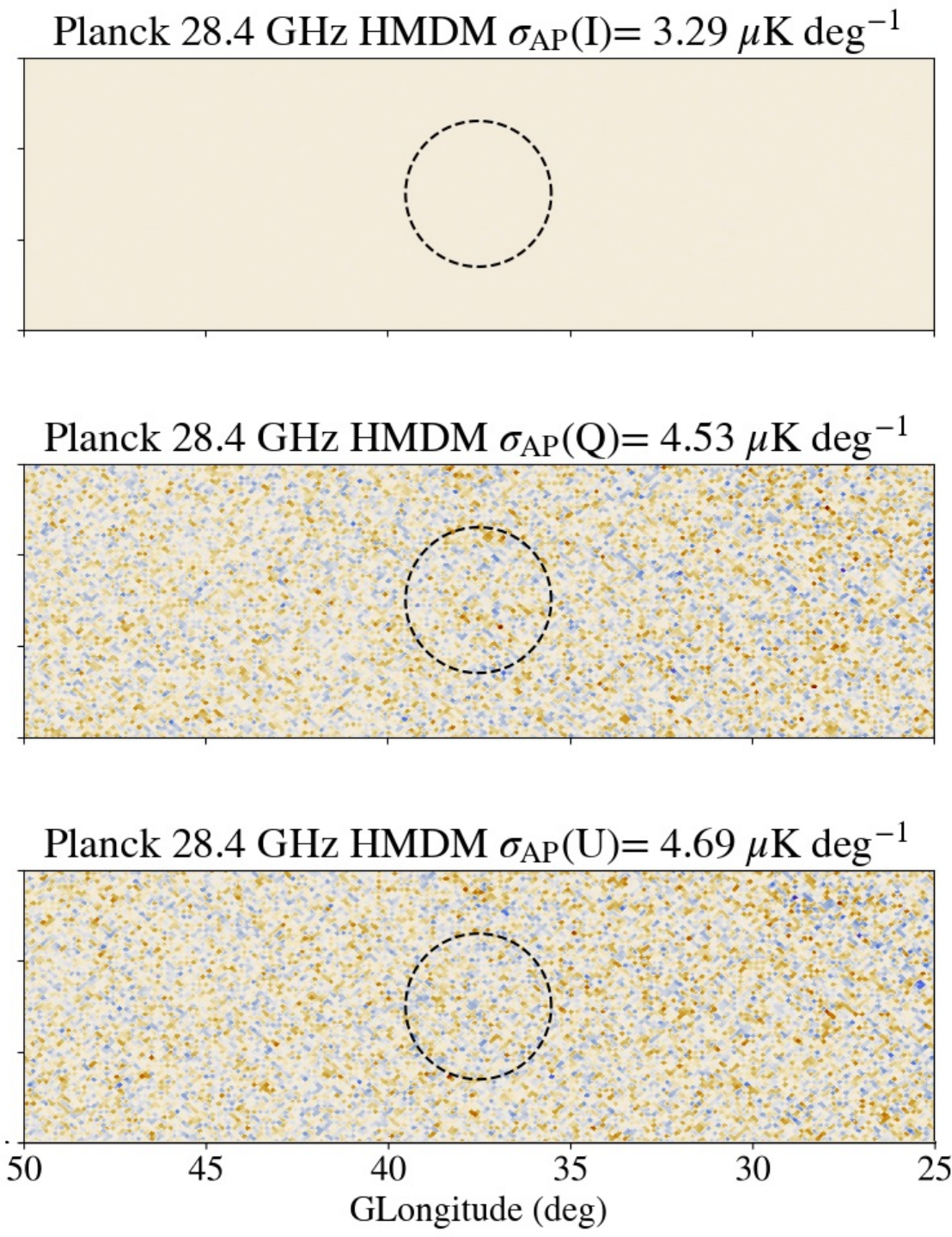}
        \includegraphics[width=0.07\linewidth, clip, trim=44.15cm 0 0 0]{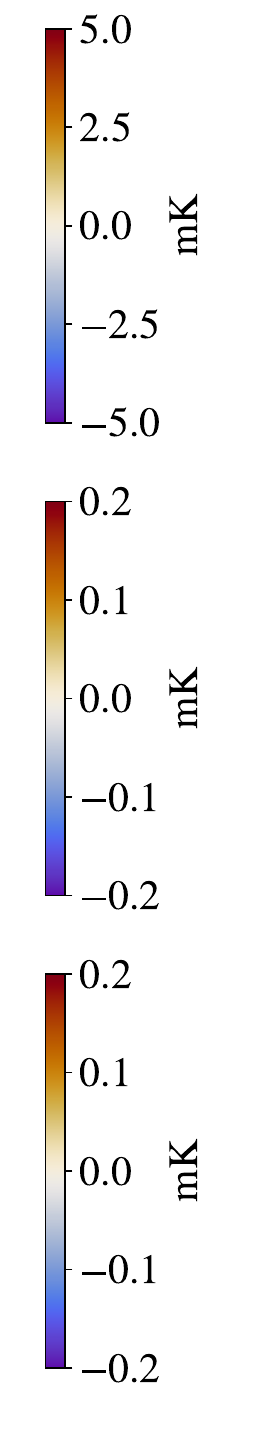}
        \caption{From top to bottom, maps for Stokes parameters $I$, $Q$ and $U$.
        Left: HMDM from the \changes{two TGI pixels} combined. Middle: similar plot for
        WMAP Ka HMDM (subtracting maps from years 1--4 and 5--9). Right: similar 
        plot for Planck 28.4\GHz{} ({\tt ringhalf} halfmaps). \changes{We note that the achieved intensity sensitivity for the combined TGI map is worse than that of the individual pixel~23 map because our combination is obtained using a simple average rather than inverse-variance or otherwise optimal weighting. This approach was adopted as our main goal is to obtain a representative estimate of the instrument sensitivity in polarization.}}
        \label{fig:sens_pixels_combined}
    \end{figure}

    We compute the RMS in these HMDM within a circular region with radius
    equal to 3\degr. We show these RMS estimates, in $\muK{}\deg{}^{-1}$ units, in
    Table~\ref{tab:w44_pol_sensitivities}. We show the HMDM maps in 
    Figs.~\ref{fig:sens_w44_pix23}, \ref{fig:sens_w44_pix26} and \ref{fig:sens_w44_pix63}. 
    Finally, we show in Fig.~\ref{fig:sens_pixels_combined} the HMDM from the combination of the 
    two TGI pixels, and we compare those with HMDM from both WMAP (years 1-4 minus years 5-9)   
    and \textit{Planck} (\texttt{ringhalf} halfmaps).
    We see that the intensity performance from the TGI pixels is worse than
    both WMAP and \textit{Planck}, as expected due to the significant $1/f$ noise contribution.
    However, its polarization sensitivity is already only $20\,\%$ higher than that of WMAP. The final polarization sensitivity for this TGI map is $\sim{8.3}\muK{}\deg{}^{-1}$.

    \begin{table}[]
        \centering
        \caption{$IQU$ sensitivities for the three TFGI pixels considered, from the
        analysis of the HMDM from the $l\in(25\degr, 50\degr)$ region.}
        \begin{tabular}{ccccc}
\hline \hline
  Frequency & \multirow{2}{*}{Pixel}  &  $\sigma(I)$  &  $\sigma(Q)$  &  $\sigma(U)$   \\
  (GHz) & & \multicolumn{3}{c}{($\mu$K deg$^{-1}$)} \\
\hline
% \multirow{3}{*}{31} &   23    &          112.2          &           10.6           &          10.0           \\
% &   26    &         173.9          &           13.3           &          13.3           \\
% & 23 \& 26 &          102.1         &           8.5           &          8.3           \\ % 4july2025 data
\multirow{3}{*}{31} &   23    &          \changes{60.1}          &           \changes{10.1}           &          \changes{9.6}           \\
&   26    &         \changes{116.2}          &           \changes{13.2}           &          13.4           \\
& 23 \& 26 &          \changes{64.5}         &           \changes{8.4}           &          \changes{8.2}           \\ 
41 &   63    &         \changes{80.0}          &           \changes{16.4}          &          \changes{16.4}           \\ % 9july2026 data
\hline
WMAP 33 & --- & 16.5 & 7.0 & 7.3 \\
\textit{Planck} 28.4 & --- & 3.29 & 4.5 & 4.7 \\ 
\hline
\end{tabular}
        \tablecomments{We also show the sensitivity 
        at 31\GHz{} computed from the combination of pixels 23 and 26. The sensitivity at 41\GHz{}
        is directly that from pixel 63. We also quote
        the values from the WMAP and \textit{Planck} closest bands to 31\,GHz{} (33 and
        28.4\,GHz{}, respectively) for
        comparison purposes, following the discussion in the text.}
        \label{tab:w44_pol_sensitivities}
    \end{table}
    
    From the $\Nhits$ map we compute a mean observation
    time per area $t_{\rm pix}=0.57\,\h{}\deg^{-2}$ within our test aperture. Each of the cosmological
    fields that we plan to observe with the TFGI accounts for $1200\deg^2$, so the
    total observed area would be $3600\deg^2$. One year of continuous observations of
    these (three) fields would yield a %$5.8
    $6.6\muK{}\deg^{-1}$ sensitivity, assuming 18 hours per day devoted to observing
    the fields\footnote{The remaining observing time would be devoted to calibration sources, sky dips, and additional astrophysical targets.} and a 50\,\% observing efficiency\footnote{This was the efficiency achieved by the MFI; it may be an optimistic value for the TFGI, as its systematics are not yet as stable.}.
    However, we are currently upgrading the instrument
    by installing a new ring of 12 additional pixels, increasing the number of TGI 
    and FGI pixels to 10 and 9, respectively. With TGI's 10 pixels,
    we will achieve a %$2.6
    ${2.9}\muK{}\deg^{-1}$ sensitivity on the TGI map after a year of observations.
    
\begin{figure}
        \centering
        \includegraphics[width=0.6\linewidth]{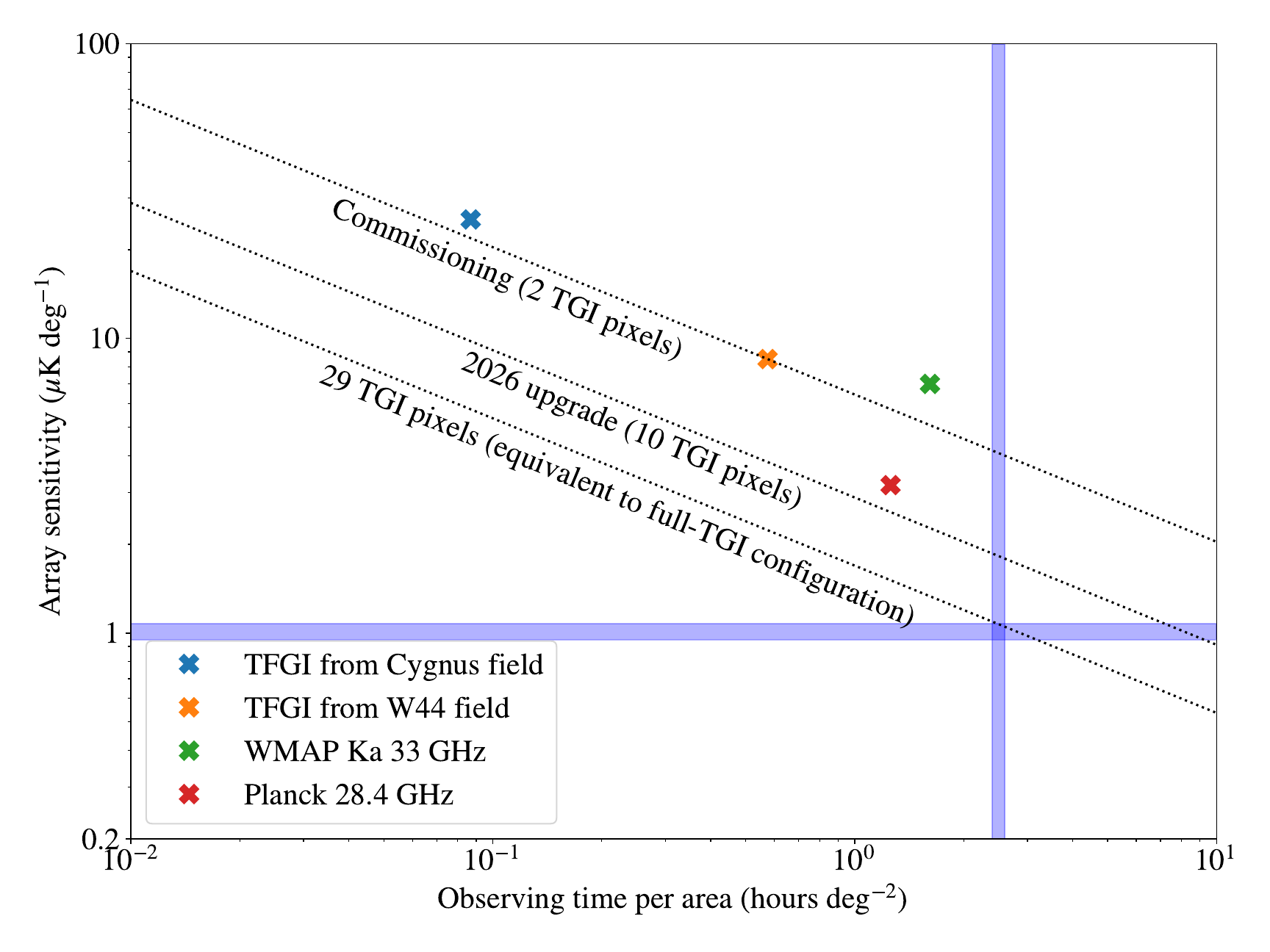}
        \caption{Dependence on the polarization sensitivity of the TFGI with
        time. We have considered three different instrument configurations: 
        that presented in this paper, with only 2 TGI pixels; one with the next
        ring of pixels already installed in the instrument, with 10 TGI pixels
        (together with 9 FGI pixels that we do not show in this
        plot), and an hypothetic one where all detectors observe at 31\GHz{}. In reality, the final configuration of the instrumentation will consist of 15 TGI detectors and 14 FGI ones. 
        Shadowed regions mark $1\muK{}\deg^{-1}$
        and 2.5\,hours\,$\deg^{-2}$ levels, representative of the final goal of the
        TFGI and the coverage expected after 1 year of continuous observations,
        respectively.}
        \label{fig:sensitivies_summary}
    \end{figure}

    From these estimates we compute the amount of time required to achieve a
    $1\muK{}\deg^{-1}$ sensitivity on these three cosmological fields, which remains the 
    main goal of the TFGI. With the 10 TGI pixels already installed
    today in the instrument, we would need an observing run of %6.6
    almost 9 years to obtain a 31\GHz{} map with such sensitivity (still assuming a 50\,\% 
    observing efficiency). Once the full array of 29 detectors is installed in the telescope, 15 would be devoted to the TGI, and 14 to the FGI. The time required to reach a sensitivity of $1\muK{}\deg^{-1}$ at 31\,GHz with 15 detectors will be \changes{5.7} years. The FGI map would take a bit longer because of the 14 detectors instead, but we expect their performance to match that from TGI
    once the secondary calibration based on the diode is implemented, as explained in Sect.~\ref{sec:cal_pol_ang}.
    In Fig.~\ref{fig:sensitivies_summary}
    we show a summary of these expected sensitivities against the observation
    time. 
    \changes{This forecast relies on the reasonable assumption that the performance of the remaining TFGI detectors will be comparable to that reported here for the three analysed pixels. Once the calibration diode is fully operational, allowing polarization-angle variations to be tracked on short timescales, the polarization efficiency is expected to improve further.}
    % \changes{This forecast relies on the reasonable assumption that the performance of the remaining TFGI detectors will be comparable to that reported here for the three analysed pixels. Once the calibration diode is fully operational, allowing polarization-angle variations to be tracked on short timescales, the polarization efficiency is expected to improve further}
    
%--------------------------------------------------------------------

\section{Conclusions}
\label{sec:conclusions}

    In this paper we presented the analysis of the commissioning data of the TFGI
    instrument, that is installed on the second QUIJOTE telescope. During this period
    only seven receivers were installed, four operating at 31\,GHz and three at 41\,GHz. The commissioning tests included 
    a characterization of the beam, an initial gain model, an assessment of the 
    polarization performance, and a first estimate of the instantaneous sensitivity. 
    The latter yielded results consistent with expectations, $\sim 350\,\muK{}\,\s^{1/2}$,
    with measured values ranging from $314$ to $376\muK{}\s^{1/2}$. 
    
    The main issue we encountered was a lower than expected polarization efficiency on
    TGI pixel 25 and FGI pixels 41 and 42. Laboratory measurements showed their
    polarization efficiencies to be at the $\sim$90\,\% level, while apparently 
    being under 50\,\% from 
    on-sky (calibrator -Tau A- observations) data. We credit this decrease to time
    variations on the differential gains between the branches of the instrument (mainly
    the amplifiers inside the BEMs, which are not cooled), which effectively translated as 
    varying polarization angles. This partly cancelled out $Q$ and $U$ when
    coadding data, and therefore worsened the polarization performance from the pixels.
    % The problem would be mitigated if the variations on the polarization angle 
    % could be perfectly traced. 
    \changes{This issue led us to exclude these three pixels from the subsequent analyses presented in this paper, which focus on polarization measurements. Their low polarization efficiency implied that including them would only marginally improve the final sensitivity, while nearly doubling the amount of data to be processed and, consequently, the complexity of the analysis. We emphasize, however, that this does not imply that these pixels were malfunctioning. Their performance in total intensity was comparable to, and in some cases even better than, that of the other pixels. Rather, the limitation arose from the lack of a calibration method in our data-processing pipeline capable of tracking variations in the polarimetric phase on short timescales. This points to a specific issue that will be} solved once the variations on the polarization angle can be perfectly traced. 
    This will be achieved with the new instrument configuration,
    expected to be installed on the telescope during the second half of 2026, which includes a polarized
    calibration diode that \changes{will inject a polarimetric signal with known properties every 30\s{}}. The
    diode will also help to improve the other results discussed in this work, such as
    \changes{the angle calibration of the rest of the pixels and the calibration of the gain}.
    Because of this, the forecasts given here should be interpreted as conservative scenarios with the potential to be further improved.
    
    After presenting the commissioning tests and issues, we described the first analyses on scientific
    data from the other three pixels: TGI pixels 23 and 26, and FGI pixel 63. These pixels
    showed a response consistent with that expected, while being stable throughout the
    commissioning period. The scientific data comprised observations of bright Galactic 
    features in the Cygnus region (for 24.1\,hours) and between $l=25\degr$ and $l=50\degr$
    (for 157.2\,hours). First, we used these data to
    show that the TFGI is able to recover the expected signal from large-scale diffuse 
    emission, with maps compatible with those from WMAP and \textit{Planck}. After that,
    we performed AP on the TFGI intensity maps for three well-known Galactic regions. We showed
    that the computed estimates were again in line with those from WMAP, \textit{Planck}
    and QUIJOTE-MFI, thus validating our gain calibration method. We used the SNR W44 as
    a cross-check of the polarization properties previously obtained with Tau A data.
    We also ran AP on the polarization maps and found consistent (with WMAP and 
    \textit{Planck}) values for the
    polarization angle in all pixels. However, while the polarized intensity ($P$) values
    of TGI pixels were good, there was a deficit of signal in the FGI one. We ascribe
    this defect to the FGI pixel being more affected by the time variations in the 
    polarization angle. W44 is fainter and its data spans a longer period of 
    time than that of Tau A, so the effect on the decreased polarization efficiency
    may be larger. This would explain why we measured a flux defect even after correcting
    for the polarization efficiency obtained from Tau A data. 
    
    Finally, we used this astrophysical dataset to build noise maps (HMDM) for these
    three pixels. We focused our analysis on the two TGI pixels available because of the
    mentioned issue on FGI observations: in any case, the much better data from TGI 
    dominated the combined map even when considering FGI.
    After combining these two TGI pixels, we found the sensitivity on the maps to be 
    $\sim{8.3}\muK\deg^{-1}$, almost matching WMAP performance in the region with 
    only 35\,\% of its effective observing time per squared degree, those being $0.57\,{\rm h}\deg^{-2}$ and 
    $1.61\,{\rm h}\deg^{-2}$, respectively.
    From this estimate, we forecasted the time the TFGI would require to achieve a
    $1\muK\deg^{-1}$ sensitivity on a combination of three different cosmological fields,
    accounting for $\sim3600\deg^2$ in total. We also considered the planned upgrades for
    the instrument: in 2026 the TFGI will be installed again at the telescope with 19
    pixels, while the final update will increase the number to 29 pixels. In this
    last case, the TFGI will need {5.7} years to achieve its goal of 
    $1\muK\deg^{-1}$ sensitivity both at 31 and 41\GHz{} on these cosmological
    fields. Such a sensitivity level
    will allow us to put a $0.05$ upper limit on the tensor-to-scalar ratio, $r$, with observations from the Northern hemisphere.

%--------------------------------------------------------------------

\begin{acknowledgements}
We thank the anonymous referee for their comments, which helped to improve this work.
We thank the staff of the Teide Observatory for invaluable assistance in the commissioning and operation of QUIJOTE.
The {\it QUIJOTE} experiment is being developed by the Instituto de Astrofisica de Canarias (IAC), the Instituto de Fisica de Cantabria (IFCA), and the Universities of Cantabria, Manchester and Cambridge. 
Partial financial support was provided by the Spanish Ministry of Science and Innovation
under the projects AYA2007-68058-C03- 01, AYA2007-68058-C03-02, 
AYA2010-21766-C03-01, AYA2010-21766-C03- 02, AYA2014-60438-P, 
ESP2015-70646-C2-1-R, AYA2017-84185-P, ESP2017- 83921-C2-1-R, 
PGC2018-101814-B-I00, PID2019-110610RB-C21, PID2020-120514GB-I00, 
IACA13-3E-2336, IACA15-BE-3707, EQC2018-004918-P, 
PID2023-150398NB-I00 and PID2023-151567NB-I00, 
the Severo Ochoa Programs SEV-2015-0548 and CEX2019-000920-S, 
the Maria de Maeztu Program MDM-2017-0765, and by the Consolider-Ingenio project CSD2010-00064 (EPI: Exploring the Physics of Inflation). 
We acknowledge support from the ACIISI, Consejeria de Economia, Conocimiento y Empleo del Gobierno de Canarias and the European Regional Development Fund (ERDF) under grant with reference ProID2020010108, and Red de Investigaci\'on RED2022-134715-T funded by MCIN/AEI/10.13039/501100011033. 
This project has received funding from the European Union's Horizon 2020 research and innovation program under grant agreement number 687312 (RADIOFOREGROUNDS), and the Horizon Europe research and innovation program under GA 101135036 (RadioForegroundsPlus). 

The author(s) wish to acknowledge the contribution of the IAC 
High-Performance Computing support team and hardware facilities to the 
results of this research.
MFT acknowledges support from the Enigmass+ research federation (CNRS,
Université Grenoble Alpes, Université Savoie Mont-Blanc); from
the Agencia Estatal de Investigación (AEI) of the Ministerio de Ciencia,
Innovación y Universidades (MCIU) and the European Social Fund (ESF)
under grant with reference PRE-C-2018-0067; and from 
grant JDC2024-055310-I funded by MICIU/AEI/10.13039/501100011033 and ESF+. DH acknowledges support from the Spanish State Research Agency (AEI) under grant number PID2022-140670NA-I00.
CITIC, as a center accredited for excellence within the Galician University System and a member of the CIGUS Network, receives subsidies from the Department of Education, Science, Universities, and Vocational Training of the Xunta de Galicia. Additionally, it is co-financed by the EU through the FEDER Galicia 2021-27 operational program (Ref. ED431G 2023/01).

We acknowledge using the 
colormap prepared by M. Petroff \citep[\url{https://mpetroff.net/2023/05/an-improved-cmb-map-colormap/,}][]{class_colormap},
and sincerely thank the effort put to concile historical tendencies on
CMB representation with inclusion and diversity in science.
\software{numpy \citep{numpy}, emcee \citep{emcee}, matplotlib \citep{matplotlib}, astropy \citep{astropy}, HEALPix \citep{healpix}, healpy \citep{healpy}}
\end{acknowledgements}

%--------------------------------------------------------------------

\bibliography{quijote,bibliography_TFGI}{}
\bibliographystyle{aasjournalv7}

\begin{appendix} %First appendix

\section{External data sets used in multifrequency analysis}
\label{appendix:SED_data}

The full set of maps used is shown in 
Table~\ref{tab:ancillary_maps}. We highlight which maps have polarization data in
addition to the intensity set, as we build both SEDs when possible.

\begin{table*}[]
    \centering
    \caption{Surveys used to build the spectral energy distributions (SED) from
    Sect.~\ref{section:galactic_regions}. \cite{nobeyama} data was only used to build
    $30^\prime$ SEDs, due to its reduced coverage span in latitude.}
    \begin{tabular}{ccccccc}
\hline \hline
\multirow{2}{*}{Telescope} & \multirow{2}{*}{Stokes} & Frequency & Calibration & FWHM & \multirow{2}{*}{Sky Coverage} & \multirow{2}{*}{Reference} \\ 
 & & (GHz) & (\%) & (arcmin) & & \\ \hline
\multirow{2}{*}{Various} & \multirow{2}{*}{$I$} & \multirow{2}{*}{0.408} & \multirow{2}{*}{10} & \multirow{2}{*}{51} & \multirow{2}{*}{All-sky} &~\cite{haslam1982} \\ 
& & & & & &~\cite{remazeilleshaslam} \\
Dwingeloo & $I$ & 0.82 & 10 & 72 & $\delta>-7\degr$ &~\cite{dwingeloo} \\ 
\multirow{2}{*}{Effelsberg} & \multirow{2}{*}{$I$} & \multirow{2}{*}{1.408} & \multirow{2}{*}{10} & \multirow{2}{*}{9.4} & $l\in[240,\ 357]\degr$ & \multirow{2}{*}{~\cite{effelsberg1, effelsberg2}} \\
 & & & & & $b\in[-4,\ 4]\degr$ & \\
\multirow{2}{*}{DRAO/Villa-Elisa} & \multirow{2}{*}{$QU$} & \multirow{2}{*}{1.41} & \multirow{2}{*}{5} & \multirow{2}{*}{36} & \multirow{2}{*}{All-sky} &~\cite{drao1} \\ 
& & & & & &~\cite{drao2} \\
\multirow{3}{*}{Stockert/Villa-Elisa} & \multirow{3}{*}{$I$} & \multirow{3}{*}{1.42} & \multirow{3}{*}{20} & \multirow{3}{*}{34.2} & \multirow{3}{*}{All-sky} &~\cite{reich1982, reich1986} \\ 
& & & & & &~\cite*{villaelisa} \\
& & & & & & ~\cite{CADE} \\
\multirow{2}{*}{HartRAO} & \multirow{2}{*}{$I$} & \multirow{2}{*}{2.326} & \multirow{2}{*}{0.20} & \multirow{2}{*}{20} & \multirow{2}{*}{$\delta<13\degr$} & \cite{hartrao1} \\
& & & & & & ~\cite{hartrao2} \\
\multirow{2}{*}{Effelsberg} & \multirow{2}{*}{$IQU$} & \multirow{2}{*}{2.695} & \multirow{2}{*}{10} & \multirow{2}{*}{4.3} & $l\in[240,\ 358]\degr$ & \multirow{2}{*}{~\cite{effelsberg3, effelsberg4}} \\
& & & & & $b\in[-5,\ 5]\degr$ & \\
\multirow{2}{*}{Stockert} & \multirow{2}{*}{$I$} & \multirow{2}{*}{2.72} & \multirow{2}{*}{10} & \multirow{2}{*}{21} & $l\in[0,\ 80]\degr$ & \multirow{2}{*}{~\cite{stockert_11cm_1, stockert_11cm_2}} \\
& & & & & $b\in[-16,\ 30]\degr$ & \\
\multirow{2}{*}{Urumqi} & \multirow{2}{*}{$IQU$} & \multirow{2}{*}{4.8} & \multirow{2}{*}{4} & \multirow{2}{*}{9.5} & $l\in[10,\ 230]\degr$ & \cite{urumqi1, urumqi2}\\
& & & & & $b\in[-5,\ 5]\degr$ & \cite{urumqi3, urumqi4}\\
\multirow{2}{*}{Nobeyama} & \multirow{2}{*}{$I$} & \multirow{2}{*}{10} & \multirow{2}{*}{10} & \multirow{2}{*}{3} & $l\in[355,\ 56]\degr$ & \multirow{2}{*}{\cite{nobeyama}} \\
& & & & & $b\in[-1.5,\ 1.5]\degr$ & \\
QUIJOTE-MFI & $IQU$ & 11.2 & 5 & 53.2 & $\delta>-32\degr$ &~\cite{mfiwidesurvey} \\ 
QUIJOTE-MFI & $IQU$ & 12.9 & 5 & 53.5 & $\delta>-32\degr$ &~\cite{mfiwidesurvey} \\ 
QUIJOTE-MFI & $IQU$ & 16.8 & 5 & 39.1 & $\delta>-32\degr$ &~\cite{mfiwidesurvey} \\ 
QUIJOTE-MFI & $IQU$ & 18.7 & 5 & 39.1 & $\delta>-32\degr$ &~\cite{mfiwidesurvey} \\ 
WMAP K 9yr  & $IQU$ & 22.8 & 3 & 51.3 & All-sky &~\cite{bennett2013} \\ 
\textit{Planck}-LFI PR3 & $IQU$ & 28.4 & 3 & 33.1 & All-sky &~\cite{planckPR3} \\ 
QUIJOTE-TFGI & $IQU$ & 31 & 10 & 21 & Patches & This work \\
WMAP Ka 9yr & $IQU$ & 33 & 3 & 39.1 & All-sky &~\cite{bennett2013} \\ 
WMAP Q 9yr & $IQU$ & 40.7 & 3 & 30.8 & All-sky &~\cite{bennett2013} \\ 
QUIJOTE-TFGI & $IQU$ & 41 & 10 & 18 & Patches & This work \\
\textit{Planck}-LFI PR3 & $IQU$ & 44.1 & 3 & 27.9 & All-sky &~\cite{planckPR3} \\ 
WMAP V 9yr & $IQU$ & 60.7 & 3 & 21.0 & All-sky &~\cite{bennett2013} \\ 
\textit{Planck}-LFI PR3 & $IQU$ & 70.4 & 3 & 13.1 & All-sky &~\cite{planckPR3} \\ 
WMAP W 9yr & $IQU$ & 93.5 & 3 & 14.8 & All-sky &~\cite{bennett2013} \\ 
\textit{Planck}-HFI PR3 & $IQU$ & 143 & 3 & 7.3 & All-sky &~\cite{planckPR3} \\ 
\textit{Planck}-HFI PR3 & $IQU$ & 353 & 3 & 4.9 & All-sky &~\cite{planckPR3} \\ 
\textit{Planck}-HFI PR3 & $I$  & 545 & 6.1 & 4.8 & All-sky &~\cite{planckPR3} \\ 
\textit{Planck}-HFI PR3 & $I$ & 857 & 6.4 & 4.6 & All-sky &~\cite{planckPR3} \\ 
COBE-DIRBE & $I$ & 1249 & 11.6 & 37.1 & All-sky &~\cite{cobe-dirbe} \\ 
COBE-DIRBE & $I$ & 2141 & 10.6 & 38.0 & All-sky &~\cite{cobe-dirbe} \\ 
COBE-DIRBE & $I$ & 2998 & 13.5 & 38.6 & All-sky &~\cite{cobe-dirbe} \\
\multirow{2}{*}{IRAS (IRIS)} & \multirow{2}{*}{$I$} & \multirow{2}{*}{3000} & \multirow{2}{*}{10} & \multirow{2}{*}{4} & \multirow{2}{*}{All-sky} & \cite{iras} \\
& & & & & & \cite{iris} \\
\hline
\end{tabular}
    \label{tab:ancillary_maps}
\end{table*}

\section{Pointing model solution}
\label{appendix:pointing_model_parameters}

Most of the parameters from the pointing model can be separated in two heavily degenerate groups, 
which effectively correct the azimuth ($P_a$, $P_c$, $P_n$) and elevation ($P_b$, $P_f$) 
positions. Observing as many different sources as possible with
different positions on the sky is mandatory to break those degeneracies and
achieve a model that works correctly. This highlights the importance of using 
astrophysical calibrators (Tau A, Cas A) together with planets (Venus and Jupiter) 
to achieve the best possible solution. The values for the pointing model parameters are
shown in Table~\ref{tab:pointing_model_parameters}. The detailed description of these
parameters, and how they transform the telescope encoder coordinates to the sky coordinates,
is available in \cite{PhD_Mateo}.

\begin{table}[]
    \centering
    \begin{tabular}{c|c} \hline \hline
    Parameter & Fitted value (degrees) \\\hline
     $P_a$ & $3.198\pm0.036$ \\
     $P_b$ & $-0.107\pm0.007$ \\
     $P_c$ & $0.047\pm0.063$ \\
     $P_n$ & $-0.118\pm0.055$ \\
     $P_f$ & $0.033\pm0.010$ \\
     $P_x$ & $-0.011\pm0.002$ \\
     $P_y$ & $-0.007\pm0.002$ \\ \hline
\end{tabular}
    \caption{Final values for the pointing model parameters for the TFGI.}
    \label{tab:pointing_model_parameters}
\end{table}

\section{Monthly gain factors and polarization angles}
\label{appendix:all_gain_calibration}

In this Appendix we show the gain factors for TGI pixels 23 and 26, plus FGI
pixel 63, computed from the coadded monthly maps: they are available in 
Tables~\ref{tab:gain_das24} to \ref{tab:gain_das25}.
We also quote the values for the calibration angles calculated
from the same monthly maps, this time for Stokes $Q$ and $U$ 
parameters, as described in Sect.~\ref{sec:cal_pol_ang}.
These are available in Tables~\ref{tab:phic_values_monthly_pix26} and 
~\ref{tab:phic_values_monthly_pix63} for pixels 26 and 63, respectively 
(the one for pixel 23 being in the article).

\begin{table*}[h]
    \centering
    \caption{Gain estimates for pixel 23, computed from the coadded monthly maps.
    Units are in mV/K. The numbers within parentheses are computed as the
    ratio (in \%) of the uncertainty with respect to the gain value.}
    \begin{tabular}{ccccc}
\hline
   Month & CH1                  & CH2                  & CH3                  & CH4                  \\
\hline
    2111 & 36.54$\pm$0.80 (2.2) & 33.22$\pm$0.58 (1.7) & 34.41$\pm$0.88 (2.6) & 37.50$\pm$1.03 (2.8) \\
    2112 & 35.88$\pm$0.24 (0.7) & 32.09$\pm$0.45 (1.4) & 33.79$\pm$0.29 (0.9) & 36.70$\pm$0.77 (2.1) \\
    2201 & 26.29$\pm$0.62 (2.4) & 23.91$\pm$0.64 (2.7) & 25.09$\pm$0.62 (2.5) & 27.44$\pm$0.68 (2.5) \\
    2202 & 25.72$\pm$0.24 (0.9) & 23.20$\pm$0.15 (0.6) & 24.50$\pm$0.27 (1.1) & 26.58$\pm$0.33 (1.2) \\
    2204 & 22.66$\pm$0.96 (4.3) & 20.52$\pm$0.91 (4.4) & 21.39$\pm$0.93 (4.4) & 23.32$\pm$1.08 (4.6) \\
    2205 & 20.96$\pm$0.37 (1.8) & 18.95$\pm$0.33 (1.8) & 19.86$\pm$0.43 (2.1) & 21.70$\pm$0.42 (1.9) \\
\hline
\end{tabular}
    \label{tab:gain_das24}
\end{table*}

\begin{table*}[h]
    \centering
    \caption{Gain estimates for pixel 26, computed from the coadded monthly maps, as per Table~\ref{tab:gain_das24}.}
    \begin{tabular}{ccccc}
\hline
   Month & CH1                   & CH2                   & CH3                   & CH4                   \\
\hline
    2111 & 30.34$\pm$0.87 (2.9)  & 31.52$\pm$0.99 (3.1)  & 28.38$\pm$0.51 (1.8)  & 30.15$\pm$0.85 (2.8)  \\
    2112 & 29.93$\pm$0.22 (0.7)  & 30.49$\pm$0.21 (0.7)  & 27.16$\pm$0.21 (0.8)  & 29.80$\pm$0.31 (1.0)  \\
    2201 & 25.55$\pm$0.57 (2.2)  & 26.07$\pm$0.58 (2.2)  & 23.32$\pm$0.26 (1.1)  & 25.47$\pm$0.71 (2.8)  \\
    2202 & 24.54$\pm$0.23 (0.9)  & 23.59$\pm$0.60 (2.5)  & 20.95$\pm$0.88 (4.2)  & 24.63$\pm$0.08 (0.3)  \\
    2204 & 31.87$\pm$3.46 (10.9) & 32.66$\pm$3.75 (11.5) & 33.13$\pm$3.60 (10.9) & 34.37$\pm$3.65 (10.6) \\
    2205 & 30.74$\pm$0.17 (0.6)  & 31.76$\pm$0.28 (0.9)  & 32.06$\pm$0.36 (1.1)  & 33.34$\pm$0.39 (1.2)  \\
\hline
\end{tabular}
    \label{tab:gain_das21}
\end{table*}

\begin{table*}[h]
    \centering
    \caption{Gain estimates for pixel 63, computed from the coadded monthly maps, as per Table~\ref{tab:gain_das24}.}
    \begin{tabular}{ccccc}
\hline
   Month & CH1                  & CH2                  & CH3                  & CH4                  \\
\hline
    2111 & 12.20$\pm$0.83 (6.8) & 12.23$\pm$0.96 (7.9) & 12.86$\pm$0.78 (6.1) & 17.62$\pm$1.23 (7.0) \\
    2112 & 12.80$\pm$0.29 (2.3) & 13.01$\pm$0.26 (2.0) & 13.43$\pm$0.44 (3.3) & 18.21$\pm$0.45 (2.5) \\
    2201 & 10.07$\pm$0.58 (5.8) & 10.37$\pm$0.47 (4.5) & 10.76$\pm$0.51 (4.8) & 14.61$\pm$0.79 (5.4) \\
    2202 & 10.30$\pm$0.41 (4.0) & 10.56$\pm$0.36 (3.4) & 10.81$\pm$0.37 (3.4) & 14.59$\pm$0.34 (2.3) \\
    2204 & 9.00$\pm$0.37 (4.1)  & 8.93$\pm$0.40 (4.5)  & 9.51$\pm$0.35 (3.7)  & 12.55$\pm$0.68 (5.4) \\
    2205 & 9.10$\pm$0.21 (2.3)  & 9.67$\pm$0.19 (2.0)  & 10.00$\pm$0.24 (2.4) & 13.13$\pm$0.27 (2.0) \\
\hline
\end{tabular}
    \label{tab:gain_das25}
\end{table*}

\begin{table}[]
    \centering
    \caption{Values for the calibration angle for pixel 26, computed from the coadded monthly maps. Units are in degrees, with uncertainties lying on the 7--13$^\circ$ range.}
    \begin{tabular}{ccccc}
\hline \hline
Month  &   CH1   &   CH2   &  CH3   &   CH4   \\ \hline
Nov21   & -47.22  &  -0.36  & -91.84 & -132.79 \\
Dec21   & -42.03  &  1.61   & -87.61 & -133.18 \\
Jan22   & -40.04  &  -8.39  & -86.81 & -129.89 \\
Feb22   & -42.57  &  -4.60  & -85.52 & -130.16 \\
Apr22   & -176.48 & -142.29 & -53.35 & -89.23  \\
May22   &  0.18   & -143.02 & -45.54 & -92.82  \\
\hline
\end{tabular}
    \label{tab:phic_values_monthly_pix26}
\end{table}

\begin{table}[]
    \centering
    \caption{Values for the calibration angle for pixel 63, computed from the coadded monthly maps. Units are in degrees, with uncertainties lying on the 10--30$^\circ$ range. In this case we ran a more strict flagging of the data due to the lower polarization efficiency, which resulted in no data from Dec21 being used.}
    \begin{tabular}{ccccc}
\hline
Month  &  CH1   &  CH2   &   CH3   &   CH4   \\
\hline
Nov21   & -48.51 & -77.45 &  6.37   & -124.79 \\  
Jan22   & -46.00 & -76.99 & 176.35  & -135.02 \\  
Feb22   & -47.49 & -82.95 & -174.46 & -125.45 \\  
Apr22   & -33.78 & -83.47 &  8.06   & -129.34 \\  
May22   & -39.00 & -78.55 & 181.28  & -126.75 \\
\hline\end{tabular}
    \label{tab:phic_values_monthly_pix63}
\end{table}

\section{Additional information from Galactic regions}
    \label{appendix:all_SED_regions}
\begin{figure}[h]
        \centering
        \includegraphics[width=0.49\linewidth, clip, trim=0 2.5cm 0 0]{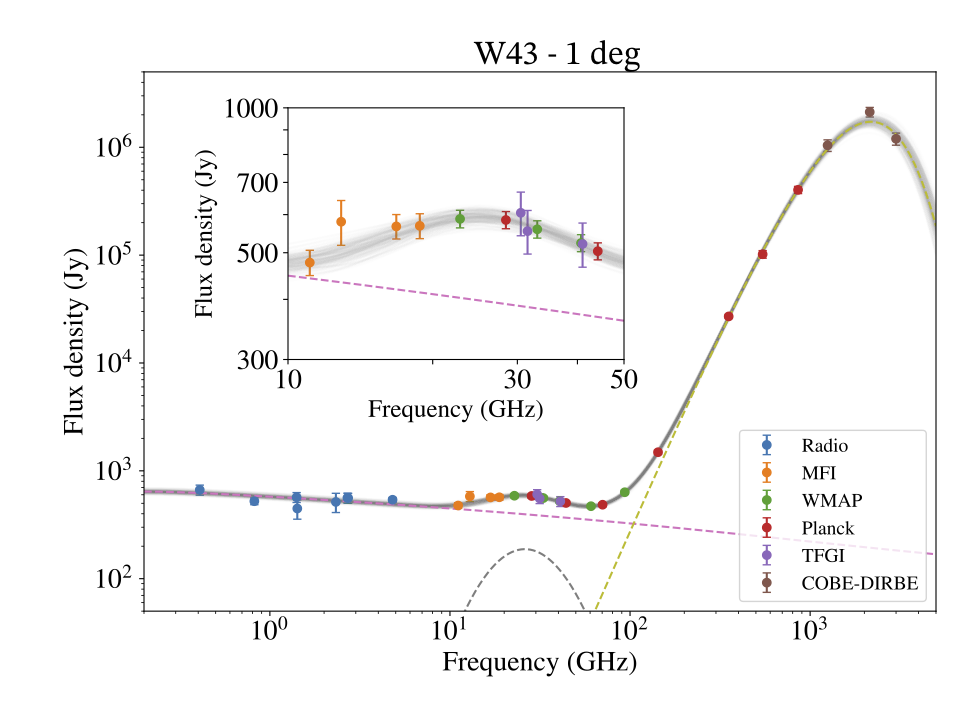}
        \includegraphics[width=0.49\linewidth, clip, trim=0 1.8cm 0 0]{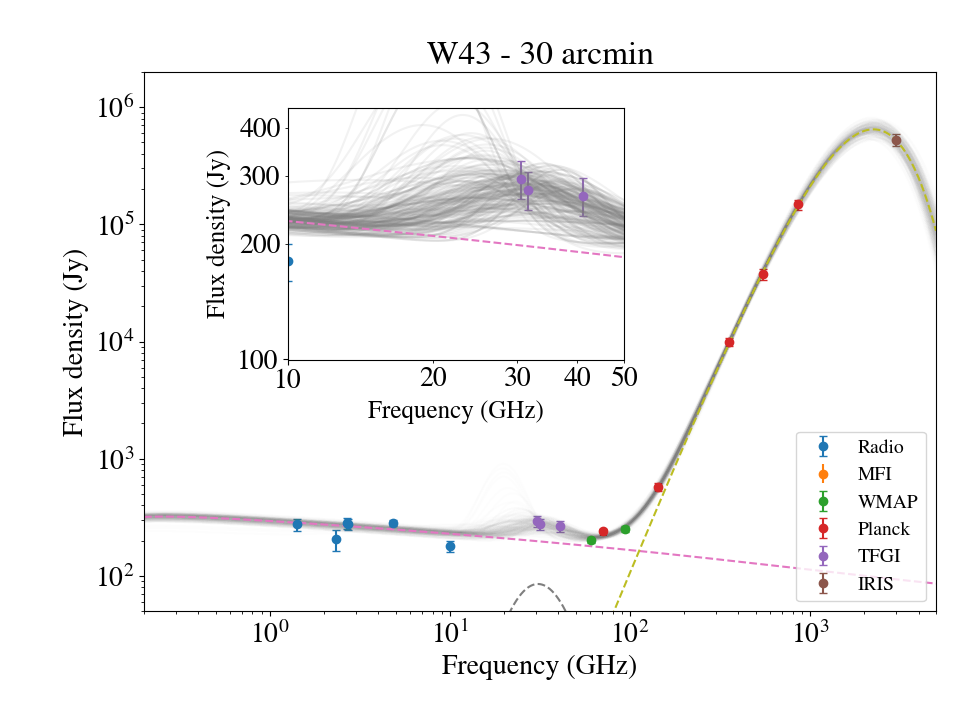}
        \includegraphics[width=0.49\linewidth, clip, trim=0 0 0 0.7cm]{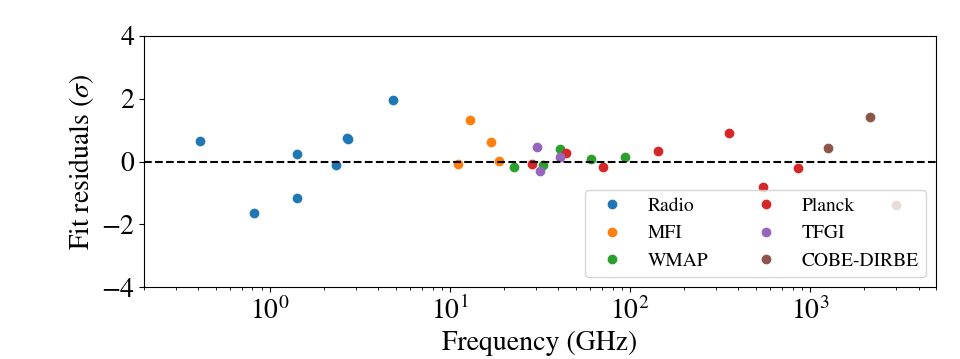}
        \includegraphics[width=0.49\linewidth, clip, trim=0 0 0 0.7cm]{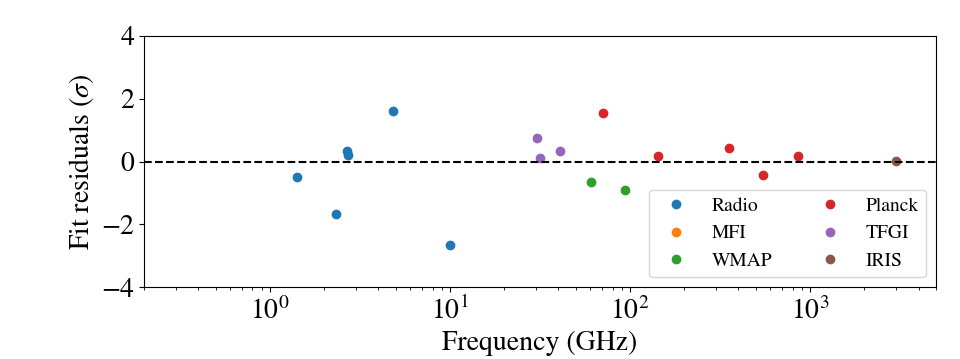}
        \caption{Left: intensity SED for W43 at 1$\degr$ resolution. Right: same SED, now at 30$^\prime$
        resolution.}
        \label{fig:w43_SED}
    \end{figure}

\begin{figure}
        \centering
        \includegraphics[width=0.49\linewidth, clip, trim=0 2.5cm 0 0cm]{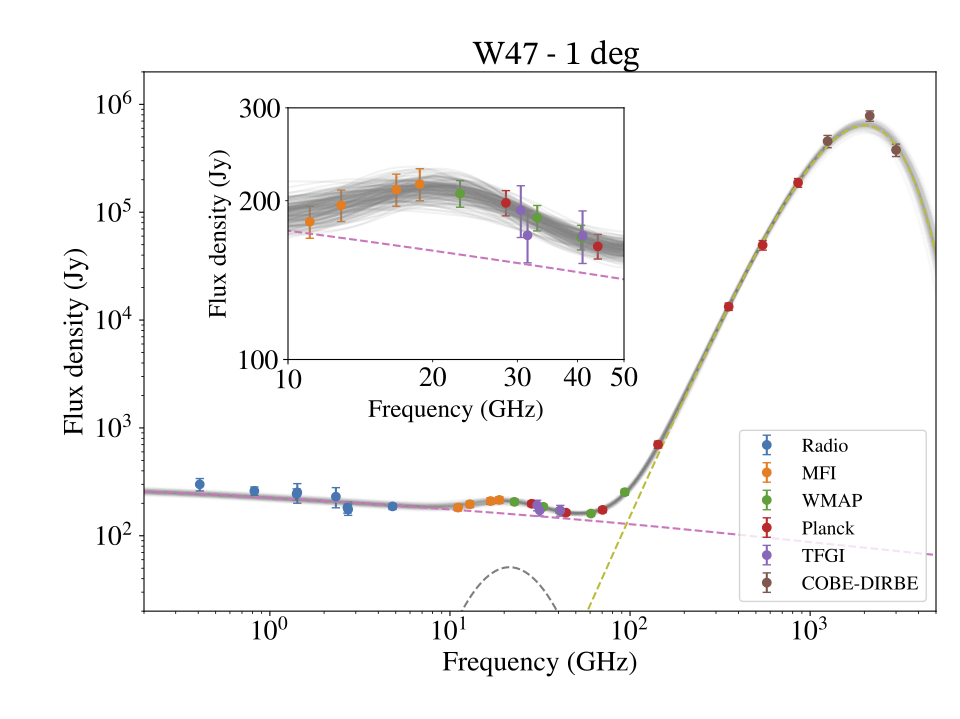}
        \includegraphics[width=0.49\linewidth, clip, trim=0 1.8cm 0 0cm]{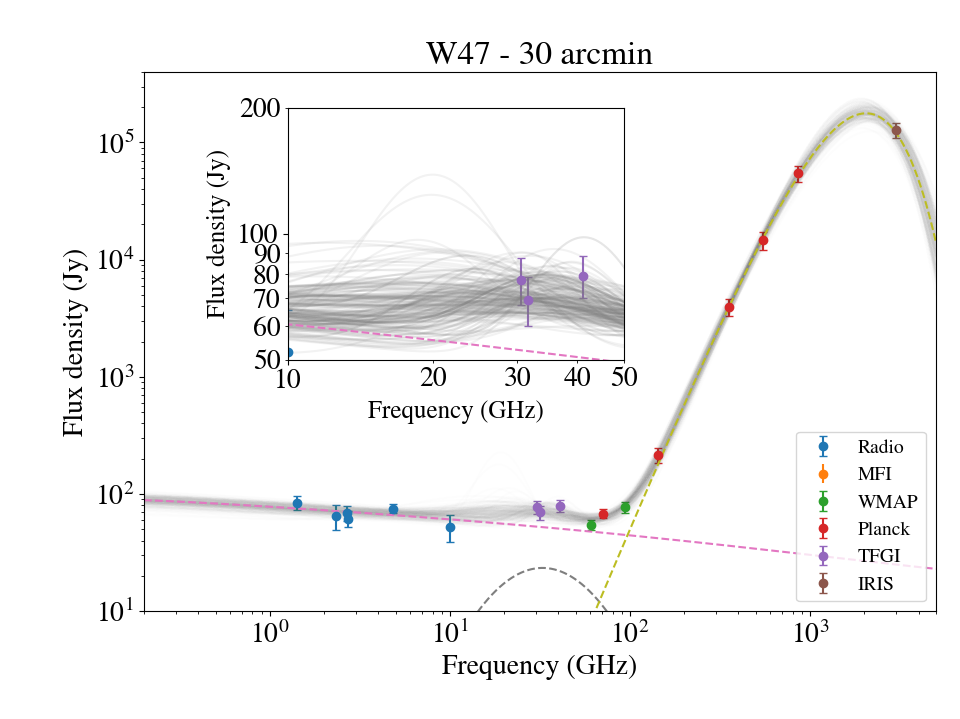}
        \includegraphics[width=0.49\linewidth, clip, trim=0 0 0 0.7cm]{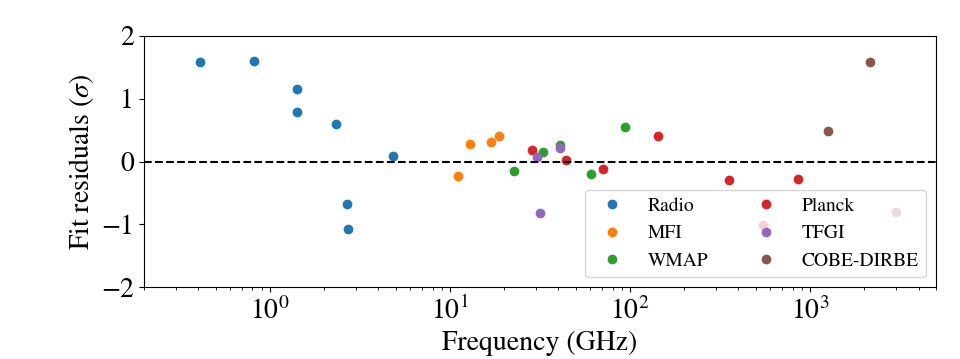}
        \includegraphics[width=0.49\linewidth, clip, trim=0 0 0 0.7cm]{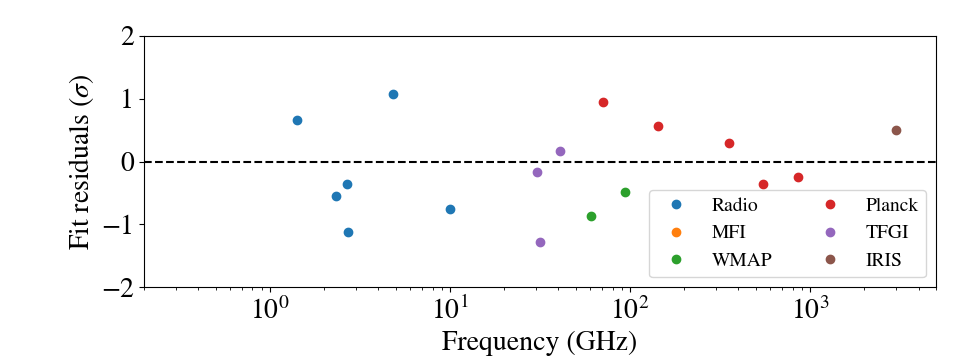}
        \caption{Left: intensity SED for W47 at 1$\degr$ resolution. Right: same SED, now at 30$^\prime$
        resolution.}
        \label{fig:w47_SED}
    \end{figure}

\begin{figure}
        \centering
        \includegraphics[width=0.48\linewidth, clip, trim=0 2cm 0 0]{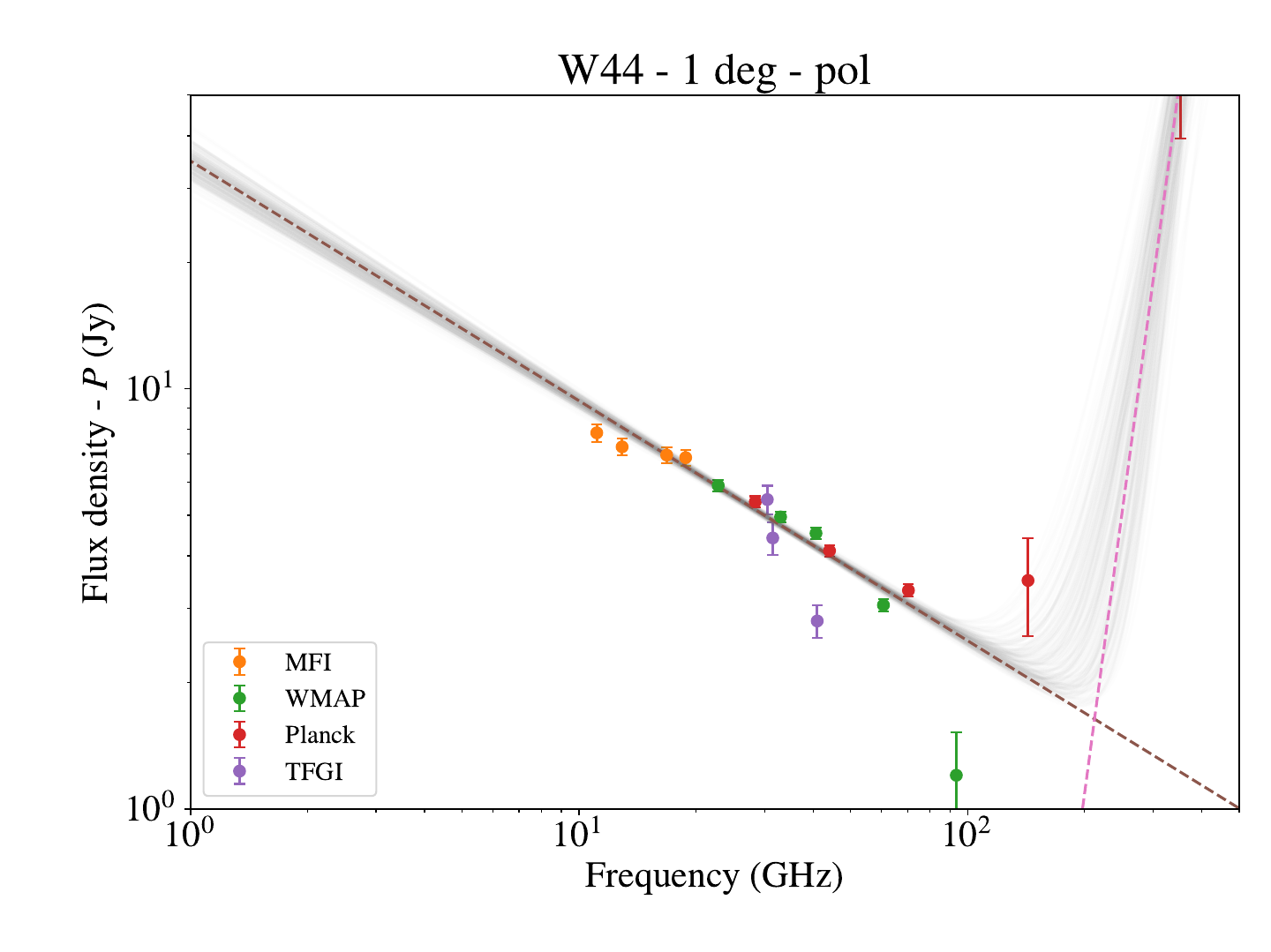}
        \includegraphics[width=0.48\linewidth, clip, trim=0 2cm 0 0]{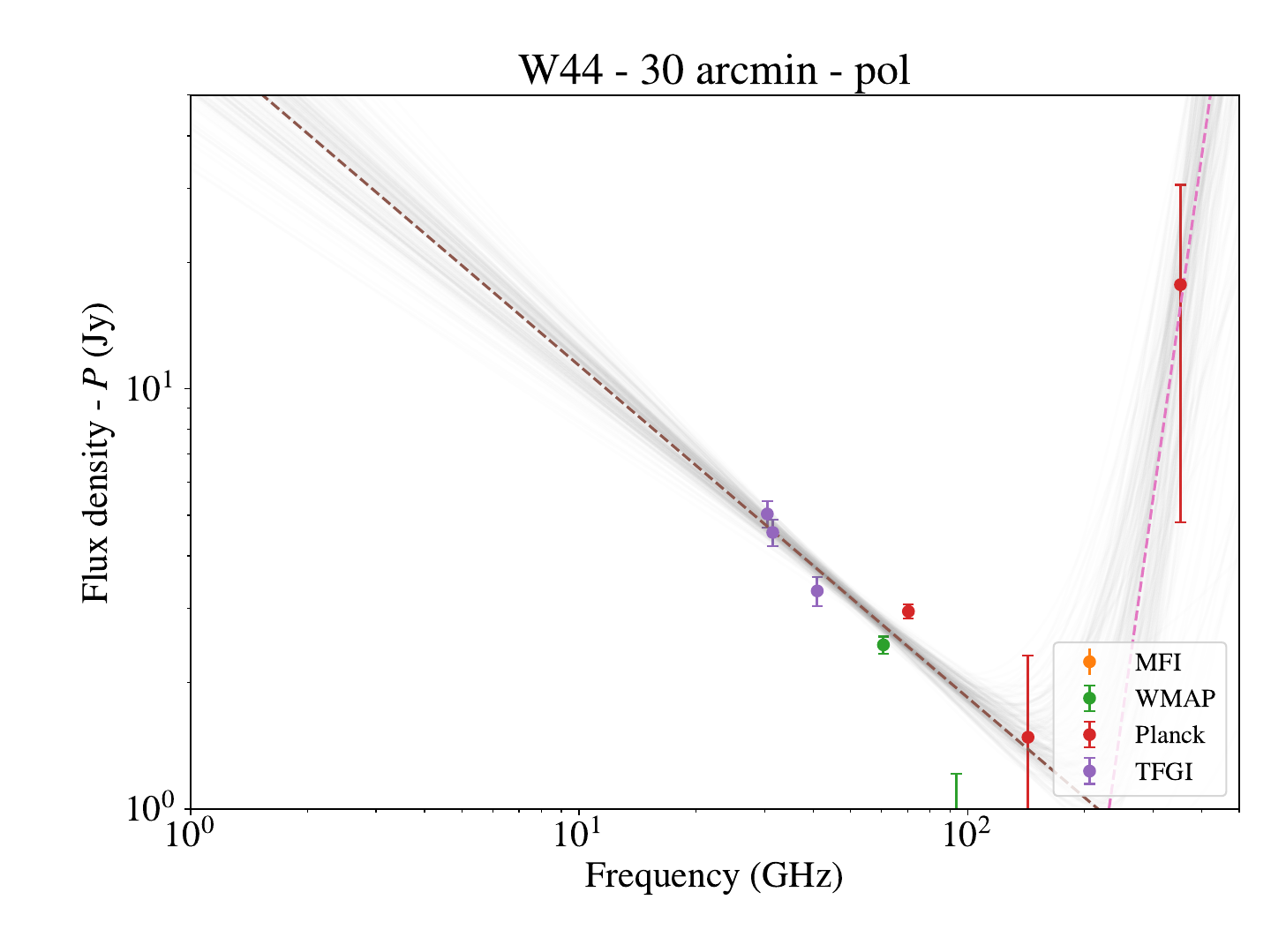}
        \includegraphics[width=0.48\linewidth, clip, trim=0 0 0 0cm]{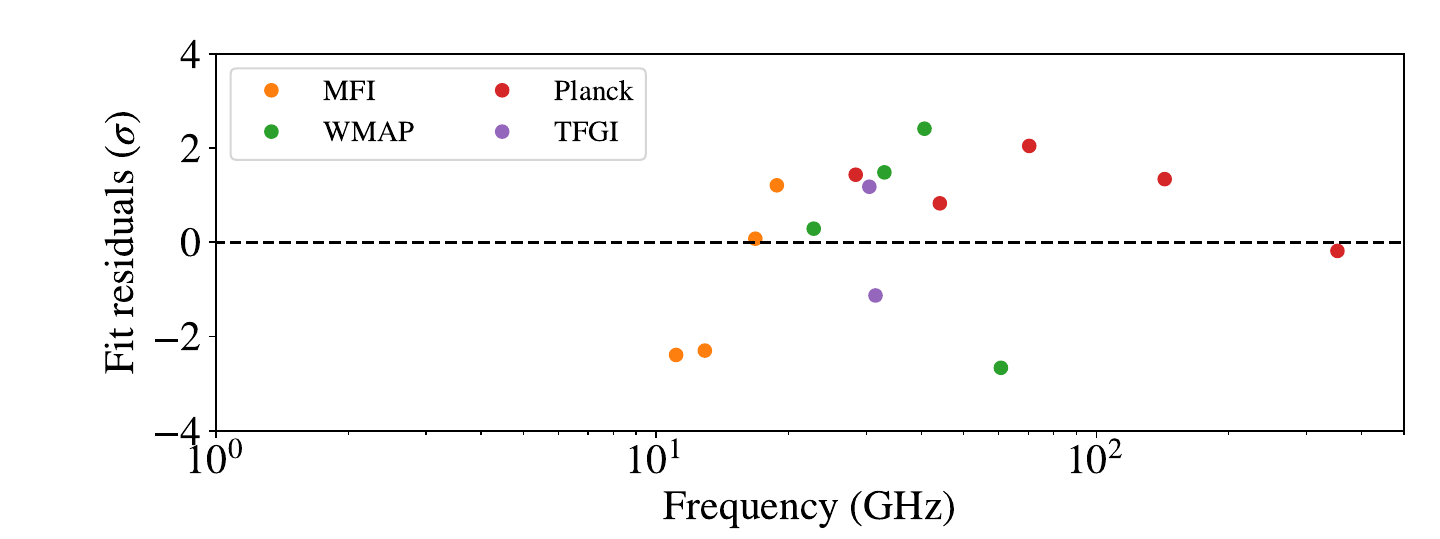}
        \includegraphics[width=0.48\linewidth, clip, trim=0 0 0 0cm]{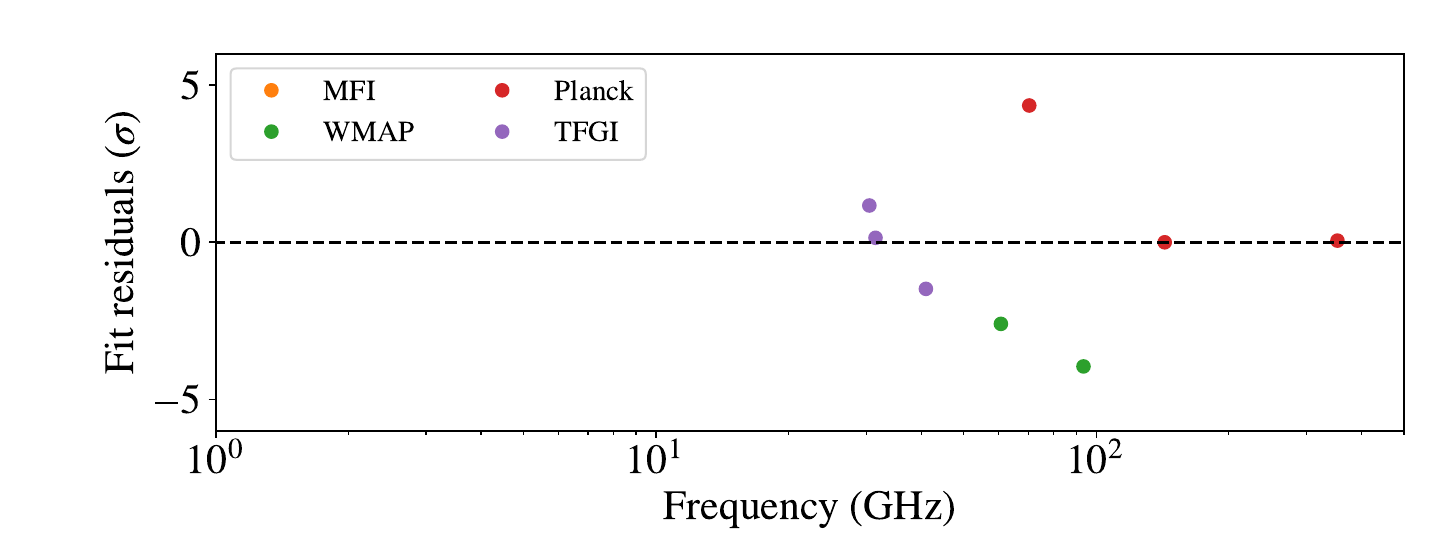}
        \caption{Polarized intensity (\textit{P}) SED for W44: left 1 degree, right 30 arcmin. Data below
        10\GHz{} were not used in any of the fits due to it being affected by Faraday rotation.
        The lack of data with resolution $\leq30^\prime$ implies an increased importance of FGI pixel 63 point in the modelling, driving the fit down.}
        \label{fig:w44_SED_P}
    \end{figure}

    In this Appendix we show the intensity SEDs from the unpolarized sources W43 and W47 in
    Figures~\ref{fig:w43_SED} and \ref{fig:w47_SED}, respectively, both at $30^\prime$ and
    $1\degr$ scales. We show the SED of polarized intensity from W44, also at $30^\prime$ and $1\degr$ scales, in Fig.~\ref{fig:w44_SED_P}. In all
    intensity cases the TFGI photometry yields consistent
    values with those expected from the rest of the surveys, and only FGI pixel 63 shows issues
    in polarization. TGI pixels have been offset with
    respect to their reference frequencies for visualization purposes. The best-fit values for the
    components (following the model from \citealt{ameplanewidesurvey}) considered are shown in
    Tables~\ref{table:w43_w44_w47_results} and \ref{table:w43_w44_w47_results_sm30arcmin}, at
    $1\degr$ and $30^\prime$ scales, respectively.

    % \begin{figure}
    %     \centering
    %     \includegraphics[width=0.49\linewidth, clip, trim=0 2.625cm 0 0]{figures/galactic_regions/W44_SED.pdf}
    %     \includegraphics[width=0.49\linewidth, clip, trim=0 2.625cm 0 0]{figures/galactic_regions/W44_SED_sm30arcmin_noAME.pdf}
    %     \includegraphics[width=0.49\linewidth, clip, trim=0 0 0 0.7cm]{figures/galactic_regions/W44_residuals_notitle.pdf}
    %     \includegraphics[width=0.49\linewidth, clip, trim=0 0 0 0.7cm]{figures/galactic_regions/W44_residuals_notitle_sm30arcmin_noAME.pdf}
    %     \caption{Left: intensity SED for W44 at 1$\degr$ resolution. Right: same SED, now at 
    %     30$^\prime$ resolution. The comparison between the two points to both AME and
    %     free-free being less important when the aperture is smaller. This is explained
    %     by the SNR being better resolved with respect to its surrounding region, so its
    %     synchrotron component is more dominant.}
    %     \label{fig:w44_SED}
    % \end{figure}
    
    \begin{table*}
        \centering
        \caption{Best fit parameters for the three Galactic regions considered (W43, W44 and W47)
        at $1\degr$ resolution. The W43 results are consistent with those from 
        \cite{perseus_w43_rhooph}, with differences arising mainly because of the C-BASS
        4.76\GHz{} data being used in their analysis. This addition decreases the amount
        of expected free-free in favour of AME. This is also one of the main implications
        of considering QUIJOTE-MFI (10--20\GHz{}) data in the fits, versus not doing so 
        (as was shown in \citealt{ameplanewidesurvey} for the Galactic plane).}
        \begin{tabular}{cccc}
\hline
            Parameters            &         W43          &         W44         &         W47         \\
\hline
%      $S_{\rm 1\,GHz}$ (Jy)       &         ---          & $299.149\pm68.425$  &         ---         \\
% $\beta_{\rm syn}$ (pc cm$^{-6}$) &         ---          &  $-0.442\pm0.104$   &         ---         \\
%       $EM$ (pc cm$^{-6}$)        & $4330.551\pm134.733$ & $884.640\pm498.589$ & $1733.550\pm66.797$ \\
%        $A_{\rm AME}$ (Jy)        &  $186.786\pm16.745$  &  $75.290\pm15.757$  &  $51.550\pm9.102$   \\
%      $\nu_{\rm AME}$ (GHz)       &   $26.021\pm1.372$   &  $22.355\pm1.921$   &  $21.343\pm1.701$   \\
%          $W_{\rm AME}$           &   $0.471\pm0.054$    &   $0.522\pm0.161$   &   $0.447\pm0.098$   \\
%     $\tau_{353}$ (10$^{-4}$)     &   $4.269\pm0.319$    &   $2.687\pm0.207$   &   $2.413\pm0.199$   \\
%            $\beta_d$             &   $1.853\pm0.097$    &   $1.771\pm0.114$   &   $1.814\pm0.114$   \\
%            $T_d$ (K)             &   $21.788\pm1.108$   &  $20.118\pm1.007$   &  $20.221\pm1.020$   \\
%  $\Delta  T_{\rm CMB}$ ($\mu$K)  & $333.977\pm164.495$  & $13.288\pm136.679$  &  $-9.734\pm92.412$  \\
      $S_{\rm 1\,GHz}$ (Jy)       &         ---          & $299\pm68$  &         ---         \\
 $\beta_{\rm syn}$ (pc cm$^{-6}$) &         ---          &  $-0.44\pm0.10$   &         ---         \\
       $EM$ (pc cm$^{-6}$)        & $4331\pm135$ & $885\pm499$ & $1734\pm67$ \\
        $A_{\rm AME}$ (Jy)        &  $187\pm17$  &  $75\pm16$  &  $51.6\pm9.1$   \\
      $\nu_{\rm AME}$ (GHz)       &   $26.0\pm1.4$   &  $22.4\pm1.9$   &  $21.3\pm1.7$   \\
          $W_{\rm AME}$           &   $0.471\pm0.054$    &   $0.52\pm0.16$   &   $0.447\pm0.098$   \\
     $\tau_{353}$ (10$^{-4}$)     &   $4.27\pm0.32$    &   $2.69\pm0.21$   &   $2.41\pm0.20$   \\
            $\beta_d$             &   $1.853\pm0.097$    &   $1.77\pm0.11$   &   $1.81\pm0.11$   \\
            $T_d$ (K)             &   $21.8\pm1.1$   &  $20.12\pm1.0$   &  $20.2\pm1.0$   \\
  $\Delta  T_{\rm CMB}$ ($\mu$K)  & $334\pm165$  & $13\pm137$  &  $-9\pm92$  \\
\hline
\end{tabular}
        \label{table:w43_w44_w47_results}
    \end{table*}
    
    \begin{table*}
        \centering
        \caption{Best fit parameters for the three Galactic regions considered (W43, W44 and W47)
        at $30^\prime$ resolution. The absence of data between 10--20\GHz{} (QUIJOTE-MFI)
        significantly impacts the recovery of the AME parameters, which are no longer well
        constrained.}
        \begin{tabular}{ccccc}
\hline
            \multirow{2}{*}{Parameters}            &          \multirow{2}{*}{W43}          &  \multicolumn{2}{c}{W44}                  &         \multirow{2}{*}{W47}          \\
 & & With AME & Without AME & \\
\hline
%      $S_{\rm 1\,GHz}$ (Jy)       &          ---          &    $199.225\pm37.800$     &  $155.310\pm26.058$  &         ---          \\
% $\beta_{\rm syn}$ (pc cm$^{-6}$) &          ---          &     $-0.723\pm0.204$      &   $-0.591\pm0.068$   &         ---          \\
%       $EM$ (pc cm$^{-6}$)        & $7262.918\pm1321.732$ &    $818.893\pm484.847$    & $1258.089\pm396.257$ & $2108.242\pm142.688$ \\
%        $A_{\rm AME}$ (Jy)        &   $46.849\pm33.078$   &     $20.757\pm24.284$     &         ---          &   $22.113\pm7.384$   \\
%      $\nu_{\rm AME}$ (GHz)       &   $35.449\pm13.637$   &     $15.130\pm13.182$     &         ---          &   $30.772\pm9.288$   \\
%          $W_{\rm AME}$           &    $0.898\pm0.550$    &      $0.873\pm0.410$      &         ---          &   $0.602\pm0.357$    \\
%     $\tau_{353}$ (10$^{-4}$)     &    $4.077\pm1.747$    &      $1.194\pm0.932$      &   $1.202\pm0.887$    &   $2.378\pm0.396$    \\
%            $\beta_d$             &    $1.691\pm0.397$    &      $1.623\pm0.639$      &   $1.741\pm0.520$    &   $1.726\pm0.255$    \\
%            $T_d$ (K)             &   $24.714\pm7.263$    &     $15.131\pm6.703$      &   $15.220\pm5.389$   &   $20.616\pm2.214$   \\
%  $\Delta  T_{\rm CMB}$ ($\mu$K)  &  $-85.390\pm449.400$  &   $-339.542\pm213.818$    & $-404.086\pm167.953$ & $-166.897\pm267.725$ \\
      $S_{\rm 1\,GHz}$ (Jy)       &          ---          &    $199\pm38$     &  $155\pm26$  &         ---          \\
 $\beta_{\rm syn}$ (pc cm$^{-6}$) &          ---          &     $-0.72\pm0.20$      &   $-0.591\pm0.068$   &         ---          \\
       $EM$ (pc cm$^{-6}$)        & $7263\pm1322$ &    $819\pm485$    & $1258\pm396$ & $2108\pm143$ \\
        $A_{\rm AME}$ (Jy)        &   $46\pm33$   &     $20\pm24$     &         ---          &   $22.1\pm7.4$   \\
      $\nu_{\rm AME}$ (GHz)       &   $35\pm14$   &     $15\pm13$     &         ---          &   $30.8\pm9.3$   \\
          $W_{\rm AME}$           &    $0.90\pm0.55$    &      $0.87\pm0.41$      &         ---          &   $0.60\pm0.36$    \\
     $\tau_{353}$ (10$^{-4}$)     &    $4.1\pm1.7$    &      $1.19\pm0.93$      &   $1.20\pm0.89$    &   $2.38\pm0.40$    \\
            $\beta_d$             &    $1.69\pm0.40$    &      $1.62\pm0.64$      &   $1.74\pm0.52$    &   $1.73\pm0.26$    \\
            $T_d$ (K)             &   $24.71\pm7.3$    &     $15.1\pm6.7$      &   $15.2\pm5.4$   &   $20.6\pm2.2$   \\
  $\Delta  T_{\rm CMB}$ ($\mu$K)  &  $-85\pm449$  &   $-339\pm213$    & $-404\pm167$ & $-167\pm268$ \\
\hline
\end{tabular}
        \label{table:w43_w44_w47_results_sm30arcmin}
    \end{table*}

\end{appendix}

%% This command is needed to show the entire author+affiliation list when
%% the collaboration and author truncation commands are used.  It has to
%% go at the end of the manuscript.
%\allauthors

%% Include this line if you are using the \added, \replaced, \deleted
%% commands to see a summary list of all changes at the end of the article.
%\listofchanges

\end{document}